\documentclass[journal,twocolumn]{IEEEtran}
\usepackage{color}
\usepackage{graphicx}
\usepackage{epstopdf}
\usepackage{amsmath}
\usepackage{amssymb}
\usepackage{hyperref}
\usepackage[english]{babel}
\usepackage{cite}
\usepackage{rotfloat}
\usepackage{mathtools}
\usepackage{amsmath}
\usepackage{makecell}
\usepackage{algorithm,algorithmic}
\usepackage{multirow}
\usepackage{subfigure}
\usepackage{booktabs}
\usepackage{colortbl}
\usepackage{multirow}
\usepackage{hhline}
\usepackage{stfloats}
\usepackage{multicol}
\usepackage{bbm}
\usepackage{cases}

\usepackage[font=small,skip=0pt]{caption}

\newcommand{\eqtext}[1]{\ensuremath{\stackrel{\text{#1}}{=}}}
\newcommand{\T}{{\scriptscriptstyle\mathsf{T}}}
\renewcommand{\H}{{\scriptscriptstyle\mathsf{H}}}

\newsavebox{\foobox}

\newcommand{\setA}{\mathcal{A}}

\definecolor{kugray5}{RGB}{224,224,224}

\usepackage[normalem]{ulem}
\newcommand\rsout{\bgroup\markoverwith
	{\textcolor{red}{\rule[0.5ex]{2pt}{0.8pt}}}\ULon}



\makeatletter
\newcommand{\ALOOP}[1]{\ALC@it\algorithmicloop\ #1%
	\begin{ALC@loop}}
	\newcommand{\ENDALOOP}{\end{ALC@loop}\ALC@it\algorithmicendloop}

\makeatother

\usepackage{etoolbox}
\let\mybibitem\bibitem
\renewcommand{\bibitem}[1]{%
	\ifstrequal{#1}{nature}
	{\color{blue}\mybibitem{#1}}
	{\color{black}\mybibitem{#1}}%
}

\graphicspath{ {Figures/} }

\newtheorem{remark}{Remark}
\newtheorem{lemma}{Lemma}

\DeclareCaptionLabelSeparator{periodspace}{.\quad}

\renewcommand{\figurename}{Fig.}
\addto\captionsenglish{\renewcommand{\figurename}{Fig.}}
\newcommand\nbthis{\addtocounter{equation}{1}\tag{\theequation}}

\newcommand{\norm}[1]{\left\lVert#1\right\rVert} 
\newcommand{\normshort}[1]{\lVert#1\rVert} 
\newcommand{\abs}[1]{\left|#1\right|} 
\newcommand{\absshort}[1]{\lvert#1\rvert} 

\newcommand{\tr}[1]{\mathrm{trace}\left(#1\right)} 
\newcommand{\diag}[1]{\mathrm{diag}\left\{#1\right\}} 
\newcommand{\diagshort}[1]{\mathrm{diag}\{#1\}} 
\newcommand{\blockdiag}[1]{\mathrm{blkdiag}\left\{#1\right\}} 

\newcommand{\re}[1]{\mathfrak{R}{\left(#1\right)}}

\newcommand{\reshort}[1]{\mathfrak{R}{(#1)}}

\newcommand{\mean}[1]{\mathbb{E} \left\{#1\right\}}
\newcommand{\meanshort}[1]{\mathbb{E} \{#1\}}

\newcommand{\mQ}{{\mathbf{Q}}}

\newcommand{\mH}{{\mathbf{H}}} 

\newcommand{\mA}{{\mathbf{A}}}

\newcommand{\mI}{{\mathbf{I}}}

\newcommand{\mC}{{\mathbf{C}}}
\newcommand{\mD}{{\mathbf{D}}}

\newcommand{\mG}{{\mathbf{G}}}

\newcommand{\setC}{\mathbb{C}} 
\newcommand{\setR}{\mathbb{R}}

\newcommand{\vc}{{\mathbf{c}}}

\newcommand{\vx}{{\mathbf{x}}}

\newcommand{\vr}{{\mathbf{r}}}

\newcommand{\vv}{{\mathbf{v}}}
\newcommand{\vn}{{\mathbf{n}}}
\newcommand{\vu}{{\mathbf{u}}}
 
\newcommand{\vh}{{\mathbf{h}}} 

\newcommand{\vq}{{\mathbf{q}}}

\newcommand{\vw}{{\mathbf{w}}}

\newcommand{\vd}{{\mathbf{d}}}

\newcommand{\vg}{{\mathbf{g}}}

\newcommand{\va}{\boldsymbol{\alpha}}
\newcommand{\vpsi}{\boldsymbol{\psi}}
\newcommand{\vphi}{\boldsymbol{\phi}}

\newcommand{\bPhi}{\boldsymbol{\Phi}}
\newcommand{\bUpsilon}{\boldsymbol{\Upsilon}}

\newcommand{\bPsi}{\boldsymbol{\Psi}}

\newcommand{\an}{\alpha_n}

\newcommand{\Na}{N_{\mathrm{a}}}

\newcommand{\hd}{\vh_{0,k}}

\newcommand{\hrt}{\mH_{1}} 
 
\newcommand{\hrr}{\vh_{2,k}} 
\newcommand{\gd}{\vg_{0,k}} 
\newcommand{\grt}{\mG_{1}} 
\newcommand{\grr}{\vg_{2,k}}

\newcommand{\ptmax}{p_{\mathrm{max}}^{\mathrm{uav}}} 
\newcommand{\prismax}{p_{\mathrm{max}}^{\mathrm{ris}}} 
\newcommand{\pris}{p^{\mathrm{ris}}} 
\newcommand{\puav}{p^{\mathrm{uav}}}

\newcommand{\bBt}{\{b_k[t]\}} 
\newcommand{\bVt}{\{ \vv[t] \}} 
 
\newcommand{\bat}{\{\an[t]\}}
\newcommand{\bXit}{\boldsymbol{\Xi}[t]}

\newcommand{\bV}{\vv} 
 
\newcommand{\ba}{\{\an\}}
\newcommand{\bXi}{\boldsymbol{\Xi}} 
\newcommand{\bW}{\{\vw_k\}} 
\newcommand{\bWt}{\{\vw[t]\}}

\usepackage{setspace}

\begin{document}

	\title{Fairness Enhancement of UAV Systems with Hybrid Active-Passive RIS}
	\author{
		Nhan~Thanh~Nguyen, \IEEEmembership{Member, IEEE}, Van-Dinh~Nguyen, \IEEEmembership{Senior Member, IEEE}, Hieu~Van~Nguyen, \IEEEmembership{Member, IEEE}, Qingqing Wu, \IEEEmembership{Senior Member, IEEE}, Antti~Tölli, \IEEEmembership{Senior Member, IEEE},\\ Symeon~Chatzinotas, \IEEEmembership{Fellow, IEEE}, and Markku~Juntti, \IEEEmembership{Fellow, IEEE}
		\thanks{This research was supported by Academy of Finland under 6G Flagship (grant 346208), EERA Project (grant 332362), Infotech Oulu under EEWA Project, and Research Council of Finland Fellowship (grant 354901). The work of V.-D. Nguyen  was supported in part by the VinUniversity Seed Grant Program. The work of Q. Wu was supported by NSFC 62371289, NSFC 62331022 Guangdong science and technology program under grant 2022A0505050011 and FDCT under Grant 0119/2020/A3. The work of S. Chatzinotas was supported by the Luxembourg National Research Fund via project 5G-SKY, ref. FNR/C19/IS/13713801/5G-Sky and project RISOTTI, ref. FNR/C20/IS/14773976/RISOTTI. A short version of this paper was presented in the IEEE International Conference on Communications (ICC), 2022, Seoul, South Korea. \textit{Corresponding author: Nhan Thanh Nguyen.}}
		\thanks{N. T. Nguyen, A. Tölli,  and M. Juntti are with Centre for Wireless Communications, University of Oulu, P.O.Box 4500, FI-90014, Finland (e-mail:\{nhan.nguyen, antti.tolli, markku.juntti\}@oulu.fi).}
  \thanks{V.-D. Nguyen is with the College of Engineering and Computer Science and also with the Center for Environmental Intelligence, VinUniversity, Vinhomes Ocean Park, Hanoi 100000, Vietnam (e-mail: dinh.nv2@vinuni.edu.vn).}
  \thanks{H. V. Nguyen is with the Faculty of Electronic and Telecommunication Engineering, The University of Danang, University of Science and Technology, Da Nang 50000, Vietnam (email: nvhieu@dut.udn.vn).}
  \thanks{Q. Wu is with the Department of Electronic Engineering, Shanghai Jiao Tong University, 200240, China (e-mail: qingqingwu@sjtu.edu.cn).}
  \thanks{S. Chatzinotas is with the Interdisciplinary Centre for Security, Reliability and Trust (SnT), University of Luxembourg, L-1855 Luxembourg, (email: symeon.chatzinotas@uni.lu).}
  }
	
	\maketitle
	
	\begin{abstract}
		
		We consider unmanned aerial vehicle (UAV)-enabled wireless systems where downlink communications between a multi-antenna UAV and multiple users are assisted by a hybrid active-passive reconfigurable intelligent surface (RIS). We aim at a fairness design of two typical UAV-enabled networks, namely the static-UAV network where the UAV is deployed at a fixed location to serve all users at the same time, and the mobile-UAV network which employs the time division multiple access protocol. In both networks, our goal is to maximize the minimum rate among users through jointly optimizing the UAV's location/trajectory, transmit beamformer, and RIS coefficients. The resulting problems are highly nonconvex due to a strong coupling between the involved variables. We develop efficient algorithms based on block coordinate ascend and successive convex approximation to effectively solve these problems in an iterative manner. {In particular, in the optimization of the mobile-UAV network, closed-form solutions to the transmit beamformer and RIS passive coefficients are derived.} Numerical results show that a hybrid RIS equipped with only $4$ active elements and a power budget of $0$ dBm offers an improvement of $38\%-63\%$ in minimum rate, while that achieved by a passive RIS is only about $15\%$, with the same total number of elements.
		
	\end{abstract}

	\begin{IEEEkeywords}
		Beamforming, hybrid active-passive RIS, UAV-enabled communications, trajectory design, successive convex approximation
	\end{IEEEkeywords}
	\IEEEpeerreviewmaketitle
	
	\section{Introduction}
	Unmanned aerial vehicle (UAV)-enabled communications systems have attracted growing interest in both the industry and research community thanks to their controllable mobility and cost-effectiveness \cite{wu2018joint}. However, the blockage of line-of-sight (LoS) links and severe path loss caused by long-distance transmissions and obstacles, especially in complex urban environments, may cause significant system performance degradation  \cite{cao2021reconfigurable}. A recently emerging technology called reconfigurable intelligent surfaces (RISs) can be deployed to reflect signals to favorable directions and offer passive beamforming gains to greatly improve system performance \cite{yang2020intelligent, QingqingTCOM20, Huang2018, munochiveyi2021reconfigurable, pogaku2022uav, tyrovolas2021performance, huang2020holographic}. However, the main barrier to deploying conventional RISs is that its pure passive reflecting gain may be insufficient to overcome the double path loss in reflecting channels \cite{wu2019intelligent}. This can happen in harsh environments such as when the UAV flies at a high altitude, or when multiple users are distributed at the edge of the network. A recently introduced variant of the passive RIS called the hybrid active-passive RIS, where a few elements are activated to amplify incident signals  \cite{nguyen2021hybrid}, can be a promising solution to overcome the aforementioned challenges.

	\subsection{Related Works}
	
	RIS-aided UAV communications have been widely studied recently \cite{cao2021reconfigurable, li2020reconfigurable, li2021robust, li2020sum, jiang2021reconfigurable, guo2021learning, diamanti2021energy, pan2021uav, nguyen2021reconfigurable, wang2021passive}. Specifically, Cao \textit{et al.} \cite{cao2021reconfigurable} designed an adaptive RIS-aided transmission protocol to maximize the system throughput via both numerical method and machine learning. In  \cite{li2020reconfigurable} and \cite{li2020sum}, a joint design of the UAV's trajectory and RIS passive beamforming was proposed to maximize the system throughput. The works in \cite{li2021reconfigurable,  li2021robust, guo2021learning} focused on designing the secure communication protocol for UAV systems with the presence of eavesdroppers. More specifically, Li \textit{et al.} \cite{li2021reconfigurable} developed an efficient algorithm based on alternating optimization (AO) to design a robust solution for the UAV's trajectory, RIS's passive beamforming, and transmit power of the legitimate transmitters under imperfect channel state information (CSI). On the other hand, Guo \textit{et al.} \cite{guo2021learning} proposed effective solutions to the UAV's trajectory, UAV's beamformer, and RIS coefficients based on deep reinforcement learning. A similar design was considered in \cite{li2021robust} for a complex urban environment where ideal LoS links are unavailable in the air-ground channel. Moreover, the rate performance of non-orthogonal multiple access \cite{liu2020machine,mu2021intelligent}, ultra-reliable and low-latency \cite{ranjha2020urllc}, millimeter-wave \cite{jiang2021reconfigurable}, and THz \cite{pan2021uav} RIS-aided UAV systems were also investigated.
	
	The deployment of low-power consumption RISs also brings benefits in terms of power consumption and energy efficiency to the system \cite{liu2020machine, nguyen2021reconfigurable, diamanti2021energy, jeong2020simultaneous, diamanti2021prospect}. In particular, Jeon \textit{et al.} \cite{jeong2020simultaneous} considered installing the RIS on an aerial platform to exploit rich LoS communications for an energy-efficient aerial backhaul system. Similarly, Diamanti \textit{et al.} \cite{diamanti2021prospect} considered the RIS-aided UAV-enabled integrated access and backhaul network targeting at energy-efficient operations, whereas the work \cite{samir2021optimizing} focused on minimizing the age-of-information to maintain the freshness of reflected information. In the aforementioned studies, the common task is to jointly design the UAV's trajectory/location and transmit beamforming/power allocation as well as the RIS reflecting coefficients. Due to highly coupled design variables and unit modules of RIS coefficients, the formulated problems are often solved via the block coordinate descent/ascent (BCD/BCA) and successive approximation (SCA) \cite{li2020sum, li2021reconfigurable, mu2021intelligent, ranjha2020urllc, li2021robust}. {Furthermore, leveraging deep neural networks \cite{cao2021reconfigurable, nguyen2021reconfigurable, guo2021learning, liu2020machine, yang2020federated} and deep reinforcement learning \cite{nguyen2021reconfigurable, huang2020reconfigurable, huang2021multi} can offer significant performance improvement and complexity reduction.}
	
	It is noted that all the above works considered passive RISs, which only provide passive reflecting gains. Recently, the hybrid active-passive RIS architecture has been introduced \cite{taha2019deep, alexandropoulos2020hardware, nguyen2021spectral} to overcome the inherent limitations of passive RISs, especially in harsh transmission scenarios such as in low signal-to-noise ratio (SNR) regime and/or the severe path loss in reflecting channels. The key idea of the hybrid RIS is to employ a few active elements to enable both reflecting and amplifying gains simultaneously at the RIS. As a result, the hybrid RIS is capable of mitigating the effects of the double path loss and significantly improving the system performance in terms of spectral efficiency (SE) \cite{nguyen2021spectral, nguyen2022downlink, nguyen2022spectral_cfmimo, nguyen2021hybrid_mag, shojaeifard2022mimo, nguyen2022hybrid, nguyen2022hybrid_UAV}, secrecy rate \cite{9598322, egashira2022secrecy}, harvested energy \cite{9653007}, and reliability \cite{yigit2021hybrid}. Moreover, active elements with RF chains in \cite{taha2019deep, alexandropoulos2020hardware, schroeder2020passive} can be leveraged for processing of incident signals and channel estimation at RISs.
	
	{The above discussed advantages of hybrid RISs can also be reaped by fully active RISs \cite{long2021active, khoshafa2021active, huang2020holographic}, which, however, require higher power consumption, hardware, and computational cost. Furthermore, based on the recent studies on the performance gains of the hybrid and active RISs \cite{nguyen2021hybrid, nguyen2021spectral, nguyen2021hybrid_mag, long2021active}, a small number of active elements are sufficient to ensure satisfactory improvement in the SE performance, and excessive use of active elements can cause a significant loss in the SE and EE when the active power budget is limited. Therefore, hybrid RISs can offer a good SE--EE tradeoff while maintaining reasonable hardware costs. It is noted that the hybrid RIS can serve as the conventional passive or active RISs when its number of active elements is set to zero or the number of available RIS elements, respectively.}

	\subsection{Contributions}
	In most of the previous works, RIS is often deployed on the building facade, which is in the vicinity of either BS or user equipments (UEs) to exploit LoS communications \cite{wu2019towards, wu2019intelligent, zhang2020capacity}. However, in multi-user air-ground communications systems, this favorable deployment becomes challenging and impractical. On one hand, it is difficult to guarantee the short distance between the RIS and the UAV due to the UAV's high mobility and altitude. On the other hand, when multiple UEs are distributed in a large area, the LoS channels from the RIS to all UEs cannot be guaranteed, especially in complex urban areas \cite{li2021robust}. Therefore, the performance gain offered by the passive RIS may be severely limited in UAV-enabled networks. To overcome the aforementioned challenges and motivated by the potential performance improvement of the hybrid active-passive RIS, we consider in this work hybrid RIS-aided UAV communications systems. The hybrid RIS is practically realized by a recently introduced technology called reflection amplifier \cite{landsberg2017low}. The deployment of the hybrid RIS benefits the UAV systems in the following aspects. First, without requiring a large number of elements, a single hybrid RIS with amplifying gains can efficiently compensate for the severe path loss on the reflecting channels, resulting in a significant throughout improvement. Secondly, the hybrid RIS offers a great performance gain with respect to conventional systems (without RIS or with passive RIS) in low and moderate power regimes \cite{nguyen2021hybrid}. This guarantees improved system sum throughput and max-min fairness, especially for the UAV with limited battery capacity.

	To reap the above-mentioned benefits of the hybrid RIS in UAV air-ground downlink communications, this work considers two typical UAV-enabled systems. In the first one, a tethered UAV is placed at a fixed location and serves all UEs at the same time. In the second scenario, the UAV employs the time division multiple access (TDMA) protocol to serve each UE at a time slot of a certain time period. Although the use  of TDMA may cause additional latency, it allows to fully exploit the high and controllable mobility of the UAV. Specifically, by optimizing the trajectory and user scheduling, the communication distance between the UAV and ground users can be signiﬁcantly shortened, enhancing the air-ground communications performance \cite{wu2018joint, QingqingJSAC18}. In both scenarios, we focus on the fairness design to maximize the minimum rate among all UEs. Our specific \textbf{contributions} are summarized as follows:
	\begin{itemize}
		\item We propose the deployments of the hybrid active-passive RIS to assist the air-ground communications in the two typical UAV-enabled networks mentioned above. These have the potential in mitigating the limitations of the conventional systems, i.e., the blockage of LoS links in UAV systems without RIS and the double path loss in reflecting channels with the passive RIS. The resultant signal models are significantly different from that of the conventional passive RIS. Specifically, the signals received at UEs in hybrid RIS-aided systems include not only the signal transmitted from the UAV, but also the noise and self-interference (SI) amplified from the RIS's active elements. Furthermore, we consider the case that the UAV is equipped with multiple antennas to further enhance the system performance.
		
		\item We formulate two max-min rate problems in two considered networks, where the  UAV's location or trajectory, transmit beamformers, and RIS amplifying/reflecting coefficients are jointly optimized to guarantee the system fairness. Although the formulated problems have the same fairness objective, their solutions are significantly different due to the different system models. Both problems are practical and appealing but have not been considered in the literature, according to the authors' best knowledge. In particular, they are more challenging compared to those with conventional passive RISs due to additional constraints on the power consumption of active elements and the introduction of the amplified noise/SI at these elements. Combining tools from BCA and SCA methods, we first decompose the original problems into subproblems and derive efficient approximation functions to tackle the nonconvex constraints. Two efficient iterative algorithms are then developed to obtain the locally optimal solutions to the formulated problems.
		
		\item Finally, we provide extensive simulation results to validate the effectiveness of the proposed algorithms as well as the performance of the hybrid RIS. They reveal that the hybrid RIS offers remarkable performance improvement compared to conventional systems without RIS and with passive RIS, especially in extreme conditions such as low SNR regime, severe path loss in reflecting channels, and limited power budget at the UAV. The results also show that the full potential of the hybrid RIS can be achieved by employing the TDMA protocol.
	\end{itemize}
	
	\subsection{Organization and Notations}
	
	The rest of the paper is organized as follows. In Section \ref{sec_system_model}, we present the system model and formulate the max-min rate problems. The solutions of these problems are given in Sections \ref{sec_loc_opt} and \ref{sec_tra_opt}. Numerical results are provided in Section \ref{sec_sim_result}. Finally, Section \ref{sec_conclusion} concludes the paper.
	
	\textit{Notations}: Throughout this paper, numbers, vectors, and matrices are denoted by lower-case, bold-face lower-case, and bold-face upper-case letters, respectively. Furthermore, $(\cdot)^*$, $(\cdot)^\T$, and $(\cdot)^\H$ denote the conjugate of a complex number, the transpose and the conjugate transpose of a matrix or vector, respectively. Matrix $\mI_N$ is the identity matrix of size $N \times N$, $\mathrm {diag} \{ a_1, \ldots, a_N \}$ represents a diagonal matrix with diagonal entries $a_1, \ldots, a_N$, whereas $\diag{\mA}$ returns the diagonal elements of matrix $\mA$. In addition, $\abs{\cdot}$ denotes the absolute value of a scalar, $\norm{\cdot}_{\mathcal{F}}$ denotes the Frobenius norm of a matrix, $\angle x$ represents the phase of complex number $x$, and $\mathrm{sgn}(x)$ returns the sign of real number $x$.  Finally, $\circ$ represents a Hadamard product.
	
	\section{System Model and Problem Formulations}\label{sec_system_model}
	
	We consider a UAV-enabled wireless network where a UAV serves $K$ single-antenna UEs, and the air-ground communication is assisted by a hybrid active-passive RIS installed on the building facade at a certain altitude. The UAV is equipped with $N_t$ antennas and is assumed to fly at a fixed altitude  of $z_0$ m. Let $\mathcal{K} \triangleq \{1,2,\ldots,K\}$ denote the set of UEs.
	
	\subsection{Hybrid RIS and Channel Model}
	\subsubsection{Hybrid RIS}
	The RIS is equipped with $N$ elements, among which $\Na$ elements are active. To guarantee a minimal increase in power consumption and hardware cost of the hybrid RIS,  a few active elements are employed. Let $\setA \subset \{ 1,2,\ldots,\Na\}$ be the set of the RIS active elements with $\abs{\setA}=\Na$. The RIS active elements can potentially be realized by low-power reflection amplifiers \cite{landsberg2017low}. We refer readers to \cite{long2021active, nguyen2021hybrid, landsberg2017low} for more details on the reflection amplifier-based active RIS. It is seen that the fully passive RIS (i.e. with $\Na=0$) is just a special case of the hybrid RIS. Therefore,  we will use the general term ``\textit{RIS}" for the discussion in the system model, while specific terms ``\textit{passive RIS}" or ``\textit{hybrid RIS}" are used for comparisons. 
	
	{Let $\an$ denote the coefficient associated with the $n$th element of the RIS. It can be expressed as $\an = \abs{\an} e^{j \theta_n}$, where $\theta_n$ and $\abs{\an}$ represent the phase shift and amplitude of the $n$-th coefficient of the RIS, respectively, with $\theta_n \in [0, 2\pi)$ \cite{wu2019towards}, $\abs{\an} \leq 1, n \notin \setA$, and $\abs{\an} \leq a_{\mathrm{max}}$ for $n \in \setA$.} Here, $a_{\mathrm{max}}$ is the maximum gain that an active load can provide, which is up to $40$ dB if active elements are realized by reflection amplifiers \cite{long2021active, landsberg2017low}. We note here that to achieve interference cancelation in multi-user systems, the reflection amplitude of the RIS passive elements may not necessarily be unity \cite{wu2019towards}. Let  $\bUpsilon = \diag{\alpha_1, \ldots, \alpha_{N}} \in \setC^{N \times N}$ be the diagonal matrix of the RIS coefficients. We define an additive decomposition for it as $\bUpsilon = \bPhi + \bPsi$, where $\bPsi = \mathbbm{1}^{\setA}_N \circ \bUpsilon$ and $\bPhi = \left(\mI_{N} - \mathbbm{1}^{\setA}_N\right) \circ \bUpsilon$ contain the active and passive coefficients, respectively. Here, $\mathbbm{1}^{\setA}_N$ is an $N \times N$ diagonal matrix whose non-zero elements are all unity and have positions determined by $\setA$.
	
	\subsubsection{Channel Model}
	Let $\vv$, $\vr$, and $\vu_k$ denote the locations of  UAV,  RIS, and UE $k$, respectively. We denote by $\hd \in \setC^{N_t\times 1}$, $\hrt \in \setC^{N \times N_t}$ and $\hrr \in \setC^{N\times 1}$  the channels between  UAV and UE $k$, between  UAV and  RIS, and between  RIS and UE $k$, respectively. {Furthermore, we denote by $\{ \gd^\H, \grt, \grr^\H \}$ the corresponding small-scale fading channels of $\{ \hd^\H, \hrt, \hrr^\H \}$. As a result, we can write \cite{li2021reconfigurable, cao2021reconfigurable}
    \begin{subequations}
        \label{eq_h2}
        \begin{align*}
		\hd^\H &= \zeta_0^{\frac{1}{2}} \norm{\vv - \vu_k}^{-\frac{\epsilon_0}{2}} \gd^\H,  \nbthis \label{eq_h0} \\
		\hrt &= \zeta_0^{\frac{1}{2}} \norm{\vv - \vr}^{-\frac{\epsilon_1}{2}} \grt,  \nbthis \label{eq_h1} \\
		\hrr^\H &= \zeta_0^{\frac{1}{2}} \norm{\vr - \vu_k}^{-\frac{\epsilon_2}{2}} \grr^\H, \nbthis \label{eq_h2k}
	\end{align*}
    \end{subequations}
    where $\zeta_0$ is the path loss at the reference distance of $1$ m, and $\{\epsilon_0, \epsilon_1, \epsilon_2\}$ are the corresponding path loss exponents. The effective channel between the UAV and UE $k$ can be written as
	\begin{align*}
		\vh_{k}^\H = \hd^\H + \hrr^\H \bUpsilon \hrt. \nbthis \label{eq_h_eff}
	\end{align*}}
	
	\subsection{Signal Models}
	
	\begin{figure}[t]
		\centering
		\subfigure[Static UAV]
		{
			\includegraphics[scale=0.66]{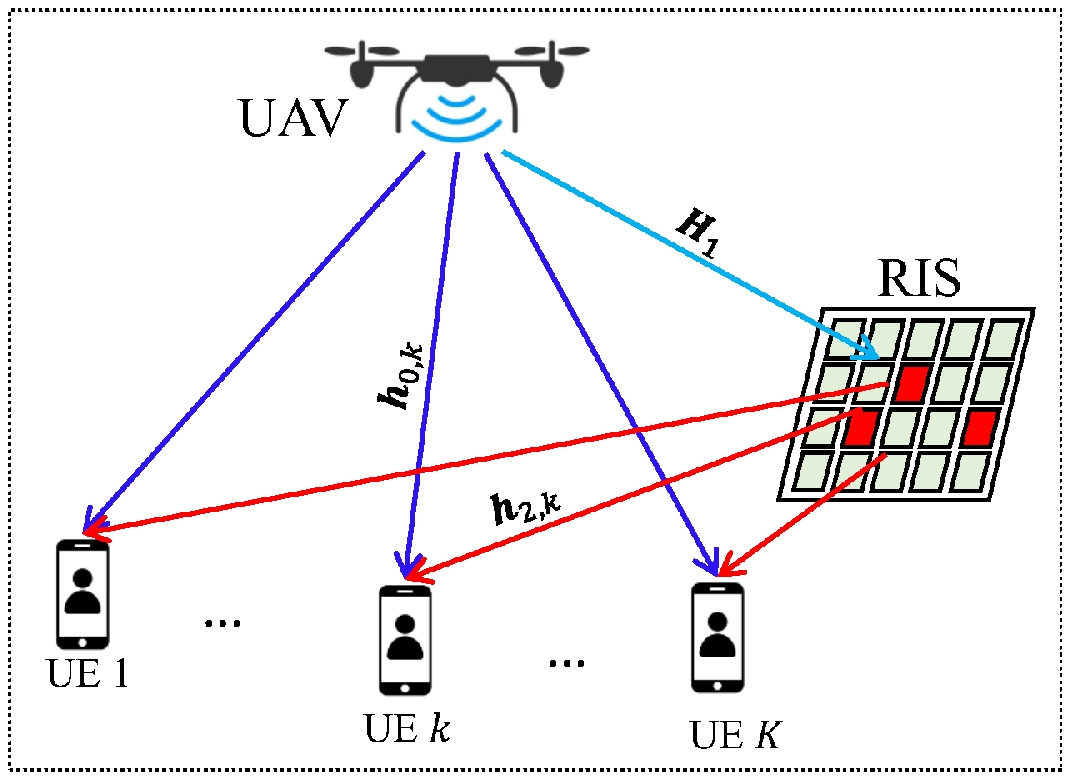}
			\label{fig_SM_loc}
		}
		\hspace{0.5cm}
		\subfigure[Mobile UAV]
		{
			\includegraphics[scale=0.66]{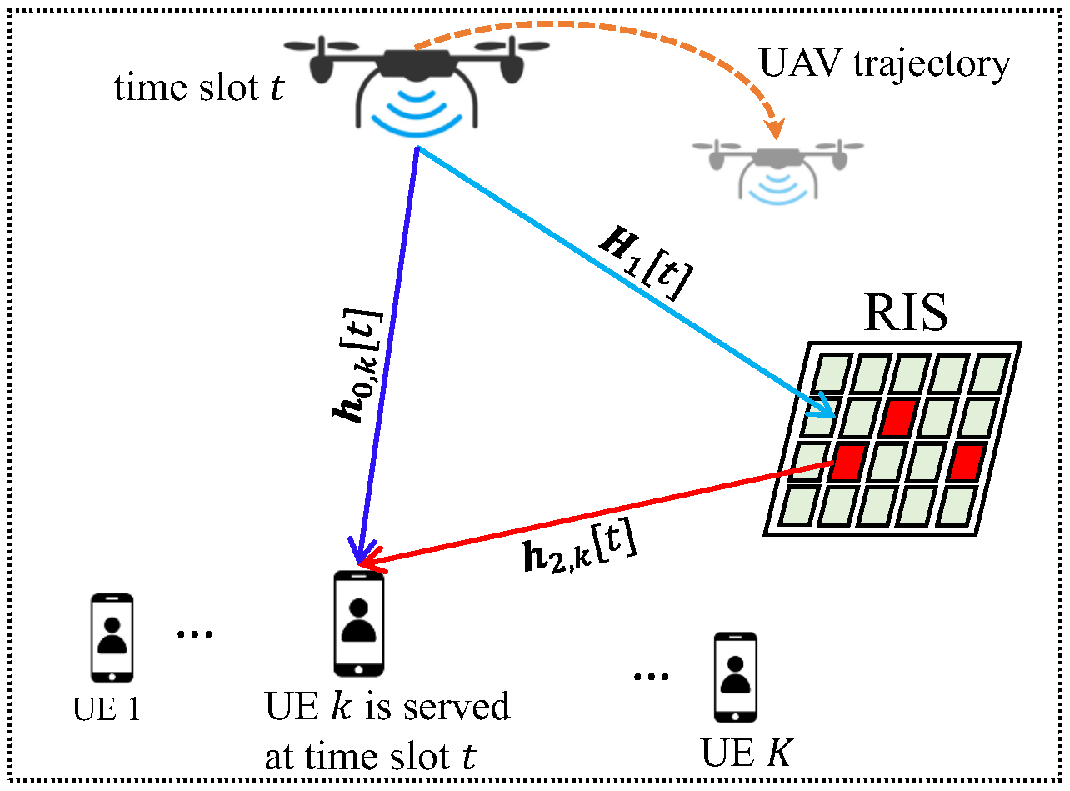}
			\label{fig_SM_tra}
		}
		\caption{(a) The UAV is deployed at a fixed location and serves all UEs at the same time. (b) The UAV flies along a trajectory and employs TDMA to serve each UE at a time slot.}
		\label{fig_system_model}
	\end{figure}
	
	In this work, we consider two typical scenarios of UAV-enabled communications systems, as illustrated in Fig.\ \ref{fig_system_model}. In  Fig.\ \ref{fig_SM_loc}, the UAV is assumed to be deployed at a fixed position and serves all  UEs at the same time \cite{mu2021intelligent}. In contrast, in  Fig.\ \ref{fig_SM_tra}, the UAV employs the TDMA transmission protocol and flies over consecutive time slots to communicate with at most one user at each time slot \cite{wu2018joint}. The signal models of these considered systems are explained in detail in the following.
	
	\subsubsection{Static UAV}
	Denote by $s_k$ and $\vw_k \in \setC^{N_t \times 1}$ the transmitted symbol and the beamforming vector intended for UE $k$, with $\meanshort{\abs{s_{k}}^2}=1$. The transmitted signal from the UAV can be expressed as $\vx = \sum_{k=1}^{K} \vw_k s_{k} \in \setC^{N_t \times 1}$. The UAV's transmit power constraint as $\puav = \sum_{k=1}^{K} \norm{\vw_k}^2 \leq \ptmax$, where $\ptmax$ is the UAV's power budget. The received signal at UE $k$ can be expressed as
	\begin{align*}
		y_k = \underbrace{\vh_{k}^\H \vw_k s_{k}}_{\text{desired signal}} + \underbrace{\sum_{j \neq k}^K \vh_k^\H \vw_j s_j}_{\text{interference}} + \underbrace{\hrr^\H \bPsi \vn_{\mathrm{r}}}_{\text{RIS effective noise}} + \underbrace{n_{\mathrm{u}}}_{\text{noise at UE}~k}, \nbthis \label{eq_signal_model_loc}
	\end{align*}
	where $n_{\mathrm{u}} \sim \mathcal{CN}(0,\sigma_{\mathrm{u}}^2)$ and $\vn_{\mathrm{r}} \sim \mathcal{CN} (\boldsymbol{0}, \mathbbm{1}^{\setA}_N \circ \sigma_{\mathrm{r}}^2  \mI_N)$ are the additive white Gaussian noise (AWGN) at UE $k$ and the total effective noise including the AWGN noise and residual SI caused by the RIS active elements, respectively. This SI is due to the dull-duplex operation of the RIS's active elements. Thus,  the term $\hrr^\H \bPsi \vn_{\mathrm{r}} + n_{\mathrm{u}}$ in  \eqref{eq_signal_model_loc} is the aggregated noise at UE $k$. Note that there are numerous sources of imperfections in active elements, and it has been shown in \cite{bharadia2014full} that the residual SI can be eliminated to be as low as 1 dB over the noise floor independent of the transmit power and the number of array elements. Therefore, we adopt the AWGN to model the RIS residual SI which is proportional to the noise power, as in \cite{malik2018optimal, nguyen2021hybrid}.
	
	\subsubsection{Mobile UAV}
	In this scenario, the UAV flies within a given period $\Delta t$, which is divided into $T$ consecutive equal time slots $\{1,\ldots,T\}$, each is of length $\delta_t = \frac{\Delta t}{T}$. Here, $\delta_t$ is assumed to be sufficiently small for which a UAV's location approximately remain unchanged even when it flies at a maximum speed $v_{\mathrm{max}}$ \cite{wu2018joint}. The trajectory of the UAV can be denoted as a sequence $\{ \vv[t] \}_{t=1}^T$, where $\vv[t] = [x_{\mathrm{v}}[t], y_{\mathrm{v}}[t], z_0], t = 1,\ldots,T,$ represents the location of the UAV at time slot $t$. Similarly, time index $[t]$ will be added to the system model to appropriately reflect the TDMA transmission. Specifically, in this scenario, $s_k[t]$ represents the transmit symbol intended for UE $k$; $\{ \hd^\H[t], \hrt[t], \hrr^\H \}$ represent the UAV-UE $k$, UAV-RIS, and RIS-UE $k$ channels, associated with small-scale fading channels $\{ \gd^\H[t], \grt[t], \grr^\H \}$, respectively; and $\{\bUpsilon[t], \bPhi[t], \bPsi[t]\}$ represent the matrices of reflecting/amplifying coefficients $\an[t] = \abs{\an[t]} e^{j \theta_n[t]}$ of the RIS at time slot $t$. We note that the channels between the RIS and UE $k$, i.e., $\hrr^\H$ and $\grr^\H$,  remains unchanged over time slots because the RIS is deployed at a fixed location, while the UEs are assumed to have slow mobility. As a result, the effective channel at time slot $t$ is expressed as
	\begin{align*}
		\vh_{k}^\H[t] = \hd^\H[t] + \hrr^\H \bUpsilon[t] \hrt[t]. \nbthis \label{eq_eff_chan_tra}
	\end{align*}
	
	Let $b_k[t] \in \{0,1\}$ indicate whether or not UE $k$ is scheduled for transmission at time slot $t$, i.e. $b_k[t] = 1$ implies that UE $k$ is served at time slot $t$; otherwise, $b_k[t] = 0$. We assume that at each time slot, the UAV serves at most one UE; thus, $\sum\nolimits_{k=1}^K b_k[t] \leq 1, \forall t$. If UE $k$ is scheduled for transmission at time slot $t$, its received signal can be modeled as
	\begin{align*}
		y_k[t] &= \vh_k^\H[t] \vw[t] s_k[t] + \left(\hrr^\H[t] \bPsi[t] \vn_{\mathrm{r}} + n_{\mathrm{u}}\right), \nbthis \label{eq_signal_model_tra}
	\end{align*}
	where $\vw[t]$ is the beamforming vector at the UAV subject to $\norm{\vw[t]}^2 \leq \ptmax, \forall t$, and $\hrr^\H[t] \bPsi[t] \vn_{\mathrm{r}} + n_{\mathrm{u}}$ is the aggregated noise at UE $k$th at time slot $t$, with $\vn_{\mathrm{r}}$ and $n_{\mathrm{u}}$ having the same distributions as those in \eqref{eq_signal_model_loc}. Here, without loss of generality, we assume the same AWGN noise $\vn_{\mathrm{r}}$ and $n_{\mathrm{u}}$ for all time slots because their distributions are unchanged. This assumption is for ease of exposition and obviously does not affect the solutions to the problems of interest. Furthermore, we assume the availability of perfect CSI for the design and optimization of both static and mobile UAV networks \cite{liu2020machine, li2020reconfigurable, li2021robust}. We recognize that this assumption limits the practicality of the design. However, in this work, we mainly focus on demonstrating the fundamental performance benefits of the hybrid RIS architecture in assisting UAV air-ground communications.

	\subsection{Problem Formulation}
	
	Our goal is to maximize the minimum rate of UEs in both scenarios discussed previously by jointly optimizing the location/trajectory and transmit beamforming of the UAV as well as the RIS beamforming coefficients. The different signal models in the two scenarios (i.e. \eqref{eq_signal_model_loc} and \eqref{eq_signal_model_tra}) lead to different optimization problems which will be elaborated next.
	
	\subsubsection{Joint UAV's Placement and Beamforming Optimization Problem}
	Based on \eqref{eq_signal_model_loc}, the achievable rate of UE $k$ (in nats/s/Hz) can be expressed as
	\begin{align*}
		\mathcal{R}_k^{\mathrm{loc}} = \log \left( 1 + \frac{\absshort{\vh_{k}^\H \vw_k}^2}{ \sum\nolimits_{j \neq k} \absshort{\vh_k^\H \vw_j}^2 + \sigma_{\mathrm{r}}^2 \normshort{\hrr^\H \bPsi}^2 + \sigma_{\mathrm{u}}^2} \right). \nbthis \label{eq_rate}
	\end{align*}
	Furthermore, note that the RIS amplifies both incident signal $\hrt \vx$ and total noise and residual SI $\vn_{\mathrm{r}}$. Thus, the transmit power of active elements of the RIS is computed as
	\begin{align*}
		\pris  &= \tr{ \bPsi \mean{\vn_{\mathrm{r}} \vn_{\mathrm{r}}^\H + \mH_1 \vx \vx^\H \mH_1^\H} \bPsi^\H}\\
            &\eqtext{(a)} \sigma_{\mathrm{r}}^2 \norm{\bPsi}_{\mathcal{F}}^2 +  \norm{ \bPsi \hrt }_{\mathcal{F}}^2 \sum_{k=1}^{K} \norm{\vw_k}^2 \eqtext{(b)} \sum_{n \in \setA} \abs{\an}^2 \xi_n, \nbthis \label{eq_pris}
	\end{align*}
	where equality (a) follows the fact that $\vn_{\mathrm{r}}$ has zero mean and is independent of $\mH_1 \vx$; in \eqref{eq_pris}, $\xi_n \triangleq \sigma_{\mathrm{r}}^2 +  \norm{\vh_{1,n}}^2 \sum_{k=1}^{K} \norm{\vw_k}^2$, with $\vh_{1,n}$ being the $n$th row of $\hrt$, and equality (b) follows the diagonal structure of $\bPsi$ whose non-zero elements are in $\setA$ only. The total transmit power at the RIS is constrained as $\pris \leq \prismax$, where $\prismax$ is the power budget of the RIS. As a result, the max-min rate problem can be formulated as:
	\begin{subequations}
		\label{prob_loc}
		\begin{align}
			(\mathcal{P}_{\mathrm{loc}}):\ \underset{\substack{\bV, \bW,\ba}}{\textrm{maximize}} \quad & \underset{k\in\mathcal{K}}{\textrm{min}} \{\mathcal{R}_k^{\mathrm{loc}}\} \nbthis \label{obj_loc} \\
			\textrm{subject to} \quad
			&\sum\nolimits_{k=1}^{K} \norm{\vw_k}^2 \leq \ptmax, \nbthis \label{cons_puav_loc} \\
			&0 \leq \theta_n \leq 2\pi, \ \forall n, \nbthis \label{cons_ris_phase_loc} \\
			&\abs{\alpha_n} \leq 1, \ \forall n \notin \setA, \nbthis \label{cons_passive_modul_loc} \\
			&\abs{\alpha_n} \leq a_{\mathrm{max}}, \ \forall n \in \setA,  \nbthis \label{cons_active_modul_loc} \\
			&\sum\nolimits_{n \in \setA} \abs{\an}^2 \xi_n \leq \prismax,  \nbthis \label{cons_pris_loc}
		\end{align}
	\end{subequations}
	where constraints \eqref{cons_ris_phase_loc}--\eqref{cons_pris_loc} are design constrains of the hybrid RIS. It is noted that in \eqref{cons_active_modul_loc}, only active elements ($n \in \setA$) can amplify the signals with amplification gains restricted by $a_{\mathrm{max}}$ \cite{long2021active}.
	
	
	\subsubsection{Joint UAV's Trajectory Planning and Beamforming Optimization}
	
	From the signal model in \eqref{eq_signal_model_tra}, the average rate (in nats/s/Hz) of UE $k$ over all $T$ time slots is given as
	\begin{align*}
		\mathcal{R}_k^{\mathrm{tra}} = \frac{1}{T} \sum_{t=1}^{T} b_k[t] \log \left( 1 + \frac{\absshort{\vh_k^\H[t] \vw[t]}^2}{\sigma_{\mathrm{r}}^2 \normshort{\hrr^\H[t] \bPsi[t]}^2 + \sigma_{\mathrm{u}}^2} \right). \nbthis \label{eq_bar_Rk}
	\end{align*}
	Following a similar derivation as in \eqref{eq_pris}, the transmit power of the RIS at time slot $t$ can be given as
	$
	\pris[t] = \sum_{n \in \setA} \abs{\an[t]}^2 \xi_n[t],
	$
	where $\xi_n[t] \triangleq \sigma_{\mathrm{r}}^2 +  \norm{\vh_{1,n}[t]}^2 \norm{\vw[t]}^2$.
	Thus, the max-min rate problem in this scenario can be formulated as
	\begin{subequations}
		\label{prob_tra}
		\begin{align}
			(\mathcal{P}_{\mathrm{tra}}): \underset{\substack{\bBt,\bVt, \\ \bWt,\bat}}{\textrm{maximize}} \quad & \underset{k\in\mathcal{K}}{\textrm{min}} \{\mathcal{R}_k^{\mathrm{tra}}\} \nbthis \label{obj_tra} \\
			\textrm{subject to} \quad 
			&\vv[1] = \vv[T],\ \nbthis \label{cons_traj_1p} \\
			&\norm{ \vv[t+1] - \vv[t] }^2 \leq d_{\mathrm{max}}^2, \nonumber \\  &\hspace{1.5cm} t = 1,\ldots,T-1, \nbthis \label{cons_traj_2p}\\
			&b_k[t] \in \{0,1\},\  \forall k, t, \nbthis \label{cons_scheduling_3p} \\
			&\sum\nolimits_{k=1}^K b_k[t] \leq 1,\  \forall t, \nbthis \label{cons_scheduling_1p} \\
			&\norm{\vw[t]}^2 \leq \ptmax,\ \forall t, \nbthis \label{cons_puav_tra} \\
			&0 \leq \theta_n[t] \leq 2\pi,\ \forall n, t, \nbthis \label{cons_ris_phase_tra} \\
			&\abs{\alpha_n[t]} \leq 1,\ n \notin \setA, \forall t, \nbthis \label{cons_passive_modul_tra} \\
			&\abs{\alpha_n[t]} \leq a_{\mathrm{max}},\ n \in \setA, \forall t, \nbthis \label{cons_active_modul_tra} \\
			&\sum_{n \in \setA} \abs{\an[t]}^2 \xi_n[t] \leq \prismax,\ \forall t. \nbthis \label{cons_pris_tra}
		\end{align}
	\end{subequations}
	In constraints \eqref{cons_traj_1p} and \eqref{cons_traj_2p}, it is required that the UAV returns to its initial location at the end of each period $\Delta t$, and the maximum horizontal distance it can travel in each time slot is $d_{\mathrm{max}} = \delta_t v_{\mathrm{max}}$. Similar to \eqref{cons_ris_phase_loc}--\eqref{cons_pris_loc}, constraints \eqref{cons_ris_phase_tra}--\eqref{cons_pris_tra} are  constrains associated with the hybrid RIS in all the time slots. 
	
	Both problems $(\mathcal{P}_{\mathrm{loc}})$ in \eqref{prob_loc} and $(\mathcal{P}_{\mathrm{tra}})$ in \eqref{prob_tra} are nonconvex, which are difficult to solve optimally. In fact, even finding a feasible point to these problems is rather challenging due to a strong coupling between the optimization variables  in the objective function as well as constraint \eqref{cons_pris_loc} and \eqref{cons_pris_tra}. Furthermore, problem  $(\mathcal{P}_{\mathrm{tra}})$ is a mixed-integer nonconvex programming due to integer constraints \eqref{cons_scheduling_1p}, \eqref{cons_scheduling_3p}. To tackle the above challenges, we first provide the following lemma with some useful approximations. Then, efficient solutions to problems $(\mathcal{P}_{\mathrm{loc}})$ and $(\mathcal{P}_{\mathrm{tra}})$ will be presented in  Sections \ref{sec_loc_opt} and \ref{sec_tra_opt}, respectively. 
	\begin{lemma}
		\label{lemma_approx}
		Consider the following concave power, quadratic, bilinear, and quadratic-over-linear functions:
	\begin{align*}
		&f_{\mathtt{pow}}(x;c) \triangleq -x^c,\ x \in \setR_{++}, c > 1 \text{~or~} c < 0,\\
		&f_{\mathtt{qua}}(\vx; \vc) \triangleq - \norm{\vx - \vc}^2,\ \vx, \vc \in \setC^n,\\
		&f_{\mathtt{bil}}(x, y,\delta_s) \triangleq \delta_s xy,\ (x,y) \in \setR_{++}^2, \delta_s = \pm 1, \\
		&f_{\mathtt{qol}}(\vx,y;\mC) \triangleq - \frac{\vx^\H \mC \vx}{y},\  \vx \in \setC^n, y \in \setR_{++}, \mC \in \setC^{n \times n},
	\end{align*}
		respectively. Their convex upper bounds can be found as \cite{nguyen2018energy, wu2018joint}:
		\begin{align*}
			&f_{\mathtt{pow}}(x;c) \leq F_{\mathtt{pow}}(x;c,x_0) \triangleq \left(c-1\right) x_0^{c} - c x_0^{c-1} x, \nbthis \label{approx_power} \\
			&f_{\mathtt{qua}}(\vx; \vc) \leq F_{\mathtt{qua}}(\vx; \vc, \vx_0)\triangleq  2(\vc - \vx_0)^\T (\vx - \vx_0) \\ 
                &\hspace{5.5cm}  - \normshort{\vx_0 - \vc}^2,  \nbthis \label{approx_norm_square}\\
			&f_{\mathtt{bil}}(x, y;1) \leq F_{\mathtt{bil}}(x, y; 1, x_0, y_0) \triangleq \frac{1}{2} \left(\frac{y_0}{x_0} x^2 + \frac{x_0}{y_0} y^2 \right), \nbthis \label{approx_bil_neg} \\
			&f_{\mathtt{bil}}(x, y;-1) \leq F_{\mathtt{bil}}(x, y; -1, x_0, y_0)\triangleq  \frac{1}{4}(x - y)^2\\ 
                &\hspace{2.25cm}  + \frac{1}{4}(x_0+y_0)^2 - \frac{1}{2}(x_0+y_0)(x+y), \nbthis \label{approx_bil_pos}\\
			&f_{\mathtt{qol}}(\vx,y) \leq F_{\mathtt{qol}}(\vx,y;\vx_0, y_0) \triangleq \frac{\vx_0^\H \mC \vx_0}{y_0^2} y - \frac{2\re{\vx_0^\H \mC \vx}}{y_0}, \nbthis \label{approx_qol}
		\end{align*}
		respectively, based on the first-order Taylor approximations around feasible points $x_0$ for \eqref{approx_power}, $\vx_0$ for \eqref{approx_norm_square}, $(x_0,y_0)$ for \eqref{approx_bil_neg} and \eqref{approx_bil_pos}, and $(\vx_0,y_0)$  for \eqref{approx_qol}, respectively. The proof is conceptually simple but requires some algebra, which is omitted in this work.
		
	\end{lemma}	
	
	\section{Joint UAV's Placement and Beamforming Optimization in $(\mathcal{P}_{\mathrm{loc}})$}
	\label{sec_loc_opt}
	
	To solve problem $(\mathcal{P}_{\mathrm{loc}})$ in \eqref{prob_loc}, we first introduce variable $\tau$ to bypass the non-smooth of \eqref{obj_loc}, which leads to the following epigraph form:
	\begin{subequations}
		\label{prob_loc_1}
		\begin{align}
			\underset{\substack{\tau, \bV, \bW,\ba}}{\textrm{maximize}} \quad & \tau \nbthis \label{obj_loc_1} \\
			\textrm{subject to} \quad 
			&\mathcal{R}_k^{\mathrm{loc}}  \geq \tau,\ \forall k, \nbthis \label{cons_min_rate_loc} \\
			&\eqref{cons_puav_loc}-\eqref{cons_pris_loc}.
		\end{align}
	\end{subequations}
     Next, we utilize the SCA framework, wherein each iteration utilizes the BCA approach to decouple \eqref{prob_loc_1} into three distinct subproblems, \textit{namely} the UAV's location optimization, and UAV and RIS beamforming designs. The detailed solutions are elaborated in the following subsections.
	
	\subsection{UAV Transmit Beamforming Design}
	With given $(\bV,\ba)$, the beamformers $\bW$ at the UAV can be obtained by solving the following problem:
	\begin{align}
		\label{subprob_BF_loc}
		\underset{\tau,\bW}{\textrm{maximize}}\ \tau,\ \textrm{subject to}\  \eqref{cons_puav_loc}, \eqref{cons_pris_loc}, \eqref{cons_min_rate_loc},
	\end{align}
	where constraint \eqref{cons_min_rate_loc} is nonconvex. To overcome this, we introduce slack variables $\{\gamma_k\}$ as the lower bound of the signal-to-interference-plus-noise ratio (SINR) term in \eqref{eq_rate} and transform \eqref{cons_min_rate_loc} to the following set of constraints:
	\begin{subequations}
		\begin{align*}
			\log (1 + \gamma_{k}) &\geq \tau,\ \forall k, \nbthis \label{cons_min_rate_BF1} \\
			\frac{\absshort{\vh_{k}^\H \vw_k}^2}{ \sum\nolimits_{j \neq k} \absshort{\vh_k^\H \vw_j}^2 + \sigma_k^2} &\geq \gamma_{k},\ \forall k, \nbthis \label{cons_min_rate_BF2}
		\end{align*}
	\end{subequations}
	where $\sigma_k^2 \triangleq \sigma_{\mathrm{r}}^2 \normshort{\hrr^\H \bPsi}^2 + \sigma_{\mathrm{u}}^2$ is independent of $\bW$. Let
	\begin{align*}
	\vw &\triangleq [\vw_1^\T, \ldots, \vw_K^\T]^\T,\\
	\hat{\mH}_k  &\triangleq \blockdiag{\boldsymbol{0}, \ldots, \boldsymbol{0}, \vh_k \vh_k^\H, \boldsymbol{0}, \ldots, \boldsymbol{0}} ,\\
	\bar{\mH}_k  &\triangleq \blockdiag{\vh_k \vh_k^\H, \ldots, \vh_k \vh_k^\H, \boldsymbol{0}, \vh_k \vh_k^\H, \ldots, \vh_k \vh_k^\H},
	\end{align*}
	with $\vw \in \setC^{KN_t \times 1}$ and $\hat{\mH}_k, \bar{\mH}_k \in \setC^{KN_t \times KN_t}$. Then, we can rewrite constraint \eqref{cons_min_rate_BF2} as $\frac{\vw^\H \hat{\mH}_k \vw}{ \vw^\H \bar{\mH}_k \vw + \sigma_k^2} \geq \gamma_{k}, \forall k$, which is equivalent to
	$
	\vw^\H \bar{\mH}_k \vw + \sigma_k^2 - f_{\mathtt{qol}}(\vw,\gamma_k;\hat{\mH}_k) \leq 0, \forall k,
	$
	where $f_{\mathtt{qol}}(\cdot;\cdot)$ is defined in Lemma \ref{lemma_approx}.
	By using approximation \eqref{approx_qol} around the feasible point $(\vw^{(i)}, \gamma_k^{(i)})$ found at iteration $i$ of the proposed iterative algorithm presented shortly, this constraint is approximated as
	\begin{align*}
		\vw^\H \bar{\mH}_k \vw + \sigma_k^2 + F_{\mathtt{qol}}(\vw,\gamma_k;\vw^{(i)}, \gamma_k^{(i)}) \leq 0,\ \forall k. \nbthis \label{cons_min_rate_BF2_2}
	\end{align*}
	Given the introduction of $\vw$, we have $\sum_{k=1}^{K} \norm{\vw_k}^2 = \norm{\vw}^2$, and constraints \eqref{cons_puav_loc} and \eqref{cons_pris_loc} become
	\begin{align*}
		\norm{\vw}^2 \leq \ptmax,\ \text{and}\
		\sum_{n \in \setA} \abs{\an}^2 \left(\sigma_{\mathrm{r}}^2 +  \norm{\vh_{1,n}}^2 \norm{\vw}^2   \right) \leq \prismax, \nbthis \label{cons_ris_power_1}
	\end{align*}
	respectively. Finally, we can approximate problem \eqref{subprob_BF_loc} by the following convex program at iteration $i$:
	\begin{align}
		\label{subprob_BF_loc_1}
		\underset{\tau,\vw, \{\gamma_k\} }{\textrm{maximize}}\quad \tau,\ \textrm{subject to}\ \eqref{cons_min_rate_BF1}, \eqref{cons_min_rate_BF2_2}, \eqref{cons_ris_power_1}.
	\end{align}

	\subsection{Optimization of the UAV's Location}
	For given $(\bW,\ba)$, the UAV's location can be optimized by solving the following problem with respect to $\{\tau,\bV\}$:
	\begin{align}
		\label{subprob_loc}
		\underset{\tau,\bV}{\textrm{maximize}}\ \tau,\ \textrm{subject to}\   \eqref{cons_pris_loc}, \eqref{cons_traj_1p}, \eqref{cons_min_rate_loc}.
	\end{align}
	The challenge in solving \eqref{subprob_loc} is that variable $\bV$ is currently hidden in all the constraints. To overcome this, we first expand the expression of $\mathcal{R}_k^{\mathrm{loc}}$ in \eqref{eq_rate} to show the role of $\vv$. Specifically, from \eqref{eq_h2} and \eqref{eq_h_eff}, we can rewrite
	\begin{align*}
		&\absshort{\vh_{k}^\H \vw_j}^2 = \absshort{\hd^\H \vw_j  + \hrr^\H \bUpsilon \hrt \vw_j}^2\\
            &=  \absshort{\hd^\H \vw_j}^2 +  \absshort{\hrr^\H \bUpsilon \hrt \vw_j}^2 + 2  \reshort{\vw_j^\H \hd \hrr^\H \bUpsilon \hrt \vw_j} \\
		&= c_{0,kj} \norm{\vv - \vu_k}^{-\epsilon_0} + c_{1,kj} \norm{\vv - \vr}^{-\epsilon_1}\\
            &\qquad + c_{2,kj} \norm{\vv - \vu_k}^{-\epsilon_0/2} \norm{\vv - \vr}^{-\epsilon_1/2}, \forall k, j, \nbthis \label{def_f_mk_v}
	\end{align*}
	where $c_{0,kj} \triangleq \zeta_0 \absshort{\gd^\H \vw_j}^2$, $c_{1,kj} \triangleq \zeta_0 \absshort{\hrr^\H \bUpsilon  \grt \vw_j}^2$, and $c_{2,kj} \triangleq 2  \zeta_0 \reshort{\vw_j^\H  \gd \hrr^\H \bUpsilon \grt \vw_j}$ are all constants with respect to $\vv$. Thus, we can rewrite $\mathcal{R}_k^{\mathrm{loc}}$ as \eqref{eq_rate_loc_2} (at the top of the next page).
	\begin{figure*}
	    \begin{align*}
		\mathcal{R}_k^{\mathrm{loc}} = \log \left( 1 + \frac{c_{0,kk} \norm{\vv - \vu_k}^{-\epsilon_0} + c_{1,kk} \norm{\vv - \vr}^{-\epsilon_1} + c_{2,kk} \norm{\vv - \vu_k}^{-\epsilon_0/2} \norm{\vv - \vr}^{-\epsilon_1/2}}{\sigma_k^2 + \underset{{j \neq k}}{\sum} c_{0,kj} \norm{\vv - \vu_k}^{-\epsilon_0} + c_{1,kj} \norm{\vv - \vr}^{-\epsilon_1} + c_{2,kj} \norm{\vv - \vu_k}^{-\epsilon_0/2} \norm{\vv - \vr}^{-\epsilon_1/2}} \right). \nbthis \label{eq_rate_loc_2}
	\end{align*}
        \hrule
	\end{figure*}
	This complicated form of $\mathcal{R}_k^{\mathrm{loc}}$ clearly exposes the nonconvexity as well as the challenges in addressing constraint \eqref{cons_min_rate_loc}. In particular, we note in \eqref{eq_rate_loc_2} that while $c_{0,kj}, c_{1,kj} > 0, \forall k, j$, $c_{2,kj}$ can be either positive or negative, making it very difficult to address the nonconvexity of \eqref{cons_min_rate_loc}. Regarding this, let us introduce set of slack variables $\boldsymbol{\mathcal{V}} \triangleq \{\{\hat{v}_{0,k}\},\{\bar{v}_{0,k}\}, \hat{v}_{1}, \bar{v}_{1}\}$ satisfying 
	\begin{subequations}
        \label{cons_v1bar}	
	\begin{align*}
		\hat{v}_{0,k}  &\leq \norm{\vv - \vu_k}^{-\epsilon_0/2}, \nbthis \label{cons_v0hat} \\
		\bar{v}_{0,k} &\geq  \norm{\vv - \vu_k}^{-\epsilon_0/2}, \forall k, \nbthis \label{cons_v0bar} \\
		\hat{v}_{1} &\leq \norm{\vv - \vr}^{-\epsilon_1/2}, \nbthis \label{cons_v1hat} \\
		\bar{v}_1 &\geq \norm{\vv - \vr}^{-\epsilon_1/2}. \nbthis \label{cons_v1bar1}	
	\end{align*}
	\end{subequations}
	Furthermore, we introduce slack variables $\{\hat{a}_{k}, \bar{a}_{kj}\}$ such that
	\begin{subequations}
		\begin{align*}
			\hat{a}_{k} &\leq 
			\begin{cases*}
				c_{0,kk} \hat{v}_{0,k}^2 + c_{1,kk} \hat{v}_{1}^2 + \abs{c_{2,kk}} \hat{v}_{0,k} \hat{v}_{1},\ \text{if~} c_{2,kk} > 0\\
				c_{0,kk} \hat{v}_{0,k}^2 + c_{1,kk} \hat{v}_{1}^2 - \abs{c_{2,kk}} \bar{v}_{0,k} \bar{v}_{1},\ \text{otherwise}
			\end{cases*},\\ &\hspace{6.5cm}\forall k, \nbthis \label{eq_ahat} \\
			\bar{a}_{kj} &\geq 
			\begin{cases*}
				c_{0,kj} \bar{v}_{0,k}^2 + c_{1,kj} \bar{v}_1^2 + \abs{c_{2,kj}} \bar{v}_{0,k}\bar{v}_1,\ \text{if~} c_{2,kj} > 0 \\
				c_{0,kj} \bar{v}_{0,k}^2 + c_{1,kj} \bar{v}_1^2 - \abs{c_{2,kj}} \hat{v}_{0,k} \hat{v}_1,\ \text{otherwise}
			\end{cases*},\\ &\hspace{6.25cm} \forall k,j
			. \nbthis \label{eq_acheck}
		\end{align*}
	\end{subequations}
	Thus, \eqref{cons_min_rate_loc} is equivalent to set of constraints \eqref{cons_v1bar}--\eqref{eq_acheck} and
	\begin{align*}
		\log\Bigl( 1 + \frac{\hat{a}_{k}}{\sigma_k^2 + \sum_{j \neq k} \bar{a}_{kj}} \Bigr)  \geq \tau, \forall k, \nbthis \label{cons_min_ratep}
	\end{align*}
	which are still nonconvex, but more tractable. To address \eqref{cons_min_ratep}, we rewrite it as $\log ( \sigma_k^2 + \hat{a}_{k} + \sum\nolimits_{j \neq k} \bar{a}_{kj})  \geq \tau +  \tilde{r}_{k}, \forall k$,
	where $\tilde{r}_{k} = \log ( \sigma_k^2 + \sum\nolimits_{j \neq k} \bar{a}_{kj})$. By applying the first-order Taylor approximation around $\bar{a}_{kj}^{(i)}$ of the concave function $\tilde{r}_{k}$, its convex upper bound can be found as
    \begin{align*}
        \tilde{r}_{k} \leq \tilde{r}_{\mathrm{ub},k}^{(i)} \triangleq \log\Big( \sigma_k^2 + \sum\nolimits_{j \neq k} \bar{a}_{kj}^{(i)} \Big) + \frac{\underset{j \neq k}{\sum} \bar{a}_{kj} - \bar{a}_{kj}^{(i)}}{\sigma_k^2 + \underset{j \neq k}{\sum} \bar{a}_{kj}^{(i)}}. 
    \end{align*}
	As a result, \eqref{cons_min_ratep} is transformed to the following convex constraint:
	\begin{align*}
		\log \Big( \sigma_k^2 + \hat{a}_{k} + \sum\nolimits_{j \neq k} \bar{a}_{kj}\Big)  \geq \tau +  \tilde{r}_{\mathrm{ub},k}^{(i)}, \forall k. \nbthis \label{cons_min_ratep_approx} 
	\end{align*}

	Next, we apply inequalities \eqref{approx_power}--\eqref{approx_bil_neg} around points $\{ \hat{v}_{0,k}^{(i)}, \bar{v}_{0,k}^{(i)}, \hat{v}_1^{(i)}, \bar{v}_1^{(i)} \}$ to approximate nonconvex constraints \eqref{cons_v1bar}--\eqref{eq_acheck} as
	\begin{subequations}
		\begin{align*}
			&\norm{\vv - \vu_k} + F_{\mathtt{pow}}(\hat{v}_{0,k};-2/\epsilon_0,\hat{v}_{0,k}^{(i)}) \leq 0,\  \forall  k, \nbthis \label{cons_v0hat_loc_approx}\\
			&F_{\mathtt{qua}}(\vv; \vu_k, \vv^{(i)}) \leq 1/F_{\mathtt{pow}}(\bar{v}_{0,k};4/\epsilon_0,\bar{v}_{0,k}^{(i)} ),\  \forall k, \nbthis \label{cons_v0check_loc_approx}\\
			&\norm{\vv - \vr}  + F_{\mathtt{pow}}(\hat{v}_1;-2/\epsilon_1, \hat{v}_1^{(i)}) \leq 0, \nbthis \label{cons_v1hat_loc_approx1}\\
			&F_{\mathtt{qua}}(\vv; \vr, \vv^{(i)}) \leq 1/F_{\mathtt{pow}}(\bar{v}_1;4/\epsilon_1, \bar{v}_1^{(i)}), \nbthis \label{cons_v1check_approx1}\\
			&\begin{cases*}
				\hat{a}_{k} + c_{0,kk} F_{\mathtt{qua}}(\hat{v}_{0,k};0,\hat{v}_{0,k}^{(i)}) + c_{1,kk} F_{\mathtt{qua}}(\hat{v}_{1};0,\hat{v}_{1}^{(i)})\\
                    \hspace{3cm} + \abs{c_{2,kk}} \hat{F}_{\mathtt{bil},kk}^{(i)} \leq 0, \text{if~} c_{2,kk} > 0 \\
				\hat{a}_{k} + c_{0,kk} F_{\mathtt{qua}}(\hat{v}_{0,k};0,\hat{v}_{0,k}^{(i)}) + c_{1,kk} F_{\mathtt{qua}}(\hat{v}_{1};0,\hat{v}_{1}^{(i)})\\
                    \hspace{3.5cm} - \abs{c_{2,kk}} \bar{F}_{\mathtt{bil},kk}^{(i)} \leq 0, \text{otherwise}
			\end{cases*},\\ &\hspace{7cm} \forall  k, \nbthis \label{eq_ahat_approx} \\
			&
			\begin{cases*}
				\bar{a}_{kj} \geq c_{0,kj} \bar{v}_{0,k}^2 + c_{1,kj} \bar{v}_1^2 + \abs{c_{2,kj}} \bar{F}_{\mathtt{bil},kj}^{(i)},\ \text{if~} c_{2,kj} > 0 \\
				\bar{a}_{kj} \geq c_{0,kj} \bar{v}_{0,k}^2 + c_{1,kj} \bar{v}_1^2 - \abs{c_{2,kj}} \hat{F}_{\mathtt{bil},kj}^{(i)},\ \text{otherwise}
			\end{cases*},\\ &\hspace{6cm} \forall k, j \neq k, \nbthis \label{eq_acheck_approx}
		\end{align*}
	\end{subequations}
	where $\hat{F}_{\mathtt{bil},kj}^{(i)} \triangleq F_{\mathtt{bil}}( \hat{v}_{0,k}, \hat{v}_{1}; \mathrm{sgn}(c_{2,kj}), \hat{v}_{0,k}^{(i)}, \hat{v}_{1}^{(i)})$ and $\bar{F}_{\mathtt{bil},kj}^{(i)} \triangleq F_{\mathtt{bil}}( \bar{v}_{0,k}, \bar{v}_{1}; \mathrm{sgn}(c_{2,kj}), \bar{v}_{0,k}^{(i)}, \bar{v}_{1}^{(i)})$, $\forall k, j$. Constraints \eqref{cons_v0hat_loc_approx}--\eqref{eq_acheck_approx} are convex assuming that $\epsilon_0, \epsilon_1 < 4$. Similarly, by writing $\xi_n = \sigma_{\mathrm{r}}^2 + c_{3,n} (\norm{\vv - \vr}^{-\epsilon_1/2})^2  $,
	where $c_{3,n} \triangleq  \zeta_0 \norm{ \vg_{1,n}}^2 \sum_{k=1}^K \norm{\vw_k}^2$ is independent of $\bV$, we can rewrite \eqref{cons_pris_loc} equivalently as set constraints \eqref{cons_v1check_approx1} and
	\begin{align*}
		\sum\nolimits_{n \in \setA} \abs{\an}^2 \Big(\sigma_{\mathrm{r}}^2 + \sum\nolimits_{k=1}^K c_{3,n} \bar{v}_{1}^2\Big) \leq \prismax. \nbthis \label{cons_pris_loc_approx}
	\end{align*}
	In summary, we can approximate problem \eqref{subprob_loc} at iteration $i$ by the following convex program:
	\begin{align}
		\label{subprob_loc_1}
		\underset{\tau,\bV,\boldsymbol{\mathcal{V}}}{\textrm{maximize}}\ \tau,\ \textrm{subject to}\ \eqref{cons_traj_1p},\eqref{cons_min_ratep_approx}, \eqref{cons_v0hat_loc_approx}-\eqref{cons_pris_loc_approx}.
	\end{align}
	
		%
		%

	\subsection{Optimization of RIS Beamforming Coefficients}
	\label{sec_opt_loc_alpha}
	
	Given $(\bV,\bW)$, $\ba$ can be optimized by solving the following problem:
	\begin{align}
		\label{subprob_ris_loc}
		\underset{\substack{\tau,\ba}}{\textrm{maximize}}\ \tau,\
		\textrm{subject to}\ \eqref{cons_min_rate_loc}, \eqref{cons_ris_phase_loc}-\eqref{cons_pris_loc}.
	\end{align}
	It is easy to see that constraints \eqref{cons_ris_phase_loc}-\eqref{cons_pris_loc} are convex with respect to $\ba$, except \eqref{cons_min_rate_loc}. Furthermore, the optimization variable $\ba$ has not been exposed in the current form of \eqref{cons_min_rate_loc}. As the first step to solve \eqref{subprob_ris_loc}, we express $\mathcal{R}_k^{\mathrm{loc}}$ as a function of $\ba$ in the following derivations. For ease of exposition, we denote $\bar{h}_{0,kj} \triangleq \hd^\H \vw_j, \forall k, j$, and $\bar{\vh}_{1,j} \triangleq \mH_1 \vw_j, \forall j$, which yields $\vh_{k}^\H \vw_k = \bar{h}_{0,kk} + \hrr^\H \bUpsilon \bar{\vh}_{1,k},\ \vh_{k}^\H \vw_j = \bar{h}_{0,kj} + \hrr^\H \bUpsilon \bar{\vh}_{1,j}$.
	Then, we can write the SINR term in \eqref{eq_rate} as
	\begin{align*}
	    \frac{\absshort{\bar{h}_{0,kk} + \hrr^\H \bUpsilon \bar{\vh}_{1,k}}^2}{\underset{{j \neq k}}{\sum} \absshort{\bar{h}_{0,kj} + \hrr^\H \bUpsilon \bar{\vh}_{1,j}}^2 + \sigma_{\mathrm{r}}^2 \normshort{\hrr^\H \bPsi}^2 + \sigma_{\mathrm{u}}^2} \triangleq \frac{N_k}{D_k}.
	\end{align*}
	Let $\va \triangleq [\alpha_1, \ldots, \alpha_N ]^\T \in \setC^{N \times 1}$, $\tilde{\mH}_{2,k} \triangleq \diagshort{\hrr^\H}  \in \setC^{N \times N}$, and $\tilde{\vh}_{12,kj} \triangleq \tilde{\mH}_{2,k} \bar{\vh}_{1,j}  \in \setC^{N \times 1}$. Then, we have
    \begin{align*}
        \bar{h}_{0,kk} + \hrr^\H \bUpsilon \bar{\vh}_{1,k} &= \bar{h}_{0,kk} + \va^\T \tilde{\vh}_{12,kk}\\
        \bar{h}_{0,kj} + \hrr^\H \bUpsilon \bar{\vh}_{1,j} &= \bar{h}_{0,kj} + \va^\T \tilde{\vh}_{12,kj},\\
        \hrr^\H \bPsi &= \va^\T \mathbbm{1}^{\setA}_N \tilde{\mH}_{2,k}.
    \end{align*}
    With several algebraic manipulations, we obtain
    \begin{align*}
        N_k &= \va^\H \mQ_{k} \va + 2 \re{\va^\H \vq_{k}} + e_{k}, \\
        D_k &= \va^\H \tilde{\mQ}_k \va + 2\re{\va^\H \tilde{\vq}_k} +  \tilde{e}_k,
    \end{align*}
    where $\mQ_{k} = \tilde{\vh}_{12,kk}^* \tilde{\vh}_{12,kk}^\T$, $\vq_{k} = \tilde{\vh}_{12,kk}^* \bar{h}_{0,kk}$, $e_{k} = \abs{\bar{h}_{0,kk}}^2$, $\tilde{\mQ}_{k} = \sigma_{\mathrm{r}}^2  \mathbbm{1}^{\setA}_N \tilde{\mH}_{2,k}^* \tilde{\mH}_{2,k}^\T \mathbbm{1}^{\setA}_N + \sum_{j \neq k} \tilde{\vh}_{12,kj}^* \tilde{\vh}_{12,kj}^\T$, $\tilde{\vq}_{k} = \sum_{j \neq k} \tilde{\vh}_{12,kj}^* \bar{h}_{0,kj}$, and $\tilde{e}_{k} = \sigma_{\mathrm{u}}^2 + \sum_{j \neq k}  \abs{\bar{h}_{0,kj}}^2$. As a result, $\va$ is exposed in the following form of $\mathcal{R}_k^{\mathrm{loc}}$:
    \begin{align*}
        \mathcal{R}_k^{\mathrm{loc}} = \log \left(1 + \frac{\va^\H \mQ_{k} \va + 2 \re{\va^\H \vq_{k}} + e_{k}}{ \va^\H \tilde{\mQ}_{k} \va + 2\re{\va^\H \tilde{\vq}_{k}} +  \tilde{e}_{k}}\right).
    \end{align*}
    With this form, we can rewrite \eqref{cons_min_rate_loc} as  
	\begin{align}
		\label{cons_min_rate_loc_1}
		\bar{r}_k \geq \tau + \tilde{r}_k,
	\end{align}
	where $\bar{r}_k \triangleq \log \Big( \normshort{\bar{\mQ}_{k}^{\frac{1}{2}} \va}^2 + 2 \re{\va^\H \bar{\vq}_{k}} + \bar{e}_{k}\Big)$, $\tilde{r}_k \triangleq \log( \va^\H \tilde{\mQ}_{k} \va +2\re{\va^\H \tilde{\vq}_{k}} +  \tilde{e}_{k})$, with $\bar{\mQ}_{k} = \mQ_{k} + \tilde{\mQ}_{k}$, $\bar{\vq}_{k} = \vq_{k} + \tilde{\vq}_{k}$, $\bar{e}_k = e_k + \tilde{e}_k$, and note that $\bar{\mQ}_{k}$ is positive semidefinite. These two are nonconvex with respect to $\va$. However, by noting from approximation \eqref{approx_norm_square} that $\normshort{\bar{\mQ}_{k}^{\frac{1}{2}} \va}^2 = - f_{\mathtt{qua}}(\bar{\mQ}_{k}^{\frac{1}{2}} \va; \boldsymbol{0}) \geq - F_{\mathtt{qua}}(\bar{\mQ}_{k}^{\frac{1}{2}} \va; \boldsymbol{0}, \bar{\mQ}_{k}^{\frac{1}{2}} \va^{(i)})$, a concave lower bound of $\bar{r}_k$ can be found as
	\begin{align*}
	    \bar{r}_k \geq \bar{r}_{\mathrm{lb},k}^{(i)} \triangleq \log \Big( 2 \re{\va^\H \bar{\vq}_{k}} - F_{\mathtt{qua}}(\bar{\mQ}_{k}^{\frac{1}{2}} \va; \boldsymbol{0}, \bar{\mQ}_{k}^{\frac{1}{2}} \va^{(i)}) + \bar{e}_{k}\Big).
	\end{align*}
	To tackle the nonconvexity of $\tilde{r}_k$, we introduce slack variables $\vartheta_k$ such that $\vartheta_k \geq \va^\H \tilde{\mQ}_{k} \va +2\re{\va^\H \tilde{\vq}_{k}}$. With this introduction of $\vartheta_k$, we can write $\tilde{r}_k = \log(\vartheta_k + \tilde{e}_{k})$. Thus, an upper bound of $\tilde{r}_k$ can be found as $\tilde{r}_k \leq \tilde{r}_{\mathrm{ub},k}^{(i)} \triangleq \log\Big( \vartheta_k^{(i)} + \tilde{e}_{k} \Big) + \frac{\vartheta_k - \vartheta_k^{(i)}}{\vartheta_k^{(i)} + \tilde{e}_{k}}$, based on the first-order Taylor approximation around $\vartheta_k^{(i)}$. As a result, constraint \eqref{cons_min_rate_loc_1} can be approximated by the following set of convex constraints:
	\begin{align}
		\bar{r}_{\mathrm{lb},k}^{(i)} &\geq \tau + \tilde{r}_{\mathrm{ub},k}^{(i)},\\ \forall k,\ \vartheta_k &\geq \va^\H \tilde{\mQ}_{k} \va + 2\re{\va^\H \tilde{\vq}_{k}},\ \forall k. \nbthis \label{cons_min_rate_loc_2b}
	\end{align}
	Furthermore, we can rewrite \eqref{cons_pris_loc} as
	\begin{align*}
		\va^\H \bXi \va \leq \prismax, \nbthis \label{cons_pris_loc2}
	\end{align*}
	where $\bXi = \diagshort{\tilde{\xi}_1, \ldots, \tilde{\xi}_N}$ with $\tilde{\xi}_n = \xi_n$ for $n \in \setA$, and $\tilde{\xi}_n = 0$, otherwise. Finally, problem \eqref{subprob_ris_loc} can be approximated by the following convex program at iteration $i$:
	\begin{align}
		\label{subprob_ris_loc_1}
		\underset{\substack{\tau,\va,\{\vartheta_k\}}}{\textrm{maximize}}\ \tau,\
		\textrm{subject to}\  \eqref{cons_ris_phase_loc}-\eqref{cons_active_modul_loc}, \eqref{cons_min_rate_loc_2b}, \eqref{cons_pris_loc2}.
	\end{align}

	\subsection{Overall Algorithm}
	\label{sec_opt_loc_alg}
	
	We summarize the BCA-SCA-based iterative algorithm to jointly optimize the UAV's transmit beamforming and locations as well as the RIS beamforming in Algorithm \ref{alg_opt_loc}. In Step 1, the initial solutions to $\vv^{(0)}$, $\{\vw_k^{(0)}\}$, $\{\alpha_n^{(0)}\}$, and $\boldsymbol{\mathcal{V}}^{(0)}$ are generated. In Steps 2--7, subproblems \eqref{subprob_BF_loc_1},  \eqref{subprob_loc_1}, and \eqref{subprob_ris_loc_1} are alternatively solved and $\vv^{(i)}$, $\{\vw_k^{(i)}\}$, $\{\alpha_n^{(i)}\}$, and $\boldsymbol{\mathcal{V}}^{(i)}$ are updated after each iteration until the objective value converges. The complexities required to solve subproblems \eqref{subprob_BF_loc_1},  \eqref{subprob_loc_1} and \eqref{subprob_ris_loc_1} are $\mathcal{O}(\sqrt{2}K^{3.5} N_t^3)$, $\mathcal{O}(8K^4)$, and  $\mathcal{O}(\sqrt{2N+2K} (N+K)^3)$, respectively. Therefore, the total complexity of Algorithm \ref{alg_opt_loc} is
	\begin{align*}
	    \mathcal{C}_{\mathrm{loc}} = \mathcal{O}\left( \mathcal{I}_{\mathrm{loc}} \left(\sqrt{2}K^{3.5} N_t^3 + 8K^4 + \sqrt{2} (N+K)^{3.5}\right) \right),
	\end{align*}
	where $\mathcal{I}_{\mathrm{loc}}$ is the number of iterations in Algorithm \ref{alg_opt_loc} until convergence. Considering that $N \gg K, N_t$, the complexity of this algorithm can be approximated as  $\mathcal{O}(N^{3.5})$, which is required for the optimization of RIS coefficients. 
	
	\begin{algorithm}[t]
		\small
		\caption{Proposed Iterative Algorithm to Solve Problem $(\mathcal{P}_{\mathrm{loc}})$ in \eqref{prob_loc}}
		\label{alg_opt_loc}
		\begin{algorithmic}[1]
			\STATE \textbf{Initialization:} Set $i=0$. Generate initial values $\vv^{(0)}$, $\{\vw_k^{(0)}\}$ and $\{\alpha_n^{(0)}\}$. Initialize $\{\gamma_k^{(0)}\}$, $\{\vartheta_k^{(0)}\}$, and  $\boldsymbol{\mathcal{V}}^{(0)}$ by considering equalities of  \eqref{cons_min_rate_BF2}, \eqref{cons_min_rate_loc_2b}, and \eqref{cons_v1bar}, respectively.
			
			\REPEAT
			
			\STATE Solve problem \eqref{subprob_BF_loc_1} with given $\{\vv^{(i)}\}$, $\alpha_n^{(i)}$, and $\{\gamma_k^{(i)}\}$ to obtain solutions $\{\vw_k^{\star}\}$ and $\{\gamma_k^{\star}\}$. Set $\vw_k^{(i+1)} = \vw_k^{\star}$ and $\gamma_k^{(i+1)} = \gamma_k^{\star}, \forall k$.
			
			\STATE Solve problem \eqref{subprob_loc_1} with given $\{\vw_k^{(i+1)}\}$, $\{\an^{(i)}\}$, and $\boldsymbol{\mathcal{V}}^{(i)}$ to obtain solutions $\vv^{\star}$ and $\boldsymbol{\mathcal{V}}^{\star}$. Set $\vv^{(i+1)} = \vv^{\star}$ and $\boldsymbol{\mathcal{V}}^{(i+1)} = \boldsymbol{\mathcal{V}}^{\star}$.
			
			\STATE Solve problem \eqref{subprob_ris_loc_1} with given $\{\vw_k^{(i+1)}\}$, $\vv^{(i+1)}$, and $\{\vartheta_k^{(i)}\}$ to obtain solutions $\va^{\star}$ and $\{\vartheta_k^{\star}\}$. Set $\va^{(i+1)} = \va^{\star}$ and $\gamma_k^{(i+1)} = \gamma_k^{\star}, \forall k$.
			
			\STATE Update $i=i+1$.
			\UNTIL convergence.
		\end{algorithmic}
	\end{algorithm}
	
	{Algorithm \ref{alg_opt_loc} solve original problem \eqref{prob_loc} via convex sub-problems \eqref{subprob_BF_loc_1}, \eqref{subprob_loc_1}, and \eqref{subprob_ris_loc_1}, combining that the early update of the partial set of variables in each iteration guarantees feasible points and non-decreasing objective values. Owing to the concreteness and convexity of the feasible sets in the sub-problems, the proposed algorithm remains the convergence of the inner approximation method. Let us denote by $\tau_{\vw}(\vw^{(i)},\vv^{(i)},\va^{(i)})$ and $\tau_{\vw}^{(i)}(\vw^{(i)},\vv^{(i)},\va^{(i)})$ the objective values of the original subproblem \eqref{subprob_BF_loc} and the approximate convex subproblem \eqref{subprob_BF_loc_1} obtained at iteration $i$ with respect to $\vw$, respectively. Following the SCA principles \cite{beck2010sequential}, it is true that
    \begin{align*}
        &\tau_{\vw}(\vw^{(i+1)},\vv^{(i)},\va^{(i)}) \geq \tau_{\vw}^{(i)}(\vw^{(i+1)},\vv^{(i)},\va^{(i)}) \\
        &\qquad \geq \tau_{\vw}^{(i)}(\vw^{(i)},\vv^{(i)},\va^{(i)}) = \tau_{\vw}(\vw^{(i)},\vv^{(i)},\va^{(i)}).
    \end{align*}
    We note that $\tau_{\vw}^{(i)}(\vw^{(i+1)},\vv^{(i)},\va^{(i)}) $ $ >  \tau_{\vw}^{(i)}(\vw^{(i)},\vv^{(i)},\va^{(i)})$ and $\tau_{\vw}(\vw^{(i+1)},\vv^{(i)},\va^{(i)}) > \tau_{\vw}(\vw^{(i)},\vv^{(i)},\va^{(i)})$ whenever $\vw^{(i)} \neq \vw^{(i+1)}$. Similarly for problems \eqref{subprob_loc} and \eqref{subprob_ris_loc}, we can prove that
    \begin{align*}
        &\tau_{\vv}(\vw^{(i+1)},\vv^{(i)},\va^{(i)}) = \tau^{(i)}_{\vv}(\vw^{(i+1)},\vv^{(i)},\va^{(i)})\\
        &\tau_{\vv}(\vw^{(i+1)},\vv^{(i+1)},\va^{(i)}) > \tau_{\vv}(\vw^{(i+1)},\vv^{(i)},\va^{(i)}),
    \end{align*}
    as far as $\vv^{(i)}\neq \vv^{(i+1)}$, and $\tau_{\va}(\vw^{(i+1)},\vv^{(i+1)},\va^{(i)}) = \tau^{(i)}_{\va}(\vw^{(i+1)},\vv^{(i+1)},\va^{(i)})$ and 
	$\tau_{\va}(\vw^{(i+1)},\vv^{(i+1)},\va^{(i+1)}) > \tau_{\va}(\vw^{(i+1)},\vv^{(i+1)},\va^{(i)})$ as far as $\va^{(i)}\neq \va^{(i+1)}$. Thus, we can readily show that $\tau(\vw^{(i+1)},\vv^{(i+1)},\va^{(i+1)})$ $> \tau(\vw^{(i)},\vv^{(i)},\va^{(i)})$, which indicates that the objective value of problem \eqref{prob_loc} is non-decreasing after each iteration. Because it is upper bounded by a finite value, Algorithm \ref{alg_opt_loc} is guaranteed to converge. Utilizing the non-decreasing property of the objective values over iterations, the convergence is determined based on the difference between the objective values in successive iterations. More specifically, the algorithm converges if $\tau(\vw^{(i+1)},\vv^{(i+1)},\va^{(i+1)}) - \tau(\vw^{(i)},\vv^{(i)},\va^{(i)})<\varepsilon$, where $\varepsilon$ is the convergence stopping criteria.}

	\section{Joint UAV's Trajectory Planning and Beamforming Optimization in $(\mathcal{P}_{\mathrm{tra}})$}
	\label{sec_tra_opt}
	
	Similar to \eqref{prob_loc_1}, $(\mathcal{P}_{\mathrm{tra}})$ in \eqref{prob_tra} has the following epigraph form:
	\begin{subequations}
		\label{prob_tra_1}
		\begin{align}
			\underset{\substack{\tau, \bBt,\bVt,\\ \bWt,\bat}}{\textrm{maximize}} \quad & \tau \nbthis \label{obj_loc_2} \\
			\textrm{subject to} \quad 
			&\frac{1}{T} \sum_{t=1}^{T}  b_k[t] \mathcal{R}_k^{\mathrm{tra}}[t] \geq \tau, \forall k, \nbthis \label{cons_min_rate} \\
			&\eqref{cons_traj_1p}-\eqref{cons_pris_tra},
		\end{align}
	\end{subequations}
	where
	\begin{align*}
		\mathcal{R}_k^{\mathrm{tra}}[t] = \log \left( 1 + \frac{\absshort{\vh_k^\H[t] \vw[t]}^2}{\sigma_{\mathrm{r}}^2 \normshort{\hrr^\H[t] \bPsi[t]}^2 + \sigma_{\mathrm{u}}^2} \right). \nbthis \label{def_Rtra_k_t}
	\end{align*}
	To solve \eqref{prob_tra_1}, we leverage the BCA framework and consider four sub-problems, including the UE scheduling,  UAV's trajectory,  beamforming optimization, and hybrid RIS beamforming design. However, compared to problem \eqref{prob_loc_1}, the numbers of variables in each subproblems of \eqref{prob_tra_1} significantly increase, which are linear with $T$. Therefore, a direct application of SCA would result in an extremely high computational complexity. To overcome this challenge as well as the mathematical difficulties of \eqref{prob_tra_1}, we provide an efficient solution with low complexity in the following subsections.
	
	\subsection{UE Scheduling Design}
	We first aim at solving $\bBt$ while fixing $(\bVt,\bWt,\bat)$. To bypass the difficulty of binary constraints \eqref{cons_scheduling_3p}, we relax binary variables $\bBt$ into continuous one, i.e. $b_k[t] \in [0,1], \forall k,t$ \cite{wu2018joint}. Thus, the UE scheduling can be optimized by solving the following convex linear program:
	\begin{align}
		\label{prob_scheduling}
		\underset{\tau, \bBt}{\textrm{maximize}}\ \tau,\ \textrm{subject to}\ \eqref{cons_scheduling_3p}, \eqref{cons_scheduling_1p}, \eqref{cons_min_rate}, 0 \leq b_k[t] \leq 1,\ \forall k,t.
	\end{align}
	
	\subsection{Optimization of UAV Beamforming}
	
	Given $(\bBt,\bVt,\bat)$, the optimal UAV beamformers $\bWt$ are found by solving
	\begin{align}
		\label{problem_power}
		\underset{\tau,\bWt}{\textrm{maximize}}\ \tau,\
		\textrm{subject to}\  \eqref{cons_puav_tra}, \eqref{cons_pris_tra}, \eqref{cons_min_rate}.
	\end{align}
    Towards an appealing application, we will find a closed-form solution to this problem that provides more insights into the system design and requires much lower complexity. To this end, we rewrite constraint \eqref{cons_min_rate} as
	$
	\frac{1}{T} \sum_{t=1}^{T}  b_k[t] \log \left( 1 + \frac{\absshort{\vh_k^\H[t] \vw[t]}^2}{\sigma_k^2[t]} \right) \geq \tau, \forall k,
	$
	where $\sigma_k^2[t] \triangleq \sigma_{\mathrm{r}}^2 \normshort{\hrr^\H[t] \bPsi[t]}^2 + \sigma_{\mathrm{u}}^2$ is constant with respect to $\bWt$. The left hand side (LHS) of this constraint monotonically increases with $\absshort{\vh_k^\H[t] \vw[t]}^2$. Thus, the optimal beamforming vector admits the following form
	\begin{align*}
		\vw[t]^{\star} = \sqrt{p[t]^{\star}} \frac{\vh_k[t]}{\norm{\vh_k[t]}},\ \forall t, \nbthis \label{eq_w_sol}
	\end{align*}
	where $p[t]^{\star}$ is the optimal transmit power of the UAV at time slot $t$  and $p[t]^{\star} \leq \ptmax$. We note the fact that $\norm{\vw[t]^{\star}}^2 = p[t]^{\star}$ and rewrite constraints \eqref{cons_min_rate}, \eqref{cons_puav_tra}, and \eqref{cons_pris_tra} equivalently as
	\begin{align*}
	&\frac{1}{T} \sum_{t=1}^{T}  b_k[t] \log \left( 1 + \frac{p[t]^{\star} \norm{\vh_k}^2}{\sigma_k^2[t]} \right) \geq \tau, \forall k,\\ 
	&0 \leq p[t]^{\star} \leq \ptmax, \forall t,\\
	&\sum_{n \in \setA} \abs{\an[t]}^2 \left(\sigma_{\mathrm{r}}^2 +  \norm{\vh_{1,n}[t]}^2 p[t]^{\star} \right) \leq \prismax, \forall t, 
	\end{align*}
	respectively. These reveal the optimal power allocation
	\begin{align*}
		p^{\star} [t] = \min \left\{ \ptmax, \frac{\prismax - \sum_{n \in \setA} \abs{\an[t]}^2 \sigma_{\mathrm{r}}^2}{\sum_{n \in \setA} \norm{\vh_{1,n}[t]}^2} \right\},\ \forall t. \nbthis \label{eq_p_sol}
	\end{align*}
	Finally, $\{\vw[t]^{\star}\}$ is reconstructed by \eqref{eq_w_sol}.

	\subsection{Optimization of the UAVs' Trajectory}
	With given $(\bBt,\bWt,\bat)$, the UAV's trajectory can be optimized by solving the following problem with respect to $\{\tau,\bVt\}$:
	\begin{align}
		\label{subprob_tra}
		\underset{\tau,\bVt}{\textrm{maximize}}\ \tau,\
		\textrm{subject to}\  \eqref{cons_traj_1p}, \eqref{cons_traj_2p}, \eqref{cons_pris_tra},
		\eqref{cons_min_rate},
	\end{align}
	where constraints \eqref{cons_traj_2p},  \eqref{cons_pris_tra} and \eqref{cons_min_rate} are nonconvex. Furthermore, $\bVt$ is hidden in all these constraints. To tackle these challenges, we transform constraints \eqref{cons_pris_tra} and \eqref{cons_min_rate} into equivalent constraints but more tractable forms.
	
	First, we expand the expression of $\mathcal{R}_k^{\mathrm{tra}}[t]$ in \eqref{eq_rate} to show its dependence on $\bV$. By a similar derivation as in \eqref{def_f_mk_v}, we obtain
	\begin{align*}
	    \absshort{\vh_{k}^\H[t] \vw[t]}^2 &= \bar{c}_{0,k}[t] \norm{\vv[t] - \vu_k}^{-\epsilon_0} + \bar{c}_{1,k}[t] \norm{\vv[t] - \vr}^{-\epsilon_1}\\
     &\qquad + \bar{c}_{2,k}[t] \norm{\vv[t] - \vu_k}^{\frac{-\epsilon_0}{2}} \norm{\vv[t] - \vr}^{\frac{-\epsilon_1}{2}},
	\end{align*}
	where
        \begin{align*}
            \bar{c}_{0,k}[t] &\triangleq \zeta_0 \absshort{\gd^\H[t] \vw[t]}^2,\\
            \bar{c}_{1,k}[t] &\triangleq \zeta_0 \absshort{\hrr^\H \bUpsilon[t]  \grt[t] \vw[t]}^2,\\
            \bar{c}_{2,k}[t] &\triangleq 2  \zeta_0 \re{\vw_k^\H[t]  \gd[t] \hrr^\H \bUpsilon[t] \grt[t] \vw[t]},
        \end{align*}
        are constants with respect to variable $\vv[t]$. Thus, $\mathcal{R}_k^{\mathrm{tra}}[t]$ in \eqref{eq_bar_Rk} can be rewritten as
	\begin{align*}
		\mathcal{R}_k^{\mathrm{tra}}[t] = &\log \Big( 1 + \rho_k[t] \Big(\bar{c}_{0,k}[t] \norm{\vv[t] - \vu_k}^{-\epsilon_0}\\
        &\qquad + \bar{c}_{1,k}[t] \norm{\vv[t] - \vr}^{-\epsilon_1} \\
		&\qquad + \bar{c}_{2,k}[t] \norm{\vv[t] - \vu_k}^{-\epsilon_0/2} \norm{\vv[t] - \vr}^{-\epsilon_1/2}\Big) \Big),
	\end{align*}
	where $\rho_k[t] \triangleq \left(\sigma_{\mathrm{r}}^2 \normshort{\hrr^\H \bPsi[t]}^2 + \sigma_{\mathrm{u}}^2\right)^{-1}$ is constant with $\bVt$.
	It is observed that although $\mathcal{R}_k^{\mathrm{tra}}[t]$ has a simpler form than $\mathcal{R}_k^{\mathrm{loc}}$ in \eqref{eq_rate_loc_2}, it inherits the mathematically intractable challenges of $\mathcal{R}_k^{\mathrm{loc}}$. Specifically, it is still a very complicated function of $\{\vv[t]\}$ with the note that while $\bar{c}_{0,k}[t] > 0$, $\bar{c}_{1,k}[t] > 0, \forall k, t$, the sign of $\bar{c}_{2,k}[t]$ varies depending on $k$ and $t$. We apply a similar approach presented in problem \eqref{subprob_loc} to tackle constraint \eqref{cons_min_rate}. Specifically, let us introduce slack variables $\{\hat{v}_{0,k}[t]\}$ and $\{\hat{v}_{1}[t]\}$, satisfying 
	\begin{align*}
		\hat{v}_{0,k}[t]  &\leq \norm{\vv[t] - \vu_k}^{-\epsilon_0/2},\ \forall k, t,\\
		\hat{v}_{1}[t] &\leq  \norm{\vv[t] - \vr}^{-\epsilon_1/2},\ \forall t, \nbthis \label{cons_v1hat_tra}	
	\end{align*}
	and $\{\bar{v}_{0,k}[t]\}$, $\{\bar{v}_{1}[t]\}$ satisfying
	\begin{align*}
		&\begin{cases*}
			\bar{v}_{0,k}[t] = \hat{v}_{0,k}[t], \text{if~} \bar{c}_{2,k}[t] \geq 0\\
			\bar{v}_{0,k}[t] \geq \norm{\vv[t] - \vu_k}^{-\epsilon_0/2}, \text{otherwise}
		\end{cases*}, \forall k, t, \\ 
		&\begin{cases*}
			\bar{v}_{1}[t] = \hat{v}_{1}[t], \text{if~} \bar{c}_{2,k}[t] \geq 0\\
			\bar{v}_{1}[t] \geq  \norm{\vv[t] - \vr}^{-\epsilon_1/2}, \text{otherwise}
		\end{cases*}, \forall t.  \nbthis \label{cons_v1check}
	\end{align*}
	$\mathcal{R}_k^{\mathrm{tra}}[t]$ is then lower bounded as
        \begin{align*}
            \mathcal{R}_k^{\mathrm{tra}}[t] &\geq \log \Big(1 + \rho_k[t] \Big(\bar{c}_{0,k}[t] (\hat{v}_{0,k}[t])^2 + \bar{c}_{1,k}[t] (\hat{v}_{1}[t])^2\\
            &\hspace{4cm}+ \bar{c}_{2,k}[t]\bar{v}_{0,k}[t] \bar{v}_{1}[t]\Big)\Big).
        \end{align*}
	To further simplify it, we introduce slack variables $\{\hat{a}_k[t]\}$ satisfying
	\begin{align*}
		\hat{a}_k[t] &\leq \begin{cases*}
			\bar{c}_{0,k}[t] (\hat{v}_{0,k}[t])^2 + \bar{c}_{1,k}[t] (\hat{v}_{1}[t])^2\\
            \hspace{1cm}+ \abs{\bar{c}_{2,k}[t]} \bar{v}_{0,k}[t] \bar{v}_{1}[t],\ \text{if~} \bar{c}_{2,k}[t] \geq 0\\
			\bar{c}_{0,k}[t] (\hat{v}_{0,k}[t])^2 + \bar{c}_{1,k}[t] (\hat{v}_{1}[t])^2\\
            \hspace{1cm} - \abs{\bar{c}_{2,k}[t]} \bar{v}_{0,k}[t] \bar{v}_{1}[t],\ \text{otherwise}
		\end{cases*},\ \forall k, t. \nbthis \label{def_f_hat}
	\end{align*}
	Thus, \eqref{cons_min_rate} is transformed to the set of constraints \eqref{cons_v1hat_tra}-\eqref{def_f_hat} and
	\begin{align*}
		\frac{1}{T} \sum_{t=1}^{T}  b_k[t] \log\Bigl( 1 + \rho_k[t] \hat{a}_k[t] \Bigr)  \geq \tau, \forall k, \nbthis \label{cons_min_ratep_tra}
	\end{align*}
	which is convex with respect to $\{\hat{a}_k[t]\}$. To handle the nonconvex constraints \eqref{cons_v1hat_tra}-\eqref{def_f_hat}, we apply the inequalities  \eqref{approx_power}--\eqref{approx_bil_pos} to convexify them as
	\begin{subequations}
		\begin{align*}
			&\norm{\vv[t] - \vu_k} + F_{\mathtt{pow}}(\hat{v}_{0,k}[t];-2/\epsilon_0,\hat{v}_{0,k}[t]^{(i)}) \leq 0, \forall k, t, \nbthis \label{cons_v0hat_tra_2}\\
			&\norm{\vv[t] - \vr} + F_{\mathtt{pow}}(\hat{v}_{1}[t];-2/\epsilon_1,\hat{v}_{1}[t]^{(i)}) \leq 0,\ \forall t, \nbthis \label{cons_v1hat_tra_2}\\
			&\begin{cases*}
				\bar{v}_{0,k}[t] = \hat{v}_{0,k}[t],\ \text{if~} \bar{c}_{2,k}[t] \geq 0 \\
				F_{\mathtt{qua}}(\vv[t]; \vu_k,\vv[t]^{(i)}) \leq 1/F_{\mathtt{pow}}(\bar{v}_{0,k}[t];4/\epsilon_0,\bar{v}_{0,k}[t]^{(i)}),
			\end{cases*}\\
                &\hspace{5.25cm}  \text{otherwise}, \forall k, t, \nbthis \label{cons_v0check_tra_2}\\
			&\begin{cases*}
				\bar{v}_{1}[t] = \hat{v}_{1}[t],\ \text{if~} \bar{c}_{2,k}[t] \geq 0\\
				F_{\mathtt{qua}}(\vv[t]; \vr,\vv[t]^{(i)}) \leq 1/F_{\mathtt{pow}}(\bar{v}_{1}[t];4/\epsilon_1,\bar{v}_{1}[t]^{(i)}), 
			\end{cases*},\\ 
                &\hspace{5.5cm} \text{otherwise}, \forall  t, \nbthis \label{cons_v1check_2}\\
			&\hat{a}_k[t] \leq \bar{c}_{0,k}[t] F_{\mathtt{qua}}(\hat{v}_{0,k}[t]; 0,\hat{v}_{0,k}[t]^{(i)})\\
                &\hspace{1.5cm}+ \bar{c}_{1,k}[t] F_{\mathtt{qua}}(\hat{v}_{1}[t];0,\hat{v}_{1}[t]^{(i)}) \\
			&\hspace{1.5cm}+ \abs{\bar{c}_{2,k}[t]} F_{\mathtt{bil}} (\bar{v}_{0,k}[t],\bar{v}_{1}[t];\\
            &\hspace{2cm} \mathrm{sgn}(\bar{c}_{2,k}[t]), \bar{v}_{0,k}[t]^{(i)},\bar{v}_{1}[t]^{(i)}),\ \forall k, t, \nbthis \label{def_f_hat_2}
		\end{align*}
	\end{subequations}
	around $\{\hat{v}_{0,k}[t]^{(i)}, \hat{v}_{1}[t]^{(i)},\bar{v}_{0,k}[t]^{(i)}, \bar{v}_{1}[t]^{(i)},\hat{a}_k[t]^{(i)},\tilde{v}_{1}[t]^{(i)}\}$.
	
	To address constraint \eqref{cons_pris_tra}, we first rewrite $\xi_n[t]$ as
	$
	\xi_n[t]  = \sigma_{\mathrm{r}}^2 +  c_{3,n}[t] \left(\norm{\vv[t] - \vr}^{-\frac{\epsilon_1}{2}}\right)^2,
	$
	where $c_{3,n}[t] \triangleq \zeta_0 \norm{\vg_{1,n}[t]}^2  \sum_{k=1}^{K} \norm{\vw[t]}^2$ is independent of $\bVt$. Thus, slack variables $\{\tilde{v}_{1}[t]\}$ satisfying
	$
	\tilde{v}_{1}[t] \geq \norm{\vv[t] - \vr}^{-\epsilon_1/2}
	$
	are introduced to transform \eqref{cons_pris_loc} to the set of following convex constraints:
	\begin{align*}
		\hspace{-0.5cm}
		&F_{\mathtt{qua}}(\vv[t]; \vr,\vv[t]^{(i)}) \leq \frac{1}{F_{\mathtt{pow}}(\tilde{v}_{1}[t];4/\epsilon_1,\tilde{v}_{1}[t]^{(i)})},\\ 
		&\sum_{n \in \setA} \abs{\an[t]}^2 \left(\sigma_{\mathrm{r}}^2 +  c_{3,n}[t] (\tilde{v}_{1}[t])^2 \right) \leq \prismax, \forall t. \nbthis \label{eq_pris_approx}
	\end{align*}
	In summary, we can approximate problem \eqref{prob_tra} at iteration $i$  by the following convex program:
	\begin{align}
		\label{prob_trajectory_1}
		\underset{\tau,\bVt,\boldsymbol{\mathcal{V}}}{\textrm{maximize}}\ \tau,\
		\textrm{subject to}\
		\eqref{cons_traj_1p}, \eqref{cons_traj_2p},
		\eqref{cons_min_ratep_tra}, \eqref{cons_v0hat_tra_2}-\eqref{def_f_hat_2}, \eqref{eq_pris_approx}
	\end{align}
	where $\boldsymbol{\mathcal{V}}[t] \triangleq \{\hat{v}_{0,k}[t], \hat{v}_{1}[t],\bar{v}_{0,k}[t], \bar{v}_{1}[t],\hat{a}_k[t],\tilde{v}_{1}[t]\}$.

	\subsection{Optimization of Hybrid RIS Coefficients}
	
	Finally, we complete the solution to problem \eqref{prob_tra_1} by solving $\bat$ in the following problem:
	%
	\begin{align}
		\label{subprob_ris_tra}
		\underset{\substack{\tau,\bat}}{\textrm{maximize}}\ \tau,
		\textrm{subject to}\
		\eqref{cons_ris_phase_tra}-\eqref{cons_pris_tra}, \eqref{cons_min_rate},
	\end{align}
	with given $(\bBt, \bVt, \bWt)$. Note that constraint \eqref{cons_min_rate} is nonconvex. To examine the role of $\{ \an \}$ and to make \eqref{cons_min_rate} more tractable, we substitute \eqref{eq_h_eff} into \eqref{eq_rate}. Furthermore, by recalling that $\bUpsilon[t] = \bPhi[t] + \bPsi[t]$, we can rewrite \eqref{def_Rtra_k_t} as
	$
	\mathcal{R}_k^{\mathrm{tra}}[t] = \log \left(1 + \frac{\abs{\bar{h}_{0,k}[t] + \hrr^\H (\bPsi[t] + \bPhi[t]) \bar{\vh}_{1,k}[t]}^2}{ \sigma_{\mathrm{r}}^2 \normshort{\hrr^\H \bPsi[t]}^2 + \sigma_{\mathrm{u}}^2}\right),
	$
	where $\bar{h}_{0,k}[t] \triangleq \hd^\H[t] \vw[t]$, $\bar{\vh}_{1,k}[t] \triangleq \mH_1[t] \vw[t]$; furthermore, we note that $\tilde{\mH}_{2,k} = \diagshort{\hrr^\H}$, and $\bPhi[t]$ and $\bPsi[t]$ are diagonal matrices containing only the passive and active coefficients, respectively. Let $\vpsi[t] \triangleq \diag{\bPsi[t]}\in \setC^{N \times 1}$, $\vphi[t] \triangleq \diag{\bPhi[t]}\in \setC^{N \times 1}$ and $\tilde{\vh}_{12,k}[t] \triangleq \tilde{\mH}_{2,k} \bar{\vh}_{1,k}[t]$. Following the development in Section \ref{sec_opt_loc_alpha}, we can rewrite $\mathcal{R}_k^{\mathrm{tra}}[t]$ as
	\begin{align*}
		\mathcal{R}_k^{\mathrm{tra}}[t] = \log \left(1 + \frac{ \abs{\bar{h}_{0,k}[t] + \left(\vpsi[t] + \vphi[t]\right)^\T \tilde{\vh}_{12,k}[t]}^2}{ \sigma_{\mathrm{r}}^2 \normshort{\vpsi[t]^\T \tilde{\mH}_{2,k}}^2 + \sigma_{\mathrm{u}}^2}\right). \nbthis \label{eq_rate_k_2}
	\end{align*}

	\begin{remark}
		\label{rm_passive_coefficient}
		For the conventional passive RIS, i.e., when $\setA = \emptyset$ and $\vpsi[t] = \boldsymbol{0}$, the denominator of the SINR term in \eqref{eq_rate_k_2} is $\sigma_u^2$. Therefore, the optimal solutions to $\{ \an[t] \}$ are those maximizing the received signal power at the UE scheduled at time slot $t$, given as
		$
		\an[t]^{\star} = e^{j \theta_n[t]^{\star}},\ \text{where~} \theta_n[t]^{\star} = \angle \bar{h}_{0,k}[t] - \angle \tilde{h}_{12,kn}[t], \forall n,t,
		$
		with $\tilde{h}_{12,kn}$ being the $n$th element of $\tilde{\vh}_{12,k}$. However, for the hybrid RIS, this solution is not valid anymore due to the additional noise $\sigma_{\mathrm{r}}^2 \normshort{\vpsi[t]^\T \tilde{\mH}_{2,k}}^2$.
	\end{remark}
	
	The solution to problem \eqref{subprob_ris_tra} can be found by jointly optimizing RIS passive and active coefficients, as done for \eqref{subprob_ris_loc}. However, the resultant solution requires a complexity of at least $\mathcal{O}(N^{3.5})$, which is high due to a large value of $N$. To overcome this, based on Remark \ref{rm_passive_coefficient},  we decouple the optimization of active and passive elements of the RIS. More specifically, with given $\vpsi$, the solution to $\vphi$ is obtained by solving the following problem
	\begin{align}
		\label{problem_phi}
		\underset{\substack{\tau,\vphi}}{\textrm{maximize}}\ \tau,
		\textrm{subject to}\ 
		\eqref{cons_ris_phase_tra}, \eqref{cons_passive_modul_tra}, \eqref{cons_min_rate},
	\end{align}
	and then, the solution to $\psi$ is obtained by solving
	%
	\begin{align}
		\label{problem_psi}
		\underset{\substack{\tau,\vpsi}}{\textrm{maximize}}\ \tau,
		\textrm{subject to}\
		\eqref{cons_ris_phase_tra}, \eqref{cons_active_modul_tra}, \eqref{cons_pris_tra}, \eqref{cons_min_rate}.
	\end{align}
	This approach allows solving \eqref{subprob_ris_tra} with much lower complexity as \eqref{problem_phi} admits a closed-form solution, while the number of variables in \eqref{problem_psi}, i.e., the number of active coefficients to optimize, is just $\Na \ll N$. We elaborate these solutions in the following.
	
	\subsubsection{Solution to Problem \eqref{problem_phi}}
	It is clear from \eqref{eq_rate_k_2} that with given $\{ \vpsi[t] \}$, the optimal solutions to the passive coefficients are
	\begin{align*}
		\phi_n[t]^{\star} = e^{j \theta_n[t]^{\star}},\ \forall n \notin \setA, \forall t. \nbthis \label{eq_opt_phase_passive}
	\end{align*}
	where $\theta_n[t]^{\star} = \angle \left(\bar{h}_{0,k}[t] + \vpsi[t]^\T \tilde{\vh}_{12,k}[t]\right) - \angle \tilde{h}_{12,kn}[t]$.
	
	\subsubsection{Solution to Problem \eqref{problem_psi}}
	We first aim at tackling the nonconvexity of constraint \eqref{cons_min_rate}. Let us denote $\bar{\bar{h}}_{0,k}[t] \triangleq \bar{h}_{0,k}[t] + \vphi[t]^\T \tilde{\vh}_{12,k}[t]$. Then, by several algebraic operations, we can express the numerator and denominator of the SINR term in \eqref{eq_rate_k_2} as
	$\absshort{\bar{h}_{0,k}[t] + \left(\vpsi[t] + \vphi[t]\right)^\T \tilde{\vh}_{12,k}[t]}^2 = \absshort{\bar{\bar{h}}_{0,k}[t] + \vpsi[t]^\T \tilde{\vh}_{12,k}[t]}^2  = \vpsi[t]^\H \mD_k[t] \vpsi[t] + 2 \re{\vpsi[t]^\H \vd_k[t]} + z_k[t]
	$
	where $\mD_{k}[t] = \tilde{\vh}_{12,k}^*[t] \tilde{\vh}_{12,k}^\T[t]$, $\vd_{k}[t] = \tilde{\vh}_{12,k}^*[t] \bar{\bar{h}}_{0,k}[t]$, and $z_{k} = \absshort{\bar{\bar{h}}_{0,kk}}^2$, and 
	$\sigma_{\mathrm{r}}^2 \normshort{\vpsi[t]^\T \tilde{\mH}_{2,k}}^2 + \sigma_{\mathrm{u}}^2 = \vpsi[t]^\H \tilde{\mD}_{k} \vpsi[t] + \sigma_{\mathrm{u}}^2$,
	where $\tilde{\mD}_k \triangleq \tilde{\mH}_{2,k} \tilde{\mH}_{2,k}^\H$. Note that $\mD_k[t], \vq_k[t], \tilde{\mD}_k$, and $z_k[t]$ are all constant with respect to $\vpsi[t]$. Now, $\mathcal{R}_k^{\mathrm{tra}}[t]$ can be represented as
	\begin{align*}
		\mathcal{R}_k^{\mathrm{tra}}[t]\!&=\!\log \left(\!1 +\!\frac{\vpsi[t]^\H \mD_k[t] \vpsi[t] + 2 \re{\vpsi[t]^\H \vd_k[t]} + z_k[t]}{\vpsi[t]^\H \tilde{\mD}_{k} \vpsi[t] + \sigma_{\mathrm{u}}^2}\right)\\
        &= \bar{r}_k[t] - \tilde{r}_k[t],
	\end{align*}
	where $\bar{r}_k[t] \triangleq \log \left(\vpsi[t]^\H \bar{\mD}_k[t] \vpsi[t] + 2 \re{\vpsi[t]^\H \vd_k[t]} + \bar{z}_{k}[t]\right)$, and $\tilde{r}_k[t] \triangleq \frac{1}{T} \sum_{t=1}^{T}  b_k[t]  \log \Big(\vpsi[t]^\H$ $\tilde{\mD}_{k} \vpsi[t] + \sigma_{\mathrm{u}}^2\Big)$,
	with $\bar{\mD}_k[t] = \mD_k[t] + \tilde{\mD}_k$, and $\bar{z}_{k}[t] = z_k[t] + \sigma_{\mathrm{u}}^2$. Thus, constraint \eqref{cons_min_rate} can be rewritten as
	$
	\bar{r}_k[t] \geq \tau + \tilde{r}_k[t], \forall k, t,
	$
	which has the same form as \eqref{cons_min_rate_loc_1}. Therefore, by applying similar transformations as in \eqref{cons_min_rate_loc_1}--\eqref{cons_min_rate_loc_2b}, we can transform it to the set of convex constraints:
	\begin{align*}
		\bar{r}_{\mathrm{lb},k}^{(i)}[t] \geq \tau + \tilde{r}_{\mathrm{ub},k}^{(i)}[t],\ 
		\vartheta_{k}[t] \geq \vpsi[t]^\H \tilde{\mD}_{k} \vpsi[t],\ \forall k, t, \nbthis \label{cons_min_rate_2b}
	\end{align*}
	where
	$
	\bar{r}_{\mathrm{lb},k}^{(i)}[t] \triangleq \log \Big(2 \re{\vpsi[t]^\H \vd_k[t]} + \bar{z}_{k}[t] - F_{\mathtt{qua}}( \bar{\mD}_k^{\frac{1}{2}}[t] \vpsi[t]; \boldsymbol{0}, \bar{\mD}_k^{\frac{1}{2}}[t] \vpsi[t]^{(i)} )  \Big)$ and
	$\tilde{r}_{\mathrm{ub},k}^{(i)}[t] \triangleq \log\Big( \vartheta_k[t]^{(i)} + \sigma_{\mathrm{u}}^2 \Big) + \frac{\vartheta_k[t] - \vartheta_k[t]^{(i)}}{\vartheta_k[t]^{(i)} + \sigma_{\mathrm{u}}^2}.
	$
	Furthermore, similar to \eqref{cons_pris_loc2}, \eqref{cons_pris_tra} is equivalent to the quadratic convex constraint 
	\begin{align*}
		\vpsi[t]^\H \bXit \vpsi[t] \leq \prismax, \forall t, \nbthis \label{cons_pris_tra_1}
	\end{align*}
	where $\bXit = \diag{\tilde{\xi}_1[t], \ldots, \tilde{\xi}_N[t]}$, with $\tilde{\xi}_n[t] = \xi_n[t], n \in \setA$ and $\tilde{\xi}_n[t] = 0, n \notin \setA$.	Finally, problem \eqref{problem_psi} at iteration $i$ can be approximated by the following convex program:
	\begin{align}
		\label{problem_psi_1}
		\underset{\substack{\tau,\vpsi[t],\{\vartheta_k[t]\}}}{\textrm{maximize}}\ \tau,\
		\textrm{subject to}\ \eqref{cons_ris_phase_tra}, \eqref{cons_active_modul_tra}, \eqref{cons_min_rate_2b}-\eqref{cons_pris_tra_1}.
	\end{align}
	Once solutions $\{\phi_n[t]^{\star}\}$ and $\{\psi_n[t]^{\star}\}$ are found, $\{\alpha_n[t]^{\star}\}$ is readily obtained by
	\begin{align*}
		\alpha_n[t]^{\star} = \phi_n[t]^{\star} + \psi_n[t]^{\star}. \nbthis \label{eq_sol_alpha}
	\end{align*}
	
	\subsection{Overall Algorithm}
	The joint optimization of the UAV's transmit beamforming, trajectory planning and the RIS beamforming design is outlined in Algorithm \ref{alg_opt_tra}. Specifically, the initial solutions to $\{b_k[t]^{(0)}\}$, $\{\vw[t]^{(0)}\}$, $\{\vv[t]^{(0)}\}$, $\{\alpha_n[t]^{(0)}\}$, and $\{\boldsymbol{\mathcal{V}}[t]^{(0)}\}$ are first generated in Step 1. Then, in Steps 2--10, subproblems \eqref{prob_scheduling}, \eqref{problem_power}, \eqref{prob_trajectory_1}, \eqref{problem_phi} and \eqref{problem_psi_1} are alternatively solved. In each iteration, $\{b_k[t]^{(i)}\}$, $\{\vw[t]^{(i)}\}$, $\{\vv[t]^{(i)}\}$, $\{\alpha_n[t]^{(i)}\}$ and $\{\boldsymbol{\mathcal{V}}[t]^{(i)}\}$ are updated until convergence. We note that the solutions to \eqref{problem_power} and \eqref{problem_phi}, i.e., $\{\vw[t]^{(i)}\}$ and $\{\alpha_n[t]^{(i)}\}, \forall n \notin \setA$, are obtained based on their closed-forms with simple  mathematical operations (scalar multiplications and dot products). Therefore, most of the complexity of Algorithm \ref{alg_opt_tra} comes from solving subproblems \eqref{prob_scheduling}, \eqref{prob_trajectory_1} and \eqref{problem_psi_1}. These require complexities of $\mathcal{O}(\sqrt{2KT+K+T}K^{3} T^3)$, $\mathcal{O}(\sqrt{3KT+K+5T}(3KT+6T)^3)$, and  $\mathcal{O}(\sqrt{N(\Na+2K+T)} \Na^3 T^3)$, respectively. Therefore, the total complexity of Algorithm \ref{alg_opt_tra} is
	\begin{align*}
	    \mathcal{C}_{\mathrm{tra}} &= \mathcal{O} \big( \mathcal{I}_{\mathrm{tra}} \Big(\sqrt{2KT+K+T}K^{3} T^3 + \sqrt{3KT+K+5T} \\
     &\qquad \times (3KT+6T)^3  + \sqrt{N(\Na+2K+T)} \Na^3 T^3\big) \Big),
	\end{align*}
	where $\mathcal{I}_{\mathrm{tra}}$ is the number of iterations until convergence. It is observed that the major complexity of Algorithm \ref{alg_opt_tra} comes from the optimization over $T$ time slots.
	The monotonic convergence the objective value in Algorithm \ref{alg_opt_tra} can be shown similarly as in Algorithm \ref{alg_opt_loc}. 
	
	\begin{algorithm}[t]
		\small
		\caption{Proposed Iterative Algorithm to Solve Problem $(\mathcal{P}_{\mathrm{tra}})$ in \eqref{prob_tra}}
		\label{alg_opt_tra}
		\begin{algorithmic}[1]
			\STATE \textbf{Initialization:} Set $i=0$. Generate initial values $\{b_k[t]^{(0)}\}$, $\{\vw[t]^{(0)}\}$, $\{\vv[t]^{(0)}\}$, and $\{\psi_n[t]^{(0)}\}$. Obtain $\{\phi_n[t]^{(0)}\}$ based on \eqref{eq_opt_phase_passive} and obtain  $\{\alpha_n[t]^{(0)}\}$ based on \eqref{eq_sol_alpha}.  Initialize $\{\boldsymbol{\mathcal{V}}[t]^{(0)}\}$ and $\{\vartheta_k[t]^{(0)}\}$ by setting  \eqref{cons_v1hat_tra}--\eqref{cons_v1check} and \eqref{cons_min_rate_2b} to be equalities.
			
			\REPEAT
			
			\STATE Solve problem \eqref{prob_scheduling} for given $\{\vw[t]^{(i)}\}$, $\{\vv[t]^{(i)}\}$, $\{\alpha_n[t]^{(i)}\}$ to obtain solution $\{b_k[t]^{\star}\}$. Set $ \{b_k[t]^{(i+1)}\} = \{b_k[t]^{\star}\}$.
			
			\STATE Obtain $\{\vw[t]^{\star}\}$ based on \eqref{eq_w_sol} and \eqref{eq_p_sol}. Set $ \{\vw[t]^{(i+1)}\} = \{\vw[t]^{\star}\}$.
			
			\STATE Solve problem \eqref{prob_trajectory_1} with given $\{b_k[t]^{(i+1)}\}$, $\{\vw[t]^{(i+1)}\}$, $\{\alpha_n[t]^{(i)}\}$, $\{\boldsymbol{\mathcal{V}}[t]^{(i)}\}$ to obtain solution $\{\vv[t]^{\star}\}$ and $\{\boldsymbol{\mathcal{V}}[t]^{\star}\}$. Set $\{\vv[t]^{(i+1)}\} = \{\vv[t]^{\star}\}$ and $\{\boldsymbol{\mathcal{V}}[t]^{(i+1)}\} = \{\boldsymbol{\mathcal{V}}[t]^{\star}\}$.
			
			\STATE Obtain $\{\phi_n[t]^{\star}\}$ based on \eqref{eq_opt_phase_passive}. Set $\{\phi_n[t]^{(i+1)}\} = \{\phi_n[t]^{\star}\}$.
			
			\STATE Solve problem \eqref{problem_psi_1} with given $\{b_k[t]^{(i+1)}\}$, $\{\vw[t]^{(i+1)}\}$, $\{\vv[t]^{(i+1)}\}$, $\{\phi_n[t]^{(i+1)}\}$, $\{\vartheta_k[t]^{(i)}\}$, and $\{\psi_n[t]^{(i)}\}$  to obtain solutions $\{\vpsi[t]^{\star}\}$ and $\{\vartheta_k[t]^{\star}\}$. Set $\{\vpsi[t]^{(i+1)}\} = \{\vpsi[t]^{\star}\}$ and $\{\vartheta_k[t]^{(i+1)}\} = \{\vartheta_k[t]^{\star}\}$.
			
			\STATE Obtain $\{\alpha_n[t]^{\star}\}$ based on \eqref{eq_sol_alpha}. Set $\{\alpha_n[t]^{(i+1)}\} = \{\alpha_n[t]^{\star}\}$.
			
			\STATE Update $i=i+1$.
			\UNTIL Convergence.
		\end{algorithmic}
	\end{algorithm}
	
	\section{Simulation Results}
	\label{sec_sim_result}
	
	In this section, numerical results are provided to validate the Algorithms \ref{alg_opt_loc} and \ref{alg_opt_tra} as well as the performance of the hybrid RIS. We assume that the UAV flies at a fixed altitude of $z_0 = 100$ m, the UEs are randomly and uniformly distributed in an area of $D \times D$ $\text{m}^2$, and the RIS is deployed at $(D/2,D,50)$ in a three-dimensional (3D) coordinate system, as will be illustrated latter in Fig.\ \ref{fig_loc_tra}. We set $\delta_t = 0.1$ s and $v_{\mathrm{max}} = 50$ m/s.
	
	The small-scale fading channels $\{ \gd, \grt, \grr \}, \forall k$ are modeled as follows. First, we assume that $\gd$ follows the Rayleigh fading model, i.e., $g_{0,kt} \sim \mathcal{CN}(0,1), t = 1,\ldots,N_t$, $\forall k$. This is because the LoS links between  UAV and ground UEs are easily blocked by high buildings, while the scattering components are often extensive, especially in the low-to-medium frequency band and complex urban environments \cite{li2021reconfigurable, cao2021reconfigurable, li2020reconfigurable, ge2020joint}. In contrast, the channels between  RIS and  UEs, i.e., $\grr, \forall k$, are assumed to follow the Rician fading model \cite{li2021reconfigurable, cao2021reconfigurable} with Rician factor $\kappa$, i.e., $\grr = \sqrt{\frac{\kappa}{\kappa+1}} \grr^{\mathrm{LoS}} + \sqrt{\frac{1}{\kappa+1}} \grr^{\mathrm{NLoS}}, \forall k$, where $\grr^{\mathrm{NLoS}}  \sim \mathcal{CN}(\boldsymbol{0},\mI_N)$ represents the non-LoS (NLoS) channels, and $\grr^{\mathrm{LoS}}$ represents the deterministic LoS component of $\grr$. The channel between  UAV and  RIS, $\grt$, is modeled as LoS deterministic channel because both  UAV and RIS are located at certain heights so that the channel between them is barely blocked by any obstacles \cite{li2021reconfigurable, cao2021reconfigurable}. Furthermore, we assume a uniform planar array (UPA) of $N = N_x \times N_y$ elements for the RIS, with $\{N_x, N_y\}$ being the numbers of elements on each row and column, respectively. Thus, the $n$th elements of $\grt$ and $\grr^{\mathrm{LoS}}$ can be given as $g_{1,n} = \exp \left(j \frac{2\pi d_0}{\lambda} f\left(n, \phi_{1,}^{\mathrm{azi}}, \phi_{1,}^{\mathrm{ele}}\right)\right)$ and $g^{\mathrm{LoS}}_{2,kn} = \exp\left(j \frac{2\pi d_0}{\lambda} f \left(n, \phi_{2,k}^{\mathrm{azi}}, \phi_{2,k}^{\mathrm{ele}}) \right)\right)$, respectively \cite{zhang2020capacity}, where $d_0$ and $\lambda$ are the antenna separation and carrier wavelength, respectively; $\{ \phi_{1,}^{\mathrm{azi}}, \phi_{1,}^{\mathrm{ele}} \}$ and $\{ \phi_{2,k}^{\mathrm{azi}}, \phi_{2,k}^{\mathrm{ele}} \}$ denote the azimuth and elevation angle	
	of-arrival/departure (AoA/AoD) of the RIS associated with the UAV-RIS and RIS-UE channels, respectively; and, $f(n,\phi,\phi') \triangleq \lfloor \frac{n}{N_x} \rfloor \sin \phi \sin \phi' + \left( n - \lfloor \frac{n}{N_x} \rfloor N_x \right) \sin \phi \cos \phi'$ \cite{zhang2020capacity}. Then, $\{ \hd^\H, \hrt, \hrr^\H \}$ are obtained based on \eqref{eq_h2}.

	In the following simulations, we set {$\zeta_0 = -30$ dB}, $\{\epsilon_0, \epsilon_1, \epsilon_2\} = \{3.2,2.0,2.2\}$ and $\kappa = 10$ \cite{mu2021intelligent}. The system bandwidth is set to $20$ MHz, corresponding to $\sigma^2_{\mathrm{u}} = -80$ dBm. The total power of noise and residual SI of the RIS is computed as $\sigma^2_{\mathrm{r}} = (\eta + 1) \sigma^2_{\mathrm{u}}$, with $\eta = 1$ dB reflecting the possible residual SI caused by active elements operating in the full-duplex mode \cite{malik2018optimal, nguyen2021hybrid}. The positions of  RIS active elements are fixed to $\setA = \{1,\ldots,\Na\}$. To solve the convex subproblems, we use the  modeling toolbox YALMIP with solver MOSEK. The convergence stopping criteria is set to $\varepsilon = 10^{-4}$. {For initialization of Algorithm \ref{alg_opt_loc}, we set $\vv^{(0)} = [D/2,D/2,z_0]$, i.e., above the center of the area, $\vw_k^{(0)} = \sqrt{\frac{\ptmax}{K}} \frac{\gd}{\normshort{\gd}}, \forall k$, and $\an^{(0)} = r e^{j \theta_n^{(0)}}$, with $\theta_n^{(0)}$ being randomly generated, $\forall n$, and $r$ satisfying \eqref{cons_pris_loc}. For Algorithm \ref{alg_opt_tra}, $\{b_k[t]\}$ are simply initialized as $b_1[t]^{(0)} = 1, \forall t$ and $b_j[t]^{(0)} = 0, \forall t, j > 1$, i.e., UE $1$ is scheduled for initialization, and thus, $\vw[t]^{(0)}$ is set to $\sqrt{\ptmax} \frac{\vg_{0,1}[t]}{\normshort{\vg_{0,1}[t]}}$. Similar to $\an^{(0)}$ in Algorithm \ref{alg_opt_loc}, we initialize $\psi_n[t]^{(0)} = r e^{j \theta_n[t]^{(0)}}$ with random phase shifts $\{\theta_n[t]^{(0)}\}$ and $r$ satisfying \eqref{cons_pris_tra}. Furthermore, $\vv[t]^{(0)}$ is initialized based on the low-complexity circular trajectory scheme with details being presented in \cite{wu2018joint}.}
	
	\renewcommand{\arraystretch}{1.1}
	\begin{table*}[t!]
		\small
		\begin{center}
			\caption{{Simulation parameters for the static- and mobile-UAV systems.}}
			\label{tab_complexity}
			\begin{tabular}{|c|c|c|c|}
					\hline
					Parameters & Values & Parameters & Values \\
					\hline
					\hline
					Deployment area ($D \times D$) & $D = \{50, 200\}$ m & Path loss at $1$ m & $\zeta_0 = -30$ dB  \\
					\hline
					RIS's location & $(D/2,D,50)$  & Path loss exponents & $\{\epsilon_0, \epsilon_1, \epsilon_2\} = \{3.2,2.0,2.2\}$\\
					\hline
					UAV's altitude & $z_0 = 100$ m  & Rician factor & $\kappa = 10$ \\
					\hline
					Time slot's length & $\delta_t = 0.1$ s  & System bandwidth & $20$ MHz \\
					\hline
					UAV's maximum speed & $v_{\max} = 50$ m/s  & Noise power & $\sigma_{\mathrm{u}}^2 = -80$ dBm \\
					\hline
			\end{tabular}
		\end{center}
	\end{table*}
	\raggedbottom
	
	\subsection{Convergence of Algorithms \ref{alg_opt_loc} and \ref{alg_opt_tra}}
	

	\begin{figure}[t]
		\centering
		\subfigure[Algorithm \ref{alg_opt_loc}]
		{
			\includegraphics[scale=0.6]{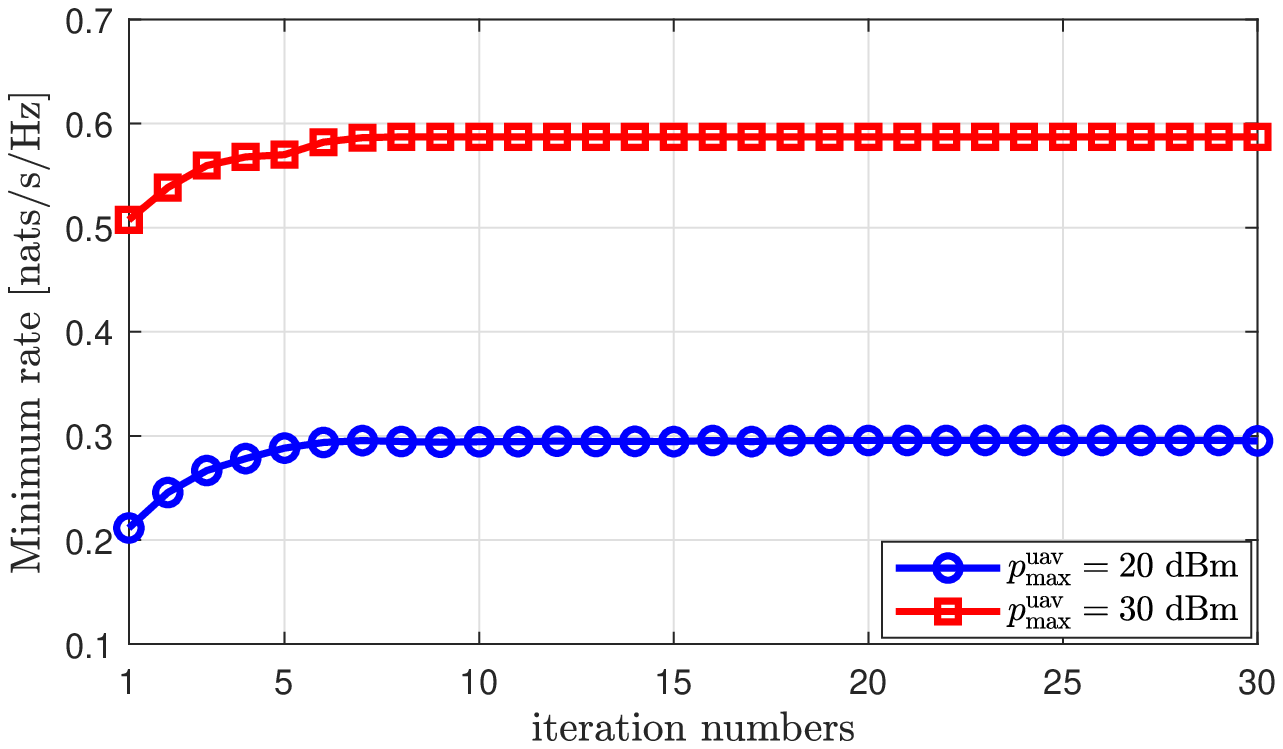}
			\label{fig_conv_A1}
		}
		\subfigure[Algorithm \ref{alg_opt_tra}, $T = 50$]
		{
			\includegraphics[scale=0.6]{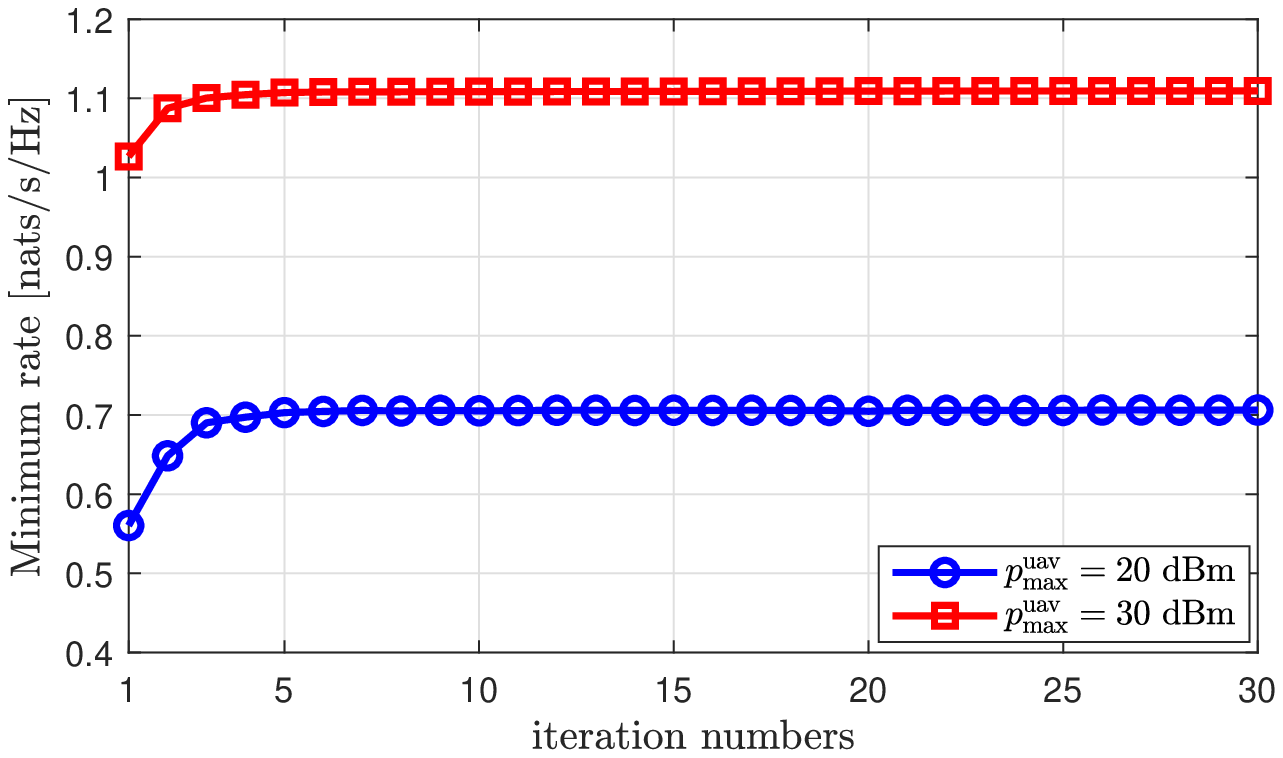}
			\label{fig_conv_A2}
		}
		\caption{Convergence of Algorithms \ref{alg_opt_loc} and \ref{alg_opt_tra} with $K = 4$, $N_t = 2$, $\Na = 2$, $N = 32$, $D = 200$ m, $\ptmax = \{20, 30\}$ dBm and $\prismax = 0$ dBm. }
		\label{fig_conv}
	\end{figure}
	
	We first show in Fig.\ \ref{fig_conv} the convergence of Algorithms \ref{alg_opt_loc} and \ref{alg_opt_tra} with $K = 4$, $N_t = 2$, $\Na = 2$, $N = 32$, $D = 200$ m, $\ptmax = \{20, 30\}$ dBm, $\prismax = 0$ dBm and $T = 50$. It is observed in Fig.\ \ref{fig_conv_A1} that Algorithm \ref{alg_opt_loc} converges after about eight to ten iterations. Whereas Algorithm \ref{alg_opt_tra} requires about five to eight iterations to converge, which is slightly faster than Algorithm \ref{alg_opt_loc}. The convergences are similar for both cases $\ptmax=\{20,30\}$ dBm. Furthermore, it is clear that Algorithm \ref{alg_opt_tra} achieves a larger minimum rate (0.7 nats/s/Hz with $\ptmax = 20$ dBm) compared to those obtained in Algorithm \ref{alg_opt_loc} (0.3 nats/s/Hz with $\ptmax = 20$ dBm). This is attributed to the fact that by employing the TDMA protocol, the high mobility of the UAV and the high active amplifying gain of the hybrid RIS can be fully exploited. {Furthermore, it is seen that Algorithms \ref{alg_opt_loc} and \ref{alg_opt_tra} can achieve relatively good performance at the first iteration, which is due to the efficient initialization of the UAV's position and trajectory. Indeed, initially deploying the UAV right above the center of the area in the static-UAV network and employing the initial circular trajectory in the mobile-UAV network \cite{wu2018joint} enables good fairness performance among the UEs. In particular, such initialization is also close to the optimized UAV locations/trajectory, as will be further seen in the next figure.}


	\subsection{Location Deployment and Trajectory of the UAV}

\begin{figure}[t]\centering
		\centering
		\subfigure[UAV's location in 20 channels]
		{
			\includegraphics[scale=0.55]{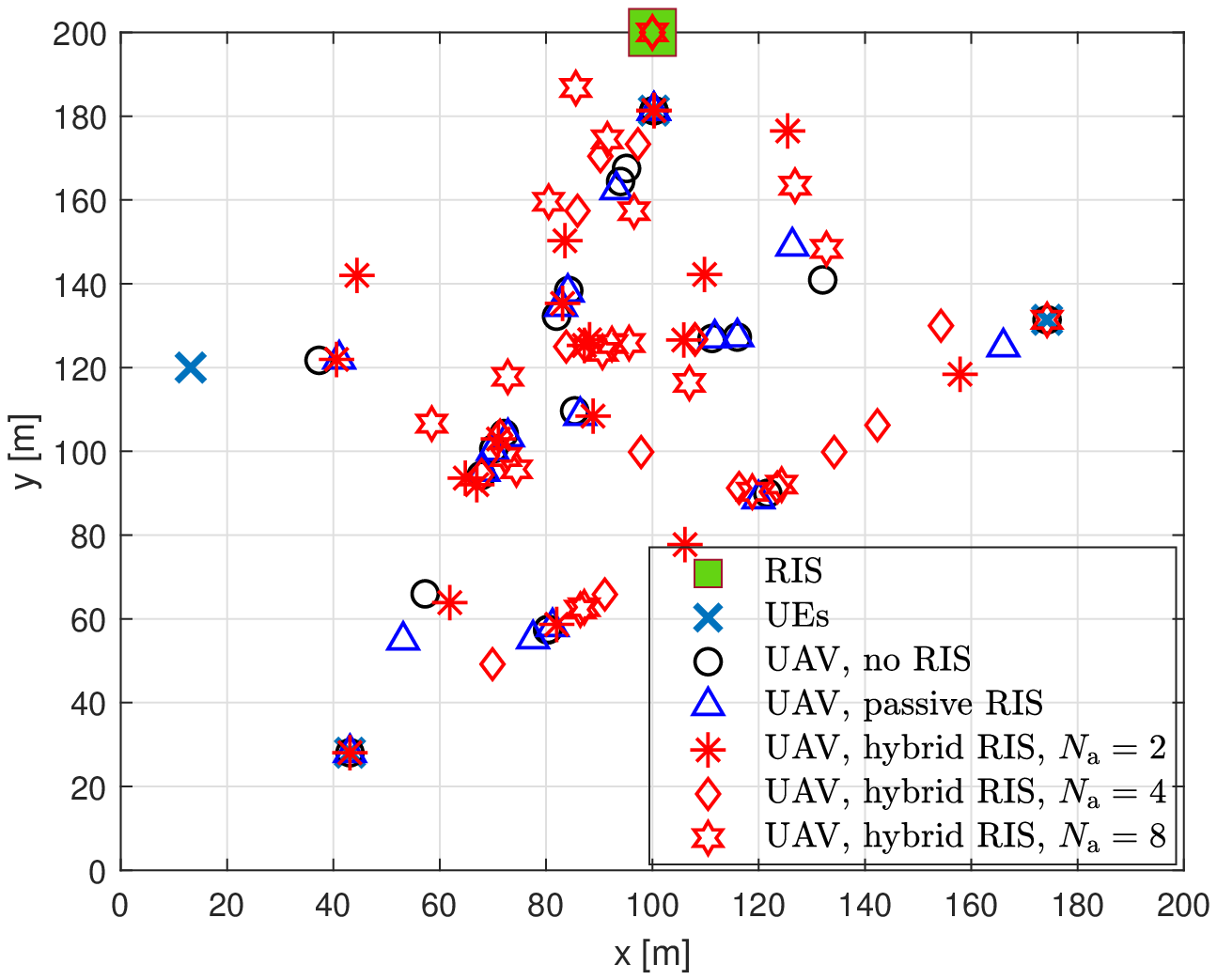}
			\label{fig_loc}
		}
		\subfigure[UAV's trajectory in $T = 50$ time slots]
		{
			\includegraphics[scale=0.55]{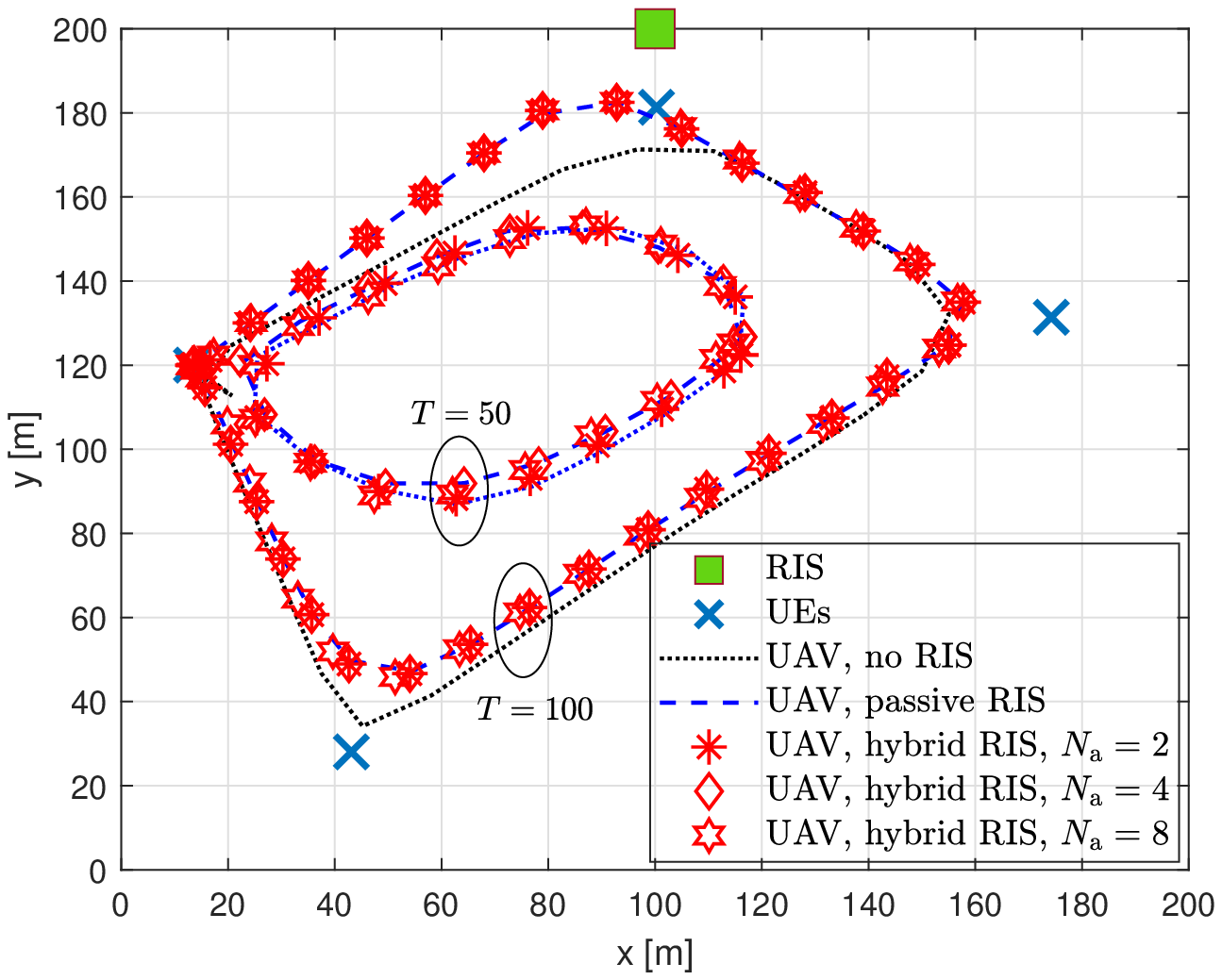}
			\label{fig_tra}
		}
		\caption{UAV's locations and trajectory with $K = 4$, $N_t = 2$, {$\Na = \{2, 4, 8\}$}, $N = 32$, $D = 200$ m, $\ptmax = 20$ dBm, and $\prismax = 0$ dBm. }
		\label{fig_loc_tra}
	\end{figure}

	In Fig.\ \ref{fig_loc_tra}, we show the optimized locations and trajectories of the UAV in the horizontal view of three systems, namely, without RIS, with the passive RIS, and with the hybrid RIS, for $D=200$, $\Na = \{2,4,8\}$, and $\ptmax = 20$ dBm; the other parameters are set the same as those in Fig.\ \ref{fig_conv}. The locations of the UAV are shown for $20$ channels in Fig.\ \ref{fig_loc}. {It is observed that without the RIS and with the passive RIS, the UAV (black circles and blue triangles) generally deploys in between the UEs, and in some cases, right above one of the UEs. However, when being aided by the hybrid RIS, the UAV (red squares, diamonds, and hexagrams) also deploys near the RIS.} There are some cases where the UAV flies right above the hybrid RISs with $\Na = \{4,8\}$. This verifies that the hybrid RIS significantly improves the reflecting channels, so that the UAV does not necessarily always deploy in between or close to the UEs to reduce the path loss on the direct channels, but it can also deploy near the RIS to enjoy the enhanced communications. In Fig.\ \ref{fig_tra}, we show the trajectories of the UAV employing the TDMA protocol with $T = \{50,100\}$ for five channels. A common observation on the trajectories of all the compared schemes is that as $T$ increases, the UAV exploits its mobility to enlarge and adjust its trajectory to move closer to all the UEs. As $T$ becomes sufficiently large, e.g., $T=100$, the UAV visit very close to all the UEs. In particular, with the presence of the passive or hybrid RISs, the UAV flies closer to the RIS compared to the case without RIS, similar to the location deployment in Fig.\ \ref{fig_loc}.
	
	\subsection{Performance Improvement of the Hybrid RIS}
	
	In this section, we show simulation results on the minimum rate versus various parameters. The positions of the UEs and the RIS are the same as those in Fig.\ \ref{fig_loc_tra}.
	
	\subsubsection{Static-UAV network without TDMA}
	
\begin{figure}[t]
		\centering
		\subfigure[$D = 200$ m]
		{
			\includegraphics[scale=0.6]{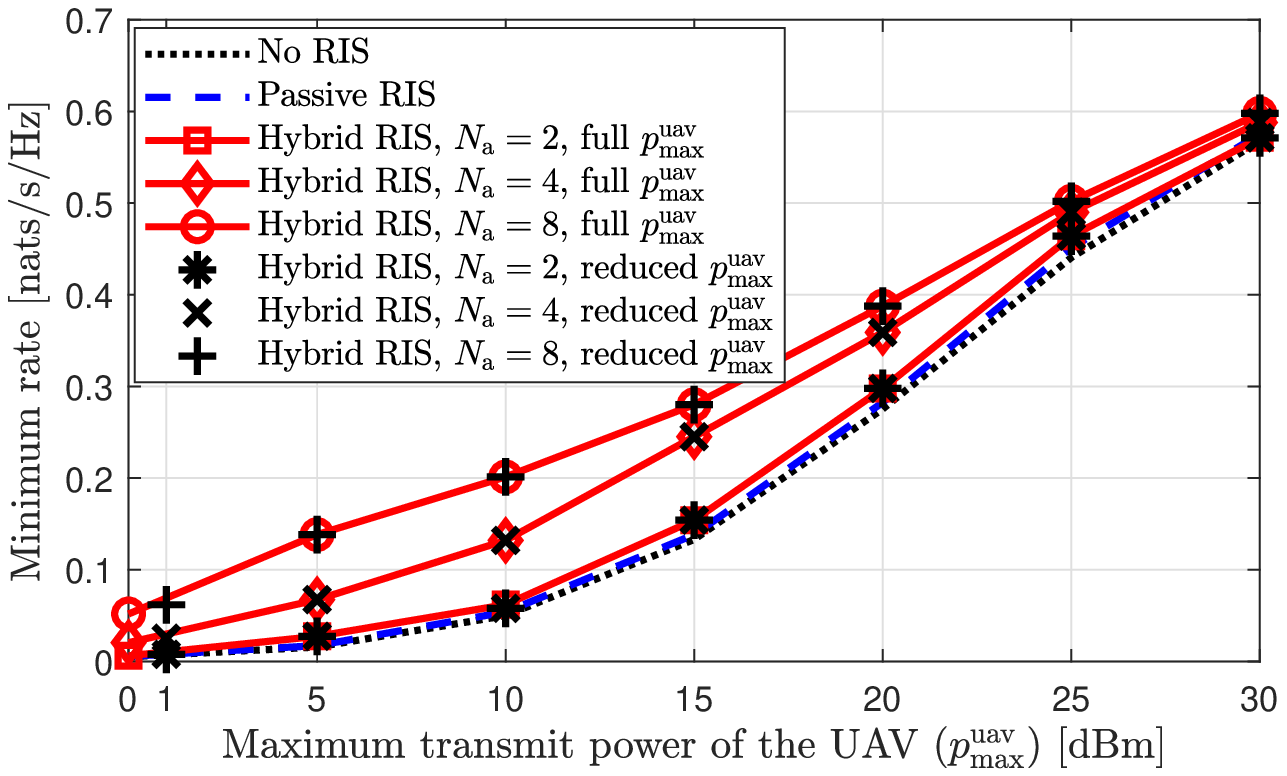}
			\label{fig_rate_vs_Pt_200}
		}
		\subfigure[$D = 50$ m]
		{
			\includegraphics[scale=0.6]{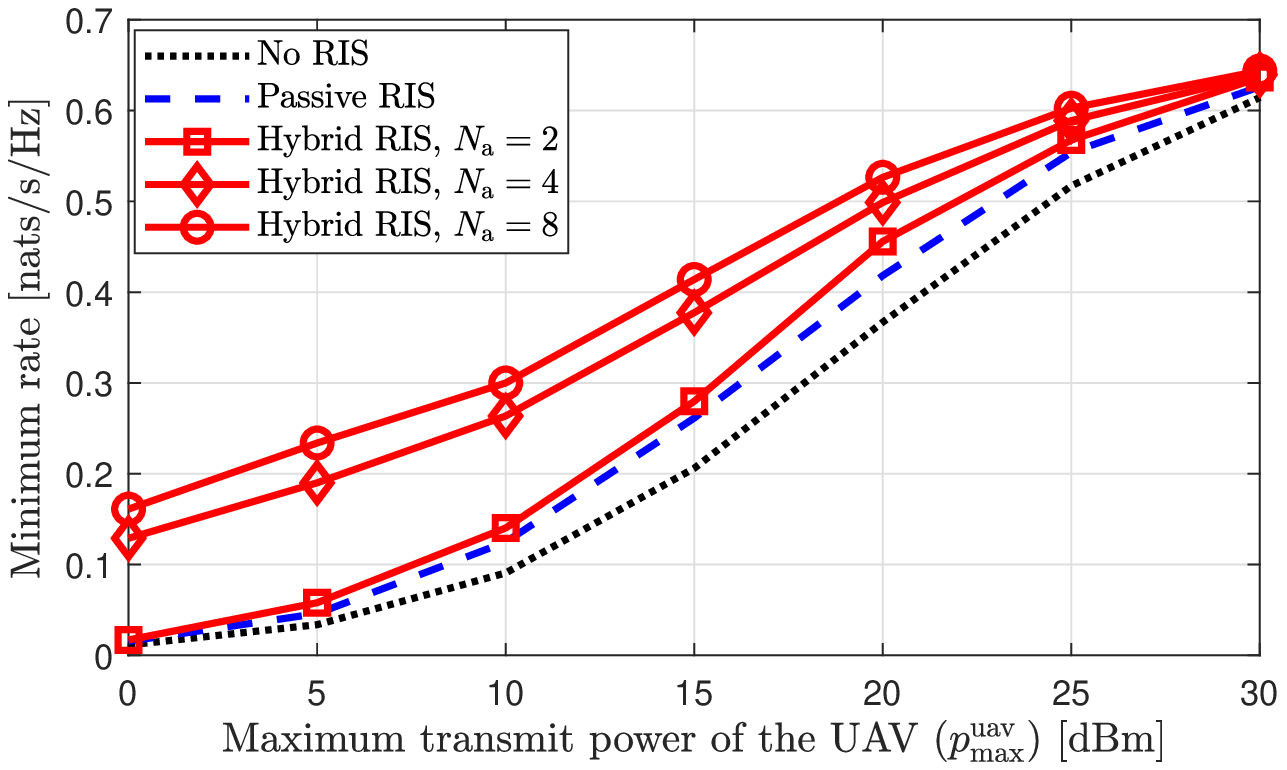}
			\label{fig_rate_vs_Pt_50}
		}
		\caption{Minimum rate of the UEs versus $\ptmax$ with $K = 4$, $N_t = 2$, $\Na = \{2,4,8\}$, $N = 32$, $D = \{50,200\}$ m, $\prismax = 0$ dBm, and $\ptmax = [0,30]$ dBm.}
		\label{fig_rate_vs_pt}
	\end{figure}

	We first show the minimum rate for $\ptmax = [0,30]$ dBm and $\Na = \{2,4,8\}$, and $D=\{200,50\}$ in Figs.\ \ref{fig_rate_vs_Pt_200} and \ref{fig_rate_vs_Pt_50}, respectively. We note that in the hybrid RIS-aided UAV system, the hybrid RIS requires an additional power budget of $\prismax$. Therefore, we also consider the case that the power budget at the UAV in this system is reduced by $\prismax$ to be $\ptmax - \prismax$ in Fig.\ \ref{fig_rate_vs_Pt_200}. We note that this is fair in the sense that all the compared schemes have the same total power budget, but it does not necessarily reflect the practical deployment. 
	From Fig.\ \ref{fig_rate_vs_pt}, the following observations are made:
	\begin{itemize}
		\item The conventional passive RIS deployed in the $200 \times 200~\text{m}^2$ (Fig.\ \ref{fig_rate_vs_Pt_200}) area provides only marginal performance improvement due to the severe double path loss on the reflecting channels. However, when the area is shrunken to $50 \times 50~\text{m}^2$ (but the UEs and RIS distribution are unchanged), its gain becomes more significant, as seen in Fig.\ \ref{fig_rate_vs_Pt_50}.
		
		\item With $\Na=2$, the hybrid RIS performs only slightly better than the passive RIS, but with $\Na = \{4,8\}$, it attains significant performance gains. For example, at $\ptmax = 20$ dBm, the hybrid RISs with $\Na=4$ achieve $\{33.33\%, 38.33\%\}$ improvement in Figs.\ \ref{fig_rate_vs_Pt_200} and \ref{fig_rate_vs_Pt_50}, respectively, while those of the passive RIS are only $\{3.70\%, 13.80\%\}$. In particular, the gain is more significant at low $\ptmax$ as power constraint \eqref{cons_pris_loc} shows that a smaller $\ptmax$ leads to larger $\abs{\an}, \forall n \in \setA$. This makes the hybrid RIS important in UAV systems since the capacity of the UAV's battery is generally limited. 
		
		\item The performance gains of the hybrid RIS with $\Na = \{4,8\}$ are still significant even when the UAV has a reduced power budget of $\ptmax - \prismax$. Specifically, it is observed in Fig.\ \ref{fig_rate_vs_Pt_200} that reducing the UAV's power budget only cause a slight performance degradation at low $\ptmax$ (we note here that because we assume $\prismax = 0$ dBm, the dotted curves must start from a point $\ptmax > 0$ dBm). This is reasonable because at moderate and high $\ptmax$ ($\geq 5$ dBm), reducing an amount of $\prismax = 0$ dBm almost does not cause any significant impact on the system performance.
	\end{itemize}
	
\begin{figure}[t]
		\centering
		\subfigure[Minimum rate vs. $\prismax$]
		{
			\includegraphics[scale=0.6]{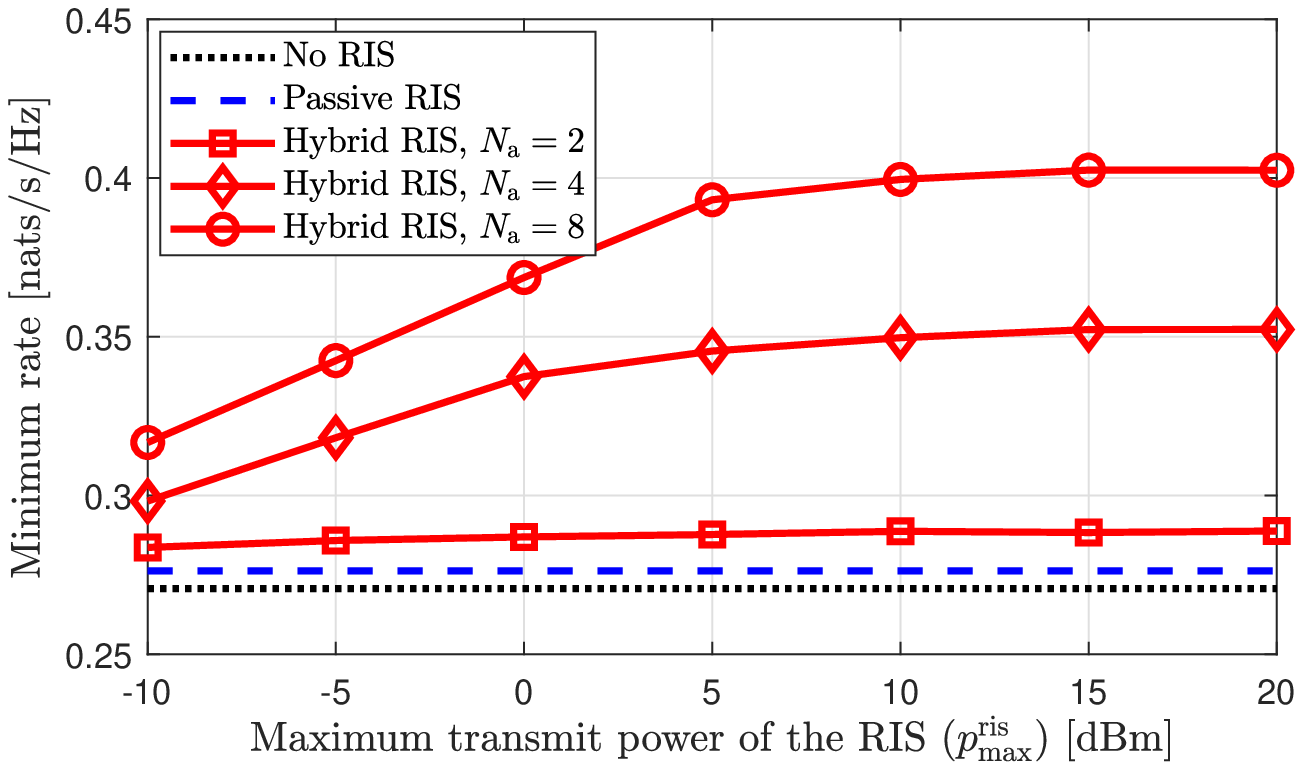}
			\label{fig_rate_vs_Pris}
		}
		\subfigure[Minimum rate vs. $N$]
		{
			\includegraphics[scale=0.6]{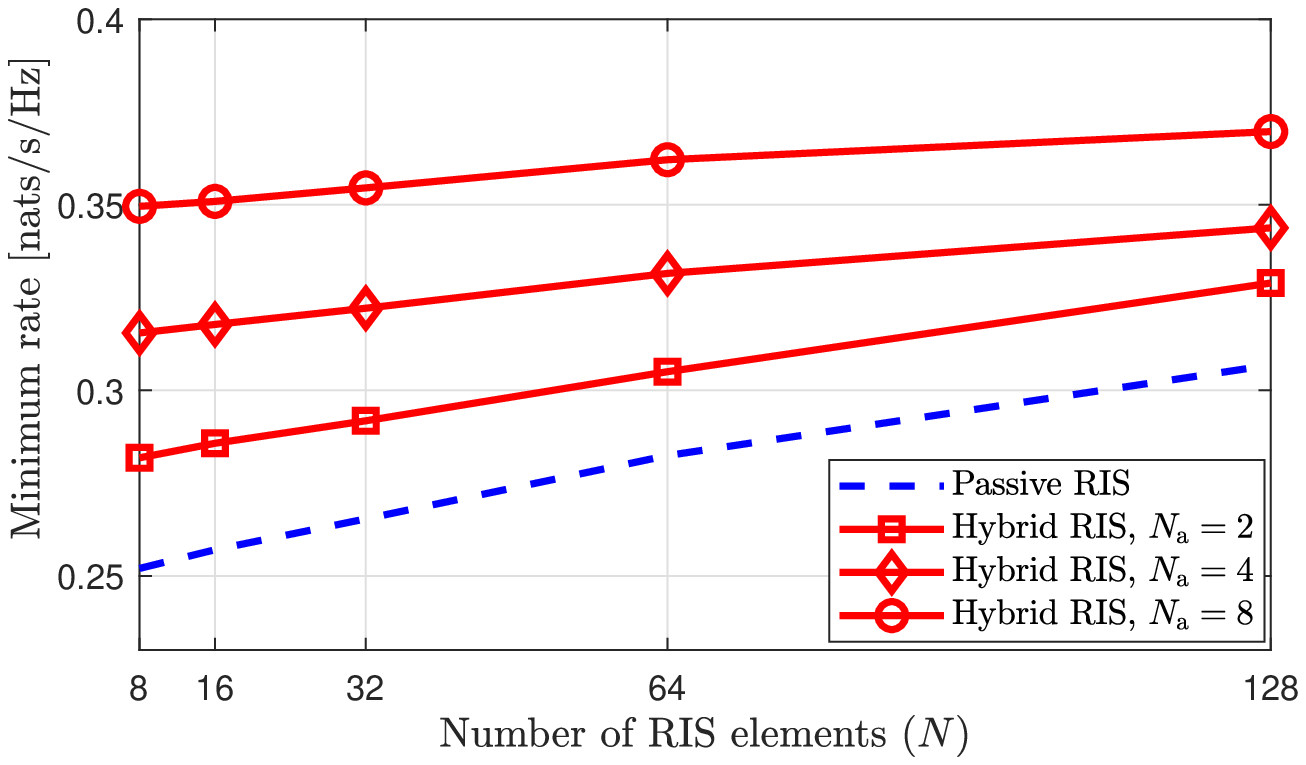}
			\label{fig_rate_vs_N}
		}
		\caption{Minimum rate of the UEs versus $\prismax$ (Fig. (a)) and versus $N$ (Fig. (b)) with $K = 4$, $N_t = 2$, $\Na = \{2,4,8\}$, $D = 200$ m, $\ptmax = 20$ dBm, (a) $N = 32$,  $\prismax \in [-10,20]$ dBm, (b) $N \in [8,128]$, $\prismax = 0$ dBm.}
		\label{fig_rate_vs_N_Pris}
	\end{figure}

	In Figs.\ \ref{fig_rate_vs_Pris} and \ref{fig_rate_vs_N}, we show the minimum rate of the hybrid RIS for different values of $\prismax$ and $N$, respectively, both with $\ptmax = 20$ dBm. It is obvious from Fig.\ \ref{fig_rate_vs_Pris} that the performance of the system without RIS and with passive RIS is unchanged with $\prismax$. The performance of hybrid RISs with $\Na = \{4,8\}$ first increases, peaks at $\prismax=5$ dBm and  then is likely to saturate at high $\prismax$. This is because the amplitudes of all the active elements are restricted by constraint \eqref{cons_active_modul_loc}. Furthermore, it should be noted that in this system, an active element with a larger amplitude does not always result in a performance improvement because it also amplifies inter-symbol interference. Regarding the performance versus $N$, it is observed in Fig.\ \ref{fig_rate_vs_N} that deploying more elements in the passive RIS results in improved performance, which is more significant than doing the same for the hybrid RIS. Specifically, as $N$ increases, the performance of the passive RIS increases faster than that of the hybrid RIS. This shows that active elements play a more important role in the RISs of smaller sizes.
	
	\subsubsection{UAV network with TDMA}

	\begin{figure}[t]
            \centering
		\subfigure[$D = 200$ m, $\prismax = 0$ dBm]
		{
			\includegraphics[scale=0.6]{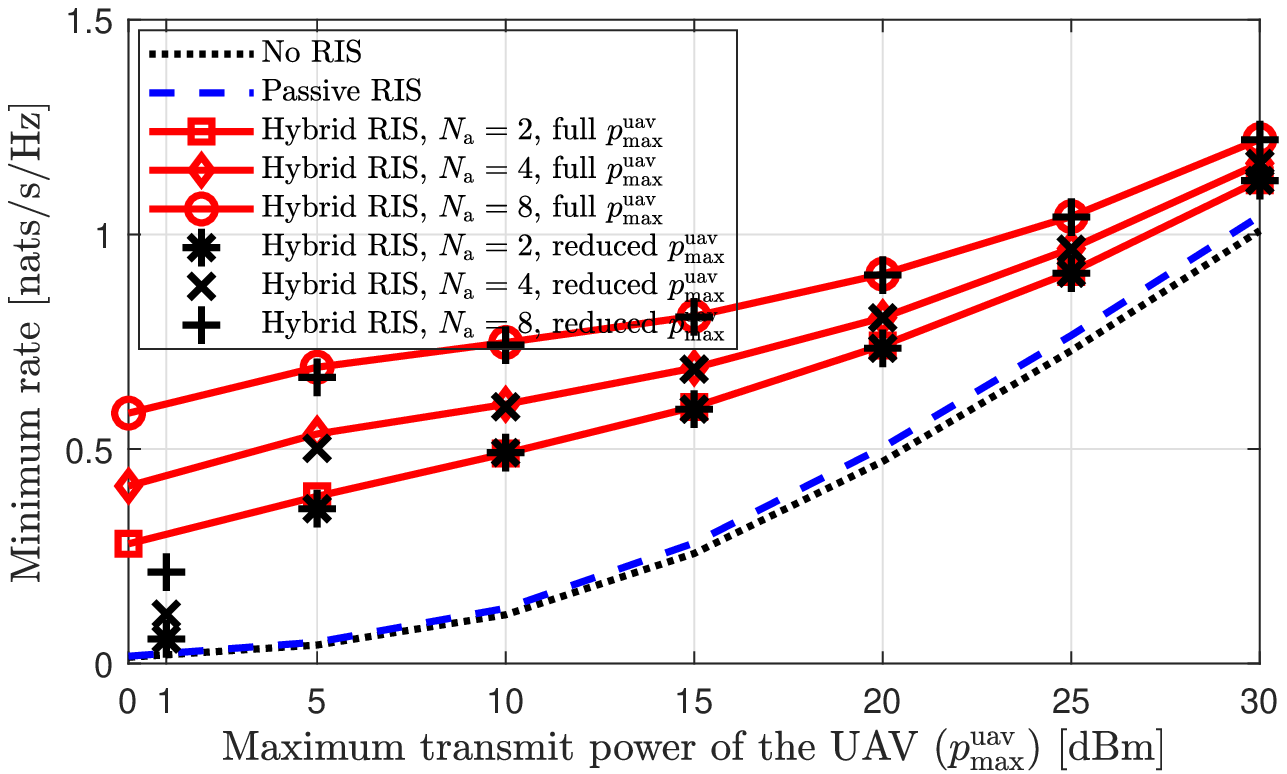}
			\label{fig_rate_vs_Pt_200_tra}
		}
		\subfigure[$D = 50$ m, $\prismax = -5$ dBm]
		{
			\includegraphics[scale=0.6]{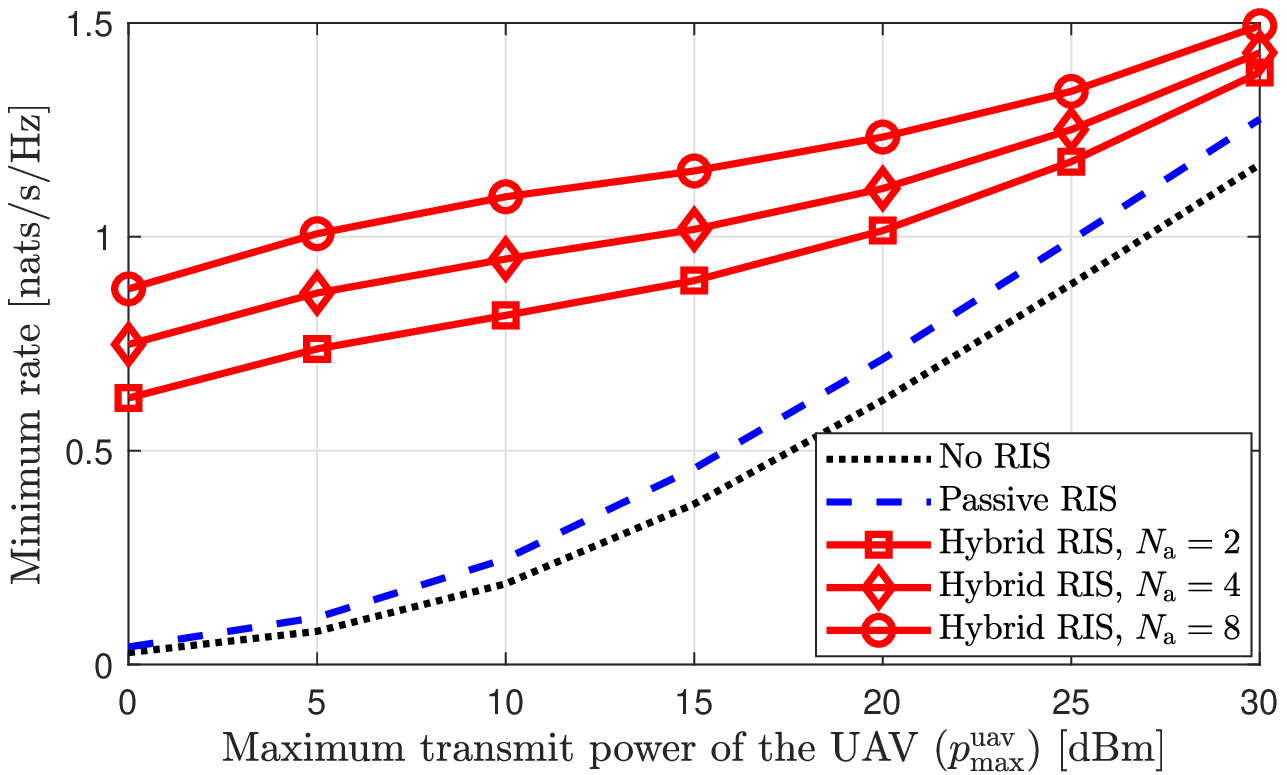}
			\label{fig_rate_vs_Pt_50_tra}
		}
		\caption{Minimum rate of the UEs versus $\ptmax$ with $K = 4$, $N_t = 2$, $\Na = \{2,4,8\}$, $N = 32$, $D = \{50,200\}$ m, $\prismax = 0$ dBm, $\ptmax = [0,30]$ dBm.}
		\label{fig_rate_vs_pt_tra}
	\end{figure}

	In this section, we investigate the performance of the network in which the UAV employs the TDMA protocol. In Fig.\ \ref{fig_rate_vs_pt_tra}, the minimum rate is shown for $T=50$, and the other parameters are the same as those in Fig.\ \ref{fig_rate_vs_pt}. Similar to Fig.\ \ref{fig_rate_vs_pt}, it is clear that the performance gain of the passive RIS is improved when it is deployed closer to the UEs (in Fig.\ \ref{fig_rate_vs_Pt_50_tra}), and the hybrid RIS with $\Na=\{4,8\}$ provides significant performance improvement, especially at low $\ptmax$. However in this scenario, i.e., with TDMA, the hybrid RIS with only $\Na=2$ active elements also performs very well to achieve up to $\{55.31\%,62.90\%\}$ improvement at $\ptmax = 20$ dBm for $D = \{200,50\}$, respectively, compared to only $\{8.51\%,14.52\%\}$ attained by the passive RIS. We note that this remarkable improvement was not seen for $\Na=2$ in Fig.\ \ref{fig_rate_vs_pt}. This is because when there is no inter-symbol interference, active elements can serve with their maximum amplifying gain to enhance the system performance. More specifically, our numerical results show that for the same power budgets, $\abs{\an}, n \in \setA$ in this scenario is much larger than those without TDMA. In particular, in the area $50 \times 50~\text{m}^2$ (Fig.\ \ref{fig_rate_vs_Pt_50_tra}), the hybrid RIS offers remarkable performance improvement with a power budget of only $-5$ dBm. It is also seen from Fig.\ \ref{fig_rate_vs_Pt_200_tra} that when the UAV transmit with a reduced power budget, i.e., $\ptmax - \prismax$, the performance loss at low $\ptmax$ is more significant than those seen in Fig.\ \ref{fig_rate_vs_pt}. This is because when TDMA is applied, the performance is mostly decided by the received signal power. However, the performance loss is negligible for $\ptmax \geq 5$ dBm. 

    \subsection{{Performance of the Hybrid RIS under Imperfect CSI}}
    
    In this section, we examine the performance of UAV systems in the presence of CSI error. We denote by $\{ \hat{\vh}_{0,k}^\H, \hat{\mH}_1, \hat{\vh}_{2,k}^H \}$ the imperfect channel estimates of $\{ \hd^\H, \hrt, \hrr^\H \}$, respectively. Here, $\hat{\vh}_{0,k}^\H$ can be modeled as $\hat{\vh}_{0,k}^\H = \hd^\H - \Delta \hd^\H$, where $\Delta \hd^\H$ represents the CSI error whose entries have distributions $\mathcal{CN}(0, \hat{\zeta}_{0,k} \epsilon_{0,k}^2)$ with $\hat{\zeta}_{0,k} = \zeta_0 \norm{\vv - \vu_k}^{-\epsilon_0}$ being the large-scale coefficient and $\epsilon_{0,k}$ being the CSI uncertainty level of $\hat{\vh}_{0,k}^\H$. Here, we assume that the CSI error only occurs in small-scale fading channels because the large-scale coefficients can be easily estimated with high accuracy  \cite{nguyen2021hybrid}. The channels $\hat{\mH}_1$ and $\hat{\vh}_{2,k}^H$ are modeled similarly.
    \begin{figure}[t]
        \centering
        \subfigure[Static-UAV system]
        {
            \includegraphics[scale=0.6]{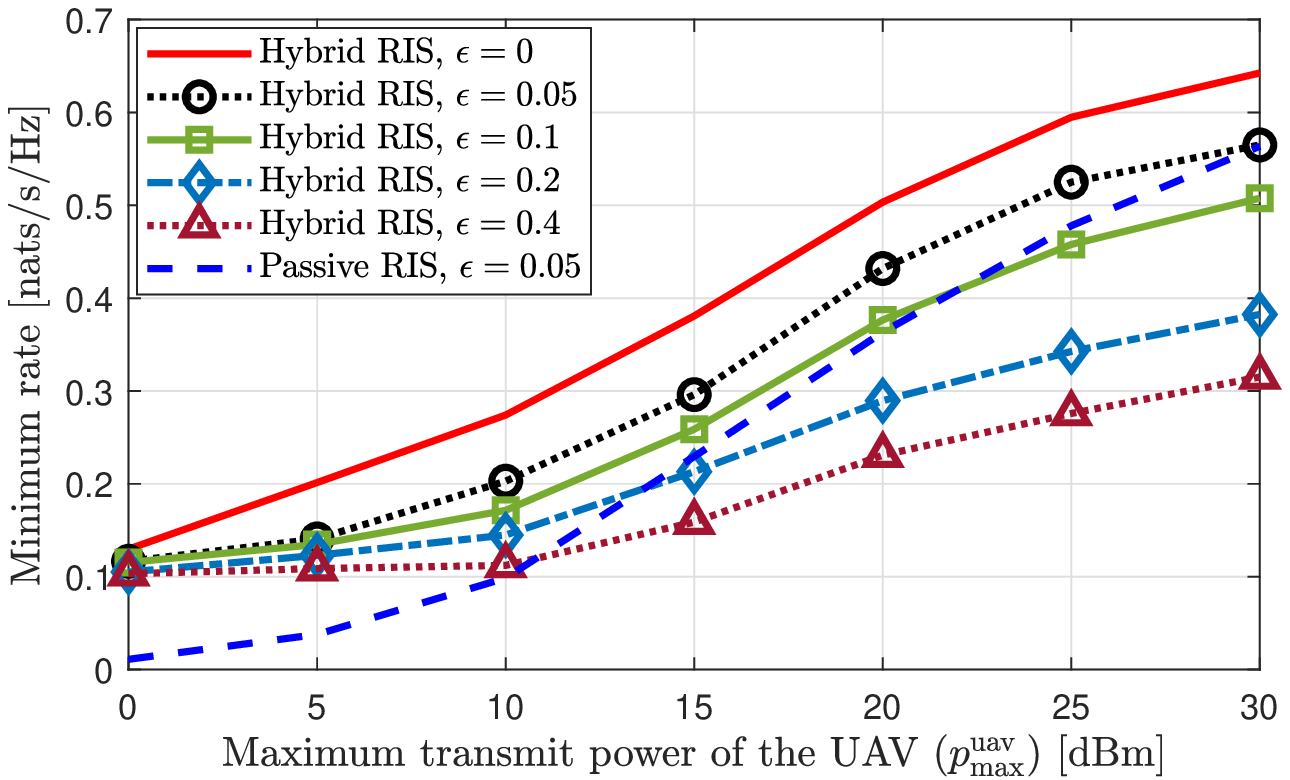}
            \label{fig_rate_vs_Pt_ICSI_loc}
        }
        \subfigure[Mobile-UAV system]
        {
            \includegraphics[scale=0.6]{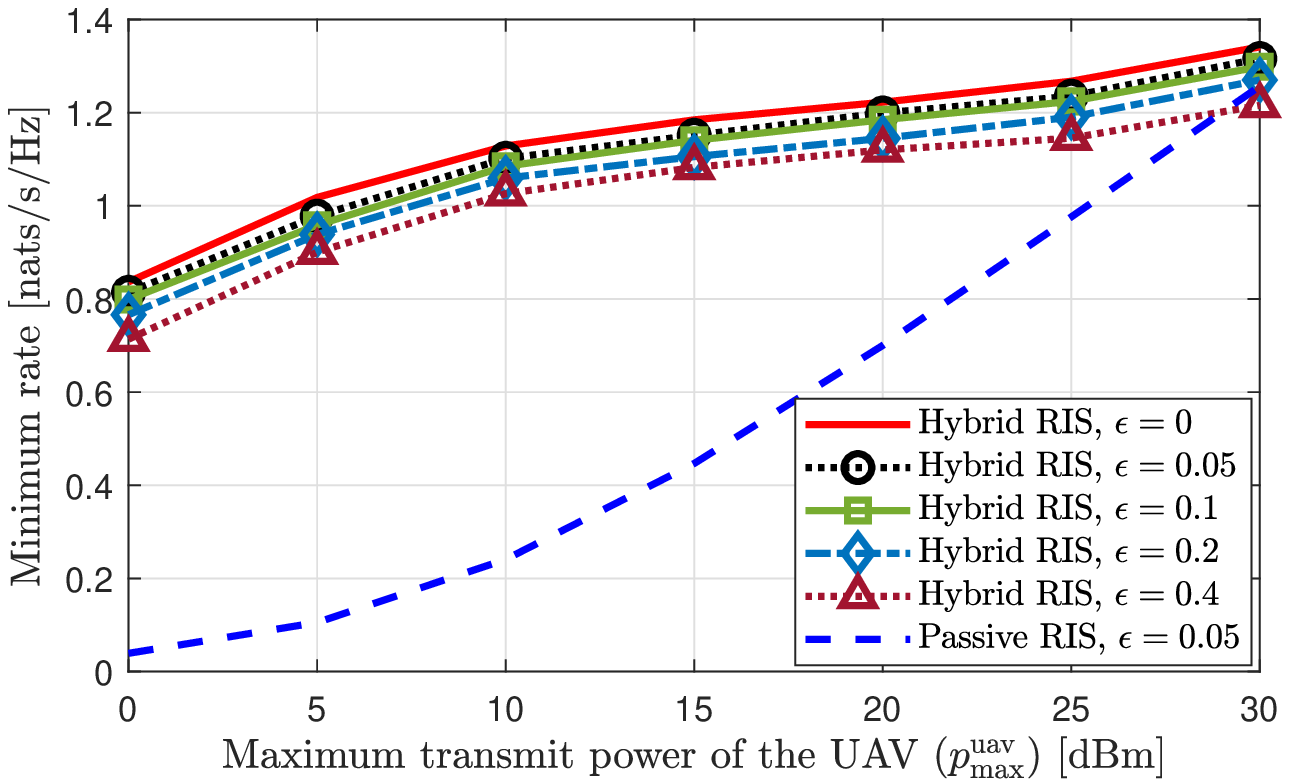}
            \label{fig_rate_vs_Pt_ICSI_tra}
        }
        \caption{{Minimum rate of the UEs in the hybrid RIS-aided static- and mobile-UAV systems in the presence of imperfect CSI with $K = 4$, $N_t = 2$, $\Na = 4$, $N = 32$, $D = 50$ m, $\prismax = 0$ dBm, $T = 50$, and $\epsilon = \{0,0.05,0.1,0.2,0.5\}$.}}
        \label{fig_rate_vs_pt_ICSI}
    \end{figure}

    In Fig.\ \ref{fig_rate_vs_pt_ICSI}, we show the minimum rate performance of the considered UAV systems aided by the hybrid and passive RISs under CSI errors. We set $K = 4$, $N_t = 2$, $\Na = 4$, $N = 32$, $D = 50$ m, $\prismax = 0$ dBm, and $T = 50$. Without loss of generality, we assume all the imperfect channel estimates have the same uncertainty level, which is set to $\epsilon \in \{0,0.05,0.1,0.2,0.4\}$. It is observed that as $\epsilon$ increases, the UAV systems have performance degradation, which is more significant in the static-UAV scenario (in Fig.\ \ref{fig_rate_vs_Pt_ICSI_loc}) than the mobile-UAV scenario (in Fig.\ \ref{fig_rate_vs_Pt_ICSI_tra}). This is because the former is vulnerable to the interference in multiuser communications. In contrast, due to the high mobility of the UAV and the TDMA protocol, the performance of the mobile-UAV network depends more on the large-scale channels rather than the erroneous small-scale ones. Furthermore, it is promising to observe that despite the loss due to the CSI errors, the hybrid RIS still offers remarkable performance improvement with respect to the passive RIS, especially at low $p_{\mathrm{uav}}^{\mathrm{max}}$. Particularly, the mobile-UAV system aided by the hybrid RIS under severe CSI errors (e.g., $\epsilon = \{0.1, 0.2, 0.4\}$) still performs far better than when being aided by the passive RIS with small CSI errors (e.g., $\epsilon = 0.05$), as seen in Fig.\ \ref{fig_rate_vs_Pt_ICSI_tra}.

    In Figs.\ \ref{fig_loc_tra}--\ref{fig_rate_vs_pt_ICSI}, we have numerically demonstrated the performance benefits of deploying hybrid RISs in two typical UAV-enabled networks, i.e., the static- and mobile-UAV networks. We remark the following important observations:
		\begin{itemize}
			\item When the systems are aided by the passive RIS or no RIS, the UAV generally flights close to the UEs. However, with the introduction of active RIS elements (i.e., when the hybrid RIS is deployed), the UAV moves toward the RIS to enjoy enhanced communications.
			\item Compared to the purely passive RIS, the hybrid one offers a significant max-min fairness enhancement, especially with a low or moderate power budget at the UAV. Thus, hybrid RISs are beneficial for practical deployments of UAVs.
			\item The performance gains of the hybrid RIS are more clearly seen when the number of active elements, i.e., $\Na$, and/or the RIS power budget, i.e., $\prismax$, increase. However, the gains are upper bounded when $\prismax$ is sufficiently large. In such scenarios, it is more beneficial to deploy the RIS with more active elements rather than with a higher power budget.
		\end{itemize}
	
	\section{Conclusion}
	\label{sec_conclusion}
	
	In this paper, we have proposed two novel hybrid RIS-assisted UAV communications systems. Toward a fairness design of these systems, our goal is to maximize the minimum rate among users through jointly optimizing the location/trajectory, and transmit beamforming of the UAV and RIS reflecting/amplifying coefficients. The formulated problems inherit the challenges of conventional passive RIS design due to non-convexity and strongly coupling variables while facing additional difficulties in the design of active coefficients under power constraints. We have developed efficient solutions to solve the considered problem by leveraging BCA and SCA approaches. Finally, we have provided extensive numerical results to demonstrate the efficacy of the proposed algorithms. They have also revealed the remarkable performance improvement of the hybrid RIS compared with existing schemes. The improvement is particularly significant when the UAV has a limited power budget, making the deployment of the hybrid RIS important because the UAV's battery capacity is generally limited. {For future studies, leveraging learning capabilities of deep neural networks and deep reinforcement learning for RIS-assisted UAV system designs is a potential extension of this work.}

	\begingroup
	\bibliographystyle{IEEEtran}
	\bibliography{IEEEabrv,Bibliography}

\begin{thebibliography}{10}
\providecommand{\url}[1]{#1}
\csname url@samestyle\endcsname
\providecommand{\newblock}{\relax}
\providecommand{\bibinfo}[2]{#2}
\providecommand{\BIBentrySTDinterwordspacing}{\spaceskip=0pt\relax}
\providecommand{\BIBentryALTinterwordstretchfactor}{4}
\providecommand{\BIBentryALTinterwordspacing}{\spaceskip=\fontdimen2\font plus
\BIBentryALTinterwordstretchfactor\fontdimen3\font minus
  \fontdimen4\font\relax}
\providecommand{\BIBforeignlanguage}[2]{{%
\expandafter\ifx\csname l@#1\endcsname\relax
\typeout{** WARNING: IEEEtran.bst: No hyphenation pattern has been}%
\typeout{** loaded for the language `#1'. Using the pattern for}%
\typeout{** the default language instead.}%
\else
\language=\csname l@#1\endcsname
\fi
#2}}
\providecommand{\BIBdecl}{\relax}
\BIBdecl

\bibitem{wu2018joint}
Q.~Wu, Y.~Zeng, and R.~Zhang, ``Joint trajectory and communication design for
  multi-{UAV} enabled wireless networks,'' \emph{{IEEE} Trans. Wireless
  Commun.}, vol.~17, no.~3, pp. 2109--2121, 2018.

\bibitem{cao2021reconfigurable}
X.~Cao, B.~Yang, C.~Huang, C.~Yuen, M.~Di~Renzo, D.~Niyato, and Z.~Han,
  ``Reconfigurable intelligent surface-assisted aerial-terrestrial
  communications via multi-task learning,'' \emph{{IEEE} J. Sel. Areas
  Commun.}, vol.~39, no.~10, pp. 3035--3050, 2021.

\bibitem{yang2020intelligent}
Y.~Yang, B.~Zheng, S.~Zhang, and R.~Zhang, ``{Intelligent reflecting surface
  meets OFDM: Protocol design and rate maximization},'' \emph{{IEEE} Trans.
  Commun.}, vol.~68, no.~7, pp. 4522--4535, 2020.

\bibitem{QingqingTCOM20}
Q.~Wu and R.~Zhang, ``Beamforming optimization for wireless network aided by
  intelligent reflecting surface with discrete phase shifts,'' \emph{IEEE
  Trans. Commun.}, vol.~68, no.~3, pp. 1838--1851, 2020.

\bibitem{Huang2018}
C.~{Huang}, A.~{Zappone}, G.~C. {Alexandropoulos}, M.~{Debbah}, and C.~{Yuen},
  ``Reconfigurable intelligent surfaces for energy efficiency in wireless
  communication,'' \emph{{IEEE} Trans. Wireless Commun.}, vol.~18, no.~8, pp.
  4157--4170, 2019.

\bibitem{munochiveyi2021reconfigurable}
M.~Munochiveyi, A.~C. Pogaku, D.-T. Do, A.-T. Le, M.~Voznak, and N.~D. Nguyen,
  ``Reconfigurable intelligent surface aided multi-user communications:
  {S}tate-of-the-art techniques and open issues,'' \emph{IEEE Access}, vol.~9,
  pp. 118\,584--118\,605, 2021.

\bibitem{pogaku2022uav}
A.~C. Pogaku, D.-T. Do, B.~M. Lee, and N.~D. Nguyen, ``{UAV}-assisted {RIS} for
  future wireless communications: {A} survey on optimization and performance
  analysis,'' \emph{IEEE Access}, vol.~10, pp. 16\,320--16\,336, 2022.

\bibitem{tyrovolas2021performance}
D.~Tyrovolas, S.~A. Tegos, P.~D. Diamantoulakis, and G.~K. Karagiannidis,
  ``Performance analysis of synergetic {UAV-RIS} communication networks,''
  \emph{arXiv preprint arXiv:2106.10034}, 2021.

\bibitem{huang2020holographic}
C.~Huang, S.~Hu, G.~C. Alexandropoulos, A.~Zappone, C.~Yuen, R.~Zhang,
  M.~Di~Renzo, and M.~Debbah, ``Holographic {MIMO} surfaces for {6G} wireless
  networks: {O}pportunities, challenges, and trends,'' \emph{{IEEE} Wireless
  Commun.}, vol.~27, no.~5, pp. 118--125, 2020.

\bibitem{wu2019intelligent}
Q.~Wu and R.~Zhang, ``Intelligent reflecting surface enhanced wireless network
  via joint active and passive beamforming,'' \emph{{IEEE} Trans. Wireless
  Commun.}, vol.~18, no.~11, pp. 5394--5409, 2019.

\bibitem{nguyen2021hybrid}
N.~T. Nguyen, Q.-D. Vu, K.~Lee, and M.~Juntti, ``Hybrid relay-reflecting
  intelligent surface-assisted wireless communications,'' \emph{{IEEE} Trans.
  Veh. Technol.}, vol.~71, no.~6, pp. 6228--6244, 2022.

\bibitem{li2020reconfigurable}
S.~Li, B.~Duo, X.~Yuan, Y.-C. Liang, and M.~Di~Renzo, ``Reconfigurable
  intelligent surface assisted {UAV} communication: Joint trajectory design and
  passive beamforming,'' \emph{{IEEE} Commun. Lett.}, vol.~9, no.~5, pp.
  716--720, 2020.

\bibitem{li2021robust}
S.~Li, B.~Duo, M.~Di~Renzo, M.~Tao, and X.~Yuan, ``Robust secure {UAV}
  communications with the aid of reconfigurable intelligent surfaces,''
  \emph{{IEEE} Trans. Wireless Commun.}, vol.~20, no.~10, pp. 6402--6417, 2021.

\bibitem{li2020sum}
J.~Li and J.~Liu, ``Sum rate maximization via reconfigurable intelligent
  surface in {UAV} communication: Phase shift and trajectory optimization,'' in
  \emph{Proc. Int. Conf. on Commun. and Networking in China}, 2020, pp.
  124--129.

\bibitem{jiang2021reconfigurable}
L.~Jiang and H.~Jafarkhani, ``Reconfigurable intelligent surface assisted
  mmwave {UAV} wireless cellular networks,'' in \emph{Proc. IEEE Int. Conf.
  Commun.}, 2021, pp. 1--6.

\bibitem{guo2021learning}
X.~Guo, Y.~Chen, and Y.~Wang, ``Learning-based robust and secure transmission
  for reconfigurable intelligent surface aided millimeter wave {UAV}
  communications,'' \emph{{IEEE} Wireless Commun. Lett.}, vol.~10, no.~8, pp.
  1795--1799, 2021.

\bibitem{diamanti2021energy}
M.~Diamanti, M.~Tsampazi, E.~E. Tsiropoulou, and S.~Papavassiliou, ``Energy
  efficient multi-user communications aided by reconfigurable intelligent
  surfaces and {UAVs},'' in \emph{Proc. IEEE Int. Conf. Smart Computing}, 2021,
  pp. 371--376.

\bibitem{pan2021uav}
Y.~Pan, K.~Wang, C.~Pan, H.~Zhu, and J.~Wang, ``{UAV}-assisted and intelligent
  reflecting surfaces-supported terahertz communications,'' \emph{{IEEE}
  Wireless Commun. Lett.}, vol.~10, no.~6, pp. 1256--1260, 2021.

\bibitem{nguyen2021reconfigurable}
K.~K. Nguyen, S.~R. Khosravirad, D.~B. Da~Costa, L.~D. Nguyen, and T.~Q. Duong,
  ``Reconfigurable intelligent surface-assisted multi-{UAV} networks: Efficient
  resource allocation with deep reinforcement learning,'' \emph{{IEEE} J. Sel.
  Topics Signal Process.}, 2021.

\bibitem{wang2021passive}
D.~Wang, Y.~Zhao, Y.~He, X.~Tang, L.~Li, R.~Zhang, and D.~Zhai, ``Passive
  beamforming and trajectory optimization for reconfigurable intelligent
  surface-assisted {UAV} secure communication,'' \emph{Remote Sensing},
  vol.~13, no.~21, p. 4286, 2021.

\bibitem{li2021reconfigurable}
J.~Li, S.~Xu, J.~Liu, Y.~Cao, and W.~Gao, ``Reconfigurable intelligent surface
  enhanced secure aerial-ground communication,'' \emph{{IEEE} Trans. Wireless
  Commun.}, vol.~69, no.~9, pp. 6185--6197, 2021.

\bibitem{liu2020machine}
X.~Liu, Y.~Liu, and Y.~Chen, ``Machine learning empowered trajectory and
  passive beamforming design in {UAV-RIS} wireless networks,'' \emph{{IEEE} J.
  Sel. Areas Commun.}, vol.~39, no.~7, pp. 2042--2055, 2021.

\bibitem{mu2021intelligent}
X.~Mu, Y.~Liu, L.~Guo, J.~Lin, and H.~V. Poor, ``Intelligent reflecting surface
  enhanced {multi-UAV NOMA} networks,'' \emph{{IEEE} J. Sel. Areas Commun.},
  vol.~39, no.~10, pp. 3051--3066, 2021.

\bibitem{ranjha2020urllc}
A.~Ranjha and G.~Kaddoum, ``{URLLC} facilitated by mobile {UAV} relay and
  {RIS}: {A} joint design of passive beamforming, blocklength, and {UAV}
  positioning,'' \emph{{IEEE} Internet Things J.}, vol.~8, no.~6, pp.
  4618--4627, 2020.

\bibitem{jeong2020simultaneous}
C.~Jeong and S.~H. Chae, ``Simultaneous wireless information and power transfer
  for multiuser {UAV}-enabled iot networks,'' \emph{{IEEE} Internet Things J.},
  vol.~8, no.~10, pp. 8044--8055, 2020.

\bibitem{diamanti2021prospect}
M.~Diamanti, P.~Charatsaris, E.~E. Tsiropoulou, and S.~Papavassiliou, ``The
  prospect of reconfigurable intelligent surfaces in integrated access and
  backhaul networks,'' \emph{{IEEE} Trans. Green Commun. Network.}, vol.~6,
  no.~2, pp. 859--872, 2021.

\bibitem{samir2021optimizing}
M.~Samir, M.~Elhattab, C.~Assi, S.~Sharafeddine, and A.~Ghrayeb, ``Optimizing
  age of information through aerial reconfigurable intelligent surfaces: {A}
  deep reinforcement learning approach,'' \emph{{IEEE} Trans. Veh. Technol.},
  vol.~70, no.~4, pp. 3978--3983, 2021.

\bibitem{yang2020federated}
K.~Yang, Y.~Shi, Y.~Zhou, Z.~Yang, L.~Fu, and W.~Chen, ``Federated machine
  learning for intelligent {IoT} via reconfigurable intelligent surface,''
  \emph{IEEE Network}, vol.~34, no.~5, pp. 16--22, 2020.

\bibitem{huang2020reconfigurable}
C.~Huang, R.~Mo, and C.~Yuen, ``Reconfigurable intelligent surface assisted
  multiuser {MISO} systems exploiting deep reinforcement learning,''
  \emph{{IEEE} J. Sel. Areas Commun.}, vol.~38, no.~8, pp. 1839--1850, 2020.

\bibitem{huang2021multi}
{Huang \textit{et al.}}, ``Multi-hop {RIS}-empowered terahertz communications:
  {A DRL}-based hybrid beamforming design,'' \emph{{IEEE} J. Sel. Areas
  Commun.}, vol.~39, no.~6, pp. 1663--1677, 2021.

\bibitem{taha2019deep}
A.~Taha, M.~Alrabeiah, and A.~Alkhateeb, ``{Deep learning for large intelligent
  surfaces in millimeter wave and massive MIMO systems},'' in \emph{IEEE Global
  Commun. Conf. (GLOBECOM)}, 2019.

\bibitem{alexandropoulos2020hardware}
G.~C. Alexandropoulos and E.~Vlachos, ``A hardware architecture for
  reconfigurable intelligent surfaces with minimal active elements for explicit
  channel estimation,'' in \emph{Proc. IEEE Int. Conf. Acoust., Speech, Signal
  Processing}, 2020.

\bibitem{nguyen2021spectral}
N.~T. Nguyen, Q.-D. Vu, K.~Lee, and M.~Juntti, ``Spectral efficiency
  optimization for hybrid relay-reflecting intelligent surface,'' \emph{Proc.
  IEEE Int. Conf. Commun. Workshop}, 2021.

\bibitem{nguyen2022downlink}
N.~T. Nguyen, V.-D. Nguyen, V.~Nguyen, H.~Q. Ngo, S.~Chatzinotas, and
  M.~Juntti, ``Downlink throughput of cell-free massive {MIMO} systems assisted
  by hybrid relay-reflecting intelligent surfaces,'' in \emph{Proc. IEEE Int.
  Conf. Commun.}, 2022.

\bibitem{nguyen2022spectral_cfmimo}
N.~T. Nguyen, V.-D. Nguyen, H.~Van~Nguyen, H.~Q. Ngo, S.~Chatzinotas, and
  M.~Juntti, ``Spectral efficiency analysis of hybrid relay-reflecting
  intelligent surface-assisted cell-free massive {MIMO} systems,'' \emph{{IEEE}
  Trans. Wireless Commun.}, vol.~22, no.~5, pp. 3397--3416, 2022.

\bibitem{nguyen2021hybrid_mag}
\BIBentryALTinterwordspacing
N.~T. Nguyen, J.~He, V.-D. Nguyen, H.~Wymeersch, D.~W.~K. Ng, R.~Schober,
  S.~Chatzinotas, and M.~Juntti, ``Hybrid relay-reflecting intelligent
  surface-aided wireless communications: Opportunities, challenges, and future
  perspectives,'' \emph{arXiv preprint arXiv:2104.02039}, 2021. [Online].
  Available: \url{https://arxiv.org/pdf/2106.10034v2.pdf}
\BIBentrySTDinterwordspacing

\bibitem{shojaeifard2022mimo}
A.~Shojaeifard, K.-K. Wong, K.-F. Tong, Z.~Chu, A.~Mourad, A.~Haghighat,
  I.~Hemadeh, N.~T. Nguyen, V.~Tapio, and M.~Juntti, ``{MIMO} evolution beyond
  {5G} through reconfigurable intelligent surfaces and fluid antenna systems,''
  \emph{Proc. IEEE}, vol. 110, no.~9, pp. 1244--1265, 2022.

\bibitem{nguyen2022hybrid}
N.~T. Nguyen, V.-D. Nguyen, Q.~Wu, A.~T{\"o}lli, S.~Chatzinotas, and M.~Juntti,
  ``Hybrid active-passive reconfigurable intelligent surface-assisted
  multi-user {MISO} systems,'' in \emph{Proc. IEEE Works. on Sign. Proc. Adv.
  in Wirel. Comms.}, 2022.

\bibitem{nguyen2022hybrid_UAV}
------, ``Hybrid active-passive reconfigurable intelligent surface-assisted
  {UAV} communications,'' in \emph{Proc. IEEE Global Commun. Conf.}, 2022, pp.
  3126--3131.

\bibitem{9598322}
K.-H. Ngo, N.~T. Nguyen, T.~Q. Dinh, T.-M. Hoang, and M.~Juntti, ``Low-latency
  and secure computation offloading assisted by hybrid relay-reflecting
  intelligent surface,'' in \emph{Int. Conf. Advanced Tech. Commun. (ATC)},
  2021, pp. 306--311.

\bibitem{egashira2022secrecy}
E.~N. Egashira, D.~P.~M. Osorio, N.~T. Nguyen, and M.~Juntti, ``Secrecy
  capacity maximization for a hybrid relay-{RIS} scheme in mmwave mimo
  networks,'' in \emph{Proc. IEEE Veh. Technol. Conf.}, 2022.

\bibitem{9653007}
S.~Ahmed, A.~E. Kamal, and M.~Y. Selim, ``Adding active elements to
  reconfigurable intelligent surfaces to enhance energy harvesting for {IoT}
  devices,'' in \emph{IEEE Military Commun. Conf. (MILCOM)}, 2021, pp.
  297--302.

\bibitem{yigit2021hybrid}
Z.~Yigit, E.~Basar, M.~Wen, and I.~Altunbas, ``Hybrid reflection modulation,''
  \emph{{IEEE} Trans. Wireless Commun.}, vol.~22, no.~6, pp. 4106--4116, 2022.

\bibitem{schroeder2020passive}
R.~Schroeder, J.~He, and M.~Juntti, ``Passive {RIS} vs. {H}ybrid {RIS}: {A}
  comparative study on channel estimation,'' in \emph{Proc. IEEE Veh. Technol.
  Conf.}, June 2021.

\bibitem{long2021active}
R.~Long, Y.-C. Liang, Y.~Pei, and E.~G. Larsson, ``Active reconfigurable
  intelligent surface aided wireless communications,'' \emph{{IEEE} Trans.
  Wireless Commun.}, vol.~20, no.~8, pp. 4962--4975, 2021.

\bibitem{khoshafa2021active}
M.~H. Khoshafa, T.~M. Ngatched, M.~H. Ahmed, and A.~R. Ndjiongue, ``Active
  reconfigurable intelligent surfaces-aided wireless communication system,''
  \emph{{IEEE} Commun. Lett.}, vol.~25, no.~11, pp. 3699--3703, 2021.

\bibitem{wu2019towards}
Q.~Wu and R.~Zhang, ``Towards smart and reconfigurable environment:
  {I}ntelligent reflecting surface aided wireless network,'' \emph{{IEEE}
  Commun. Mag.}, vol.~58, no.~1, pp. 106--112, 2019.

\bibitem{zhang2020capacity}
S.~Zhang and R.~Zhang, ``Capacity characterization for intelligent reflecting
  surface aided {MIMO} communication,'' \emph{{IEEE} J. Sel. Areas Commun.},
  vol.~38, no.~8, pp. 1823--1838, 2020.

\bibitem{landsberg2017low}
N.~Landsberg and E.~Socher, ``{A low-power 28-nm CMOS FD-SOI reflection
  amplifier for an active F-band reflectarray},'' \emph{IEEE Trans. Microw.
  Theory Techn.}, vol.~65, no.~10, pp. 3910--3921, 2017.

\bibitem{QingqingJSAC18}
Q.~Wu, J.~Xu, and R.~Zhang, ``Capacity characterization of {UAV}-enabled
  two-user broadcast channel,'' \emph{IEEE J. Sel. Areas Commun.}, vol.~36,
  no.~9, pp. 1955--1971, 2018.

\bibitem{bharadia2014full}
D.~Bharadia and S.~Katti, ``{Full duplex {MIMO} radios},'' in \emph{11th USENIX
  Symp. Netw. Syst. Design Implement. (NSDI)}, 2014, pp. 359--372.

\bibitem{malik2018optimal}
R.~Malik and M.~Vu, ``Optimal transmission using a self-sustained relay in a
  full-duplex {MIMO} system,'' \emph{{IEEE} J. Sel. Areas Commun.}, vol.~37,
  no.~2, pp. 374--390, 2018.

\bibitem{nguyen2018energy}
K.-G. Nguyen, Q.-D. Vu, L.-N. Tran, and M.~Juntti, ``Energy efficiency fairness
  for multi-pair wireless-powered relaying systems,'' \emph{{IEEE} J. Sel.
  Areas Commun.}, vol.~37, no.~2, pp. 357--373, 2018.

\bibitem{beck2010sequential}
A.~Beck, A.~Ben-Tal, and L.~Tetruashvili, ``A sequential parametric convex
  approximation method with applications to nonconvex truss topology design
  problems,'' \emph{Journal of Global Optimization}, vol.~47, no.~1, pp.
  29--51, 2010.

\bibitem{ge2020joint}
L.~Ge, P.~Dong, H.~Zhang, J.-B. Wang, and X.~You, ``Joint beamforming and
  trajectory optimization for intelligent reflecting surfaces-assisted {UAV}
  communications,'' \emph{{IEEE} Access}, vol.~8, pp. 78\,702--78\,712, 2020.

\end{thebibliography}
	\endgroup

\end{document}